\documentclass[10pt]{article}
 
 \usepackage[title,titletoc,toc]{appendix}
\usepackage[margin=1.0in]{geometry} 
\usepackage{ytableau}
\usepackage{color}
\usepackage{fullpage}
\usepackage{hyperref} 
\definecolor{darkred}{rgb}{0.8,0,0}
\definecolor{darkgreen}{rgb}{.0,.8,.0}
\hypersetup{ colorlinks=true, linkcolor=blue, citecolor=red } 
\usepackage{amsmath, amsthm, amssymb,amsfonts}
\usepackage{epsfig}
\usepackage{graphicx}
\newcommand{\tkim}{\tilde k_{1\mu}}
\newcommand{\tkin}{\tilde k_{1\nu}}
\newcommand{\tkir}{\tilde k_{1\rho}}
\newcommand{\tkis}{\tilde k_{1\sigma}}
\newcommand{\n}{\nabla}
\newcommand{\s}{\sigma }                                              
\newcommand{\SI}{\Sigma }

\newcommand{\hf}{\frac{1}{2}}

\newcommand{\nb}{{\bar {n}}}
\newcommand{\mb}{{\bar {m}}}
\newcommand{\xn}{x_{n}}
\newcommand{\xnb}{{\bar x _{n}}}

\newcommand{\xm}{x_{m}}
\newcommand{\xmb}{\bar x _{m}}

\newcommand{\e}{e^{i k_{0}Y}}                                      

\newcommand{\kim}{ k_{1\mu}}                                      
\newcommand{\kom}{ k_{0\mu}}                                      
\newcommand{\ki}{ k_{1}}
\newcommand{\yi}{ Y_{1}}
\newcommand{\yib}{ Y_{\bar 1}}  
\newcommand{\yn}{ Y_{n}}                                             
 
\newcommand{\kn}{ k_{n}}

\newcommand{\km}{ k_{m}}

\newcommand{\kt}{ k_{2}}                                             
\newcommand{\ko}{ k_{0}}                                             
\newcommand{\yim}{ Y_{1}^{\mu}}                                      
\newcommand{\yin}{ Y_{1}^{\nu}}                                      
\newcommand{\kin}{ k_{1\nu}}  
\newcommand{\kir}{ k_{1\rho}} 
\newcommand{\kis}{ k_{1\sigma}}                                    
\newcommand{\kon}{ k_{0\nu}}
\newcommand{\kor}{ k_{0\rho}}
\newcommand{\kos}{ k_{0\sigma}}                                      
\newcommand{\ktm}{ k_{2\mu}}   
 \newcommand{\ktn}{ k_{2\nu}}                                   
\newcommand{\ytm}{ Y_{2}^{\mu}}                                      
                                                              
\newcommand{\lpp}{\mbox {$e^{i\int _{c} \alpha (t)                             
k(t) \partial _{z} X(z+t) dt +ik_{0}X}$}}

\newcommand{\gvk}{ e^{i\sum _{n }k_{n}Y_{n}}}

\newcommand{\ddXz}{\frac{\delta}{\delta X(z)}}
\newcommand{\ddXzp}{\frac{\delta}{\delta X(z')}}

\newcommand{\p}{\partial}                                           
\newcommand{\pp}{\partial ^{2}} 
\newcommand{\ppp}{\partial ^{3}}

\newcommand{\li}{ \lambda_{1}} 
\newcommand{\lt}{ \lambda_{2}} 
\newcommand{\lib}{ \lambda_{\bar 1}}                                    
\newcommand{\ltb}{ \lambda_{\bar 2}}                                    
\newcommand{\eps}{ \epsilon}                                        
\newcommand{\al}{\alpha }                                             
 
\newcommand{\tY}{\tilde Y}                                 
\newcommand{\lan}{\langle}
\newcommand{\ran}{\rangle}

\newcommand{\zb}{{\bar{z}}}                                             
\newcommand{\tb}{\mbox{$\bar{t}$}}

\newcommand{\kimb}{\mbox {$ {k_{\bar1\mu}}$}}                               
\newcommand{\kinb}{\mbox {$ {k_{\bar1\nu}}$}}  
\newcommand{\kirb}{\mbox {$ {k_{\bar1\rho}}$}}                               
\newcommand{\kisb}{\mbox {$ {k_{\bar1\sigma}}$}}

\newcommand{\kib}{\mbox {$ {k_{\bar1}}$}}                                      
  
\newcommand{\ktrb}{\mbox {$ {k_{\bar 2\rho}}$}}                                      
\newcommand{\qt}{ q_{2}}                                             
\newcommand{\qi}{\mbox {$ q_{1}$}}                                             
\newcommand{\qtb}{\mbox {$ \bar{q_{2}}$}}                                      
\newcommand{\qib}{\mbox {$ \bar{q_{1}}$}}                                      
\newcommand{\qo}{ q_{0}}

\newcommand{\la}{ \lambda }                                           
\newcommand{\be}{\begin{equation}}                                             
\newcommand{\br}{\begin{eqnarray}}                                             
\newcommand{\ee}{\end{equation}}                                               
\newcommand{\er}{\end{eqnarray}}                                               
\newcommand{\eln}{\mbox {$ e^{\sum _{n}\lambda _{-n}L_{+n}}$}}

\begin{document}
\title{
\hfill\parbox{4cm}{\normalsize IMSc/2015/08/06\\
}\\
\vspace{2cm}
Exact Renormalization Group and Loop Variables: A Background Independent Approach to String Theory.
}
\author{B. Sathiapalan\\ {\em                                                  
Institute of Mathematical Sciences}\\{\em Taramani                     
}\\{\em Chennai, India 600113}}                                     
\maketitle                                                                     
\begin{abstract}   
This paper is a self contained review of the Loop Variable approach to string theory. The Exact Renormalization Group is applied to a world sheet theory describing string propagation in a general background
involving both massless and massive modes. This gives interacting equations of motion for the modes of the string. Loop variable techniques
are used to obtain gauge invariant equations. Since this method is not tied to flat space time or any particular background metric,
it is manifestly background independent. The technique can be applied to both open and closed strings. Thus gauge invariant and generally
covariant interacting equations of motion can be written for massive higher spin fields in arbitrary backgrounds.  Some explicit examples are given. 
 \end{abstract}  
 
\newpage 
\tableofcontents 
                                                             
\newpage                                                                       
\section{Introduction} 

This is a review article on the Loop Variable (LV) approach to obtaining equations of motion (EOM) for the various modes of the string. The LV method  is based
on the old idea \cite{L,CDMP,AS1,FT1,CG,DS,HLP,BM,KPP,BGKP,T} that requiring conformal invariance of the Polyakov action describing strings propagating in a background
gives EOM for the background fields. From the point of view of the Polyakov action, these background fields are generalized coupling constants and their EOM are just generalized $\beta$ functions. This technique was first applied to massless fields in the closed string where Einstein's equations were obtained. Similarly for open strings Maxwell's equations are obtained to lowest order in $\al '$. At higher orders one obtains Dirac-Born-Infeld type generalization of Maxwell's action\cite{FT}. The vertex operators describing massless gauge fields are marginal operators near zero momentum. So the beta functions do indeed give low energy equations. For a massive field such as the tachyon, vertex operators become marginal for non zero momentum (near the mass shell). The beta function calculation is a little more complicated \cite{DS} for massive modes. A related proper time formalism was developed in \cite{BSPT}.   It can be shown \cite{Poly,BSPT,BSRev,BSZ} that to all orders the equation of motion so obtained is equivalent to the S-matrix for on shell tachyons.\footnote{There is a technicality here: the EOM is proportional to the beta function, the proportionality factor being the Zamolodchikov metric \cite{Poly,Zam,BSPT,ACNY,BSZ}.} When we get to the higher spin massive modes the naive method breaks down. This is because the equations obtained by this method are not gauge invariant. 

The problem of obtaining gauge invariant EOM  is addressed by the LV formalism. A key idea in this formalism is borrowed from
BRST string field theory where it was shown that including the extra fields needed for a gauge invariant description of string theory requires
an extra world sheet field - the bosonized ghost coordinate. In the LV formalism it is found that the theory looks massless in one extra dimension and mass is obtained as the extra momentum component during dimensional reduction.  The other key ingredient is the introduction of an infinite number of parameters - a generalization of the proper time - to parametrize the space of gauge transformations. These extra proper time parameters are reminiscent of the time parameters in the KP hierarchy. There too the group $Diff(S^1)$, parametrized by an infinite number of variables plays an important role. A third ingredient is that all the vertex operators are collected together in one non local object - the loop variable. Thus unlike in string field theory where one expands in oscillators, here the expansion is in terms of vertex operators. All these ingredients put together give us a way of writing a very general two dimensional field theory. Instead of calculating the beta function for marginal operators as one does in a continuum field theory, here we apply the Exact Renormalization Group (ERG)
transformation as first defined by Wilson \cite{WK,W,P,W2}, on this cutoff field theory. The (infinite set of)  equations are found to be gauge invariant and describe all the massive fields at the interacting level. The exact equations are quadratic in the fields. If, for instance, one solves for the massive modes in terms of massless modes one obtains  low energy equations which are non polynomial in massless fields. Thus for the graviton  Einstein's equation is obtained.

The free gauge invariant equations for open strings were first obtained by this method in \cite{BSLV}. They describe massless particles, and after dimensional reduction with mass, massive particles. The problem of writing down
gauge invariant (free) equations for massless/massive tensor fields with arbitrary symmetry is thus solved. \footnote{For symmetric tensors it was solved many years back in \cite{F,SH}. } There is another generalization that is also possible: One can do all this in a background curved space time. The requirement is that not only should the equations be gauge invariant but also generally covariant. It turns out that this can also be done - but only {\em after} dimensional reduction with mass. 

The gauge transformations have a particularly simple form in this method. The loop variable is parametrized by $k_\mu(t)$ a generalized momentum coordinate:
\[
k_\mu(t) = \kom + {\kim \over t} + {\ktm\over t^2} +...{k_{n\mu}\over t^n}+...
\]
Here $\kom$ is the usual space time momentum.
The gauge transformations have the simple rescaling form
\be     \label{GT}
k_\mu(t)\to k_\mu(t) \la (t)
\ee
where $\la(t)$ contains the gauge parameters
\be
\la(t) = \la_0 + {\la_1\over t} +....+{\la_n \over t^n}+...
\ee
This suggests that gauge transformation in string theory has a geometric space time interpretations as some kind of scale transformations.
In \cite{BSLV} it was speculated that the underlying gauge principle in string theory, generalizing the coordinate invariance of Einstein's theory of gravity,  is some generalized renormalization group elevated to a gauge symmetry.

An approach to the interacting open string was developed in \cite{BSLV0,BSRev} where the interacting theory was recast as a free theory where the string was, so to speak, thickened to a band with an extra parameter. When the product of two vertex operators had the same value of this parameter it was interpreted as a higher mode vertex operator, and when the value is different, it is interpreted as an interaction term between two fields. The method gave very easily a gauge invariant interacting theory. While this was elegant from the world sheet viewpoint, from the space time perspective, the equations were very complicated. 

A more satisfactory interacting theory is obtained by applying ERG transformation \cite{BSERGopen1,BSERGopen2}. The ERG is a quadratic equation. Thus in this formalism the string field equations are automatically quadratic.  This is natural in string theory because the basic interaction is a cubic one - splitting one string into two or joining of two strings to form one. This interacting theory, for open strings, has an interesting property: unlike in string field theory, the interactions do not modify the gauge transformation rule of the fields. The interactions are written in terms of gauge invariant ``field strengths". In this sense even though the theory is interacting, it looks Abelian - when the gauge group is $U(1)$. If Chan-Paton factors are added to make the group non Abelian, then the gauge transformation rule is indeed modified. In contrast BRST string field theory \cite{SZ,WS,Wi2} looks non Abelian even when the gauge group is U(1) although there are hints that field redefinitions can modify this feature \cite{Justin} . 

The story gets a little more involved for closed strings \cite{BSERGclosed1,BSERGclose d_{2},BSERGclosed3}. This is due to the presence of the massless graviton. At the free level things work out as they should - one obtains linearized equations for the graviton with the usual ``Abelian" gauge transformation:
\[
\delta h_{\mu\nu}= \p_{(\mu} \eps_{\nu)}
\]

For the interactions, if one applies the procedure followed for open strings, one finds that everything works out correctly for a {\em massive} graviton! The gauge invariant ``field strength" 
can be written down and, in terms of them, interactions also. However this field strength is not gauge invariant if the graviton is massless.
It turns out that the resolution of this problem  is to modify the gauge transformation law to include a shift of the coordinate $X^\mu$. This relates the gauge transformation to coordinate transformations (as it should in general relativity (GR)) and modifies the gauge transformation rule for the metric fluctuation $h_{\mu\nu}$:
\be
\delta h_{\mu\nu}= \p_{(\mu} \eps_{\nu)}+  \eps ^\la h_{\mu \nu,\la}+\eps ^\la_{~,\mu}h_{\la \nu}+\eps^\la_{~,\nu}h_{\mu\la}
\ee
The tensorial rotations are the result of including $\delta X^\mu = -\eps^\mu$ in the gauge transformation. (Note: $\eps_\mu \equiv \eta_{\mu\nu}\eps^\nu$.) Thus for closed strings, interactions do indeed modify the gauge transformations and give them a non-Abelian structure.

One can then introduce a background metric and make the theory invariant under background coordinate invariance i.e. a symmetry where not only the physical metric but also the background metric transforms. This is analogous to what happens in the background field formalism for non Abelian gauge theories. This requires that
all other terms in the world sheet action be modified so that the theory is  coordinate invariant. 

At this point there are two options: If the background metric is unrelated to the physical metric, then we have to ensure that somehow the action is further modified so that there is no net dependence on the background metric. In order to achieve this  some field redefinitions of the massive modes have to be performed. However if we let the background metric be the same as the physical metric then the theory is correct as it stands. 

Once this covariantization is done, we have to study again whether there is a clash between gauge invariance and general coordinate invariance. It turns out that there is a clash for the higher spin fields if we follow the naive procedure. However, when the higher spin fields are massive, it is possible to get around this. There is a systematic procedure for modifying the equations of motion by adding higher derivative terms involving the Stuckelberg fields
coupled to the curvature tensor such that gauge symmetry is preserved. These terms are not there in flat space. Also since the Stuckelberg fields can be set to zero by a gauge transformation, the physical fields continue to have two derivative propagation equations, even in curved space time.

In summary, this  procedure gives gauge invariant and general coordinate equations for all the massive modes of the string. This formalism is thus ``background independent". Another approach to background independent formalism for string theory is described in \cite{Wi} and developed further in \cite{LW,Sh,KMM}.

There are some open questions as well:

Whether these
equations are physically equivalent to those of BRST string field theory has not been checked or investigated in detail. However there is reason to believe that they are equivalent. Being gauge invariant one can always fix gauge and obtain world sheet theory in the old covariant formalism, which is the Polyakov action in conformal gauge without any ghosts. Also imposing conformal (Weyl) invariance at the free level is equivalent to the physical state Virasoro constraints. So the theory reduces exactly to the old covariant formalism and the interacting beta function constraints should be equivalent to the on shell S-matrix, by the same arguments that  worked for the tachyon \cite{BSOC}. 

A related issue that needs to be resolved is about the world sheet action for the extra coordinate. In obtaining the free equations of motion
for the string fields, only the coincident two point function of this field is needed. However in calculating the S matrix, if one wants to reproduce the S matrix of the old covariant formalism, this coordinate should not contribute at all. This requires that the Green function vanish everywhere except at coincident points. This is a little unusual - it corresponds to an infinitely massive world sheet field. This needs to be understood better.

This paper is a self contained review of this approach and is organized as follows: Section 2 describes the beta function approach and explains the problems involved in obtaining gauge invariant equations. Section 3 describes the basic loop variable and explains how one obtains the free gauge invariant equations for the open string. Section 4 describes dimensional reduction. Section 5 reviews the ERG and  applies it to open strings in flat space. Section 6 gives a prescription for a consistent  map from loop variables to space time fields in curved space time in such a way that gauge invariance is preserved.  Section 7 describes closed strings in arbitrary backgrounds and gives the gauge invariant formulation. Section 8 describes the connection with old covariant formalism for string theory.  Section 9 contains conclusions.

\section{Free Equations for Open Strings Fields}
\setcounter{equation}{0}

In this section we explain the RG method with some examples and explain the issues. We start with the Polyakov action modified by boundary terms describing an open string background.

\be      \label{Poly}
S~ =~ {1\over4\pi \alpha'}\int _\Gamma d^2\sigma \{\partial ^\alpha X^\mu 
 \partial _\alpha X_\mu \}~+~
\int _{ \partial \Gamma} dx  \sum _i g^i V_i(x)
\ee

Here $\Gamma$ is the UHP plane (or the unit disc) and $\p \Gamma$ is the real axis (or the unit circle).
The two point function for points on the real axis is obeying Neumann boundary conditions is
\[
\lan X(z) X(w)\ran = 2\alpha' ln(z-w) 
\]
This will be regularized when $z=w$ to 
\[
\lan X^\mu(z) X^\nu(z) \ran= -2\eta^{\mu\nu} \alpha ' (ln (a) + \sigma (z))
\]
For the open string the vertex operators are all along the real axis where $z=\bar z$. So the chiral field $X(z)$ has all the oscillators
for the open string and on the real axis the vertex operators corresponding to derivatives of $X(z)$. 
The cutoff is taken to be $ae^\sigma$ where $\sigma$ is the Liouville mode.
This simple regularization prescription is all one needs for the free theory. When we write down the full ERG a more well defined regularization will be required. Even in that case, the exact form of the regulator is immaterial. Different choices correspond to different schemes and the idea of universality is that the physics of the continuum does not depend on these details. In string theory this translates
to the statement that the S-matrix will not be affected. Thus different world sheet regularization schemes must correspond to field redefinition
 of the space-time fields.
\subsection{Vertex Operators and $\beta$ functions}

Let us consider some of the simplest vertex operators in turn.

\subsubsection{Tachyon}

The vertex operator for the tachyon is just $e^{ik_0.X(z)}$. Thus we add to the action
\[
\Delta S = \int dk_0~ \phi(k_0) \int _{\p \Gamma}dz ~{1\over a e^\sigma} e^{ik_0.X(z)}
\]
We can write the vertex operator as a normal ordered operator:
\[
e^{ik_0.X(z)}=e^{-{k_0^2\over 2}\lan X(z) X(z)\ran}:e^{ik_0.X}:= e^{\alpha' k_0^2 (ln(a)+\sigma)}:e^{ik_0.X}:
\]
Thus
\[
\Delta S = \int dk_0 ~ \phi(k_0) \int _{\p \Gamma}dz ~{1\over a e^\sigma} e^{\alpha' k_0^2 (ln(a)+\sigma)}:e^{ik_0.X}:
\]
Requiring either, ${d\over d ln (a)} \Delta S=0$ (vanishing $\beta$ function) or more precisely $({d\over d\sigma}\Delta S)|_{\sigma=0}=0$
(independence from Liouville mode  i.e. Weyl invariance) gives the condition:
\be	\label{tachyonfree}
(\alpha' k_0^2 -1)\phi(k_0) =0
\ee

Note that this is the same as the Virasoro condition $[L_0, V(z)]=0$.

\subsubsection{Vector}

The question of gauge invariance arises for the vector. The vertex operator is $\kim \p_z X^\mu e^{ik_0.X(z)}$. One can think of $\kim$ as the polarization of the gauge field.  But it is more useful to proceed as follows:
In the world sheet action we add the background term 
\[
\int _{\p \Gamma} dz A_\mu (X(z)) \p_z X^\mu(z)=\int dk_0 \int dz A_\mu(\ko)e^{i\ko.X(z)}\p_zX^\mu (z)
\]

It is convenient to write $A_\mu$ as a moment:
\be   \label{psi}
A_\mu(\ko)=\int [\prod_{\nu=0}^D d\kin ] \kim \Psi[\kon,\kin, A_\nu(\ko)]
\ee
and think of $\kim$ as a generalized momentum, dual to $A_\mu$. The ``wave function" $\Psi$ has the information about the
momentum dependence of $A_\mu(\ko)$. 
It is also convenient to write the vertex operator as $e^{ik_0.X(z) + i\ki .\p_z X(z)}$ remembering to keep only terms linear in $\ki$. This can be written in terms of a normal ordered vertex operator with the Liouville mode dependence being made explicit. 
\[
e^{ik_0.X(z) + i\ki .\p_z X(z)}=e^{-\hf (\ko.\ko \lan X(z) X(z)\ran + 2 \ko. \ki \lan X(z) \p_z X(z) \ran + \ki.\ki \lan \p_z X(z) \p_z X(z)\ran)}:e^{ik_0.X(z) + i\ki .\p_z X(z)}:
\]
In our case we need keep only terms linear in $\ki$. This gives
\[
e^{-\hf \ko.\ko \lan X(z) X(z)\ran} [i:\kim \p_z X^\mu e^{ik_0.X(z)}: - \ko. \ki \lan X(z) \p_z X(z) \ran e^{i\ko.X(z)}]
\]
\[
=e^{-\hf \ko.\ko \sigma(z)} [i:\kim \p_z X^\mu e^{ik_0.X(z)}: - \ko. \ki \hf\p_z\sigma(z) e^{i\ko.X(z)}]
\]
The coefficient of $\sigma$ gives the $L_0$ condition $k_0^2=0$ and the coefficient of $\p_z \sigma$ gives the condition $[L_1, V(z)]=0$. However it is possible to combine the two as follows:
\[
{\delta\over \delta \sigma (z)}\int dz ~  e^{-\hf \ko.\ko \sigma(z)} [i:\kim \p_z X^\mu e^{ik_0.X(z)}: - \ko. \ki \hf\p_z\sigma(z) e^{i\ko.X(z)}]
=0
\]
\be		\label{vector}
=[ \ko.\ko i\kim-\ko. \ki i\kom]  :\p_z X^\mu e^{ik_0.X(z)}:=0
\ee
We have integrated by parts the $z$ derivative. This is equivalent to the freedom of adding total derivatives to the action. The net result is
a gauge invariant equation: If we replace the polarization $\kim$ by a longitudinal $\la_1\kom$ the equation is invariant. 
\be	\label{u1}
\kim \to \kim + \la_1\kom
\ee
If we replace $\kim$
by $A_\mu$ we get Maxwell's equation with gauge invariance 
\[
A_\mu(\ko)\to A_\mu (\ko)+ \kom \Lambda (\ko)
\] where $\Lambda$ is a gauge parameter which can be related to $\la_1$ if  the wave function is expanded to include the gauge parameters also. Thus let $ \Psi[\kon,\kin, A_\nu(\ko),\la _1,\Lambda(\ko)]$ be the wave function satisfying additionally 

\be \label{psi1}
\int [d\kin d\la_1]~ \la_1\Psi[\kon,\kin,\la_1, A_\nu(\ko),\Lambda(\ko)]=\Lambda(\ko)
\ee
Thus in all equations involving the generalized momenta $\kim$, the integral over these momenta and $\Psi$ can be understood, and we can convert it to field equations. Schematically we write
\[
\lan \la_1\ran = \Lambda(\ko) ~;~~\lan i\kim \ran = A_\mu(\ko)
\]
Thus
\[
\lan  \ko.\ko \kim-\ko. \ki \kom \ran = \ko^2 A_\mu - \kom \ko.A 
\]
which is Maxwell's equation in momentum space. 

The gauge invariance of this equation, as explained above corresponds to the freedom to add total derivatives to the world sheet action.
This is also the invariance associated with $L_{-1}$ of the Virasoro group. 

\subsubsection{Spin2}
At spin 2 one expects two vertex operators:$ \p_z X^\mu \p_z X^\nu e^{i\ko.X(z)}$ and $ \p_z^2 X^\mu e^{i\ko.X(z)}$. Thus one can add to the action
\[
\Delta S = \int dz~ [-\hf S_{\mu\nu}(X(z))\p_z X^\mu \p_z X^\nu + S_{2\mu}(X(z))\p_z^2 X^\mu]
\]

There are two problems:

For spin 2 and higher there should be symmetries associated with $L_{-1}$, $L_{-2}$ and higher. $L_{-1}$ symmetries correspond to adding total derivatives and is manifest. However there is no such freedom manifest in the world sheet action for $L_{-2}$ or higher. Thus one does not expect the equation obtained by the same method to have any of the required higher gauge symmetries associated with massive modes.

Furthermore if one applies the Weyl invariance constraint on  vertex operators such as $\pp X^\mu e^{i\ko.X(z)}$ one obtains
Liouville mode dependent terms such as $\kom \pp \sigma e^{i\ko.X(z)}$. Integrating by parts one obtains a contribution to the equation 
involving three powers of momenta. This does not lead to an acceptable equation of motion, which should be quadratic in derivatives. This problem gets worse at higher levels.

This the straightforward application of the vanishing $\beta$ function condition does not lead to a satisfactory equation of motion. In the next section we introduce the loop variable to solve these problems at the free level.

\section{Loop Variable for the Open String}
\setcounter{equation}{0}

\subsection{Loop Variable as a Collection of Vertex Operators}
We first describe how all the open string vertex operators can be collected in a loop variable.
Consider the following loop variable \footnote{The integral is actually$ \int_c {dt\over 2\pi i}$. We suppress the $2\pi i$ in all the equations. We use $t$ (or $s$) always to parametrize the loop} where $c$ refers to a circle about the point $z$:
\be     \label{L}
e^{i \ko  X(z) + 
i  \int _c ~dt ~ak(t) \p _z X(z+at)}
\ee
with
\[
k(t) ~=~ k_0 ~+~ {\ki \over t} ~+~ {\kt \over t^2}~+~ ....
\]
$a$ is a short distance cutoff and is useful for keeping track of the dimension of operators. When we Taylor expand the exponential in a power series we get the following terms:
\be  
e^{i \ko  X(z) + 
i  \int _c ~dt ~ak(t) \p _z X(z+at)}= e^{i \ko  X(z)}[1+ i \kim a\p _z X^\mu - \hf \kim \kin a^2\p _z X^\mu \p _z X^\nu + i\ktm a^2\pp X^\mu +...]
\ee  

We generalize $\Psi$ in the last section to be $\Psi[\ko,\ki,\kt,...,\kn,...]$ and define
\be   \label{psi2}
\lan O \ran \equiv \prod_{n=1,\infty}\int [d\kn d\la_n]O \Psi[\ko,\ki,\kt,...,\kn,...]
\ee
generalizing what was done for the first few levels in Section 2. Thus for instance
\br
\lan \kim \kin \ran &=& S_{\mu\nu}(\ko) \nonumber \\
\lan \ktm \ran &=& S_{2\mu}(\ko)
\er

\subsection{Gauge Invariant Formulation}
\subsubsection{Extra variables to parametrize gauge transformations}

The loop variable is generalized to \cite{BSLV} (We have set $a=1$ below)
\be
\lpp
\ee
$\al (t)$ can be thought of as an ``einbein" to make the loop variable reparametrization invariant in $t$. But we will not need this interpretation here. 
Let us assume the following Laurent expansion:
\be
\al (t) ~=~ 1 ~+~ {\al _1 \over t}~+~{\al _2 \over t^2} ~+~ {\al _3 \over t^3}+...
\ee
We Taylor expand $X(z+t)$ and Laurent expand $k(t),\al (t)$. It is useful to define the following combinations:

\br
Y &~=~& X ~+~ \al _1 \p _z X ~+~ \al _2 \p_z^2 X ~+~
 \al _3 {\p _z ^3 X \over 2}~+~ ...~+~{\al _n \p _z^n X\over (n-1)!}~+~...\nonumber \\
&~=~& X ~+~ \sum _{n>0} \al _n \tY _n \\
Y_1 &~=~& \p _z X ~+~ \al _1 \p_z^2 X ~+~ \al _2 {\p _z ^3 X \over 2}~+~ ...~+~{\al _{n-1} \p _z^n X\over (n-1)!}~+~...\nonumber \\
...& & ...\nonumber \\
Y_m &~=~& {\p _z^m X\over (m-1)!} ~+~ \sum _{n > m}{\al _{n-m} \p _z^n X\over (n-1)!}\\
\er
We also define $\al _0 =1$ then the $>$ signs in the summations above can be replaced by $\ge$.

In terms of these $\yn$'s we have ($Y_0\equiv Y$)
\be
\lpp = \gvk
\ee

Let us now introduce $\xn$ by the following:
\be
\al (t) = \sum _{n\ge 0} \al _n t^{-n} = e^{\sum _{m\ge 0} t^{-m} x_m}
\ee

Thus 
\br
\al _1 &=& x_1  \nonumber \\
\al _2 &=& {x_1^2 \over 2} + x_2 \nonumber \\
\al _3 &=& {x_1^3 \over 3!} + x_1x_2 + x_3
\er

They satisfy the property,
\be
{\p \al _n \over \p x_m} = \al _{n-m} , ~~ n\ge m
\ee

Using this we see that 
\be  
Y_n = {\p Y\over \p x_n}
\ee
and also
\be \label{prop1}
{\pp Y\over\p \xm \p \xn}={\p Y\over x_{n+m}}
\ee

\subsubsection{Generalizing the Liouville mode}

We will work with the $Y$'s rather than $X$ and define  $\Sigma = \lan Y(z) Y(z) \ran$. (This is
equal to the Liouville mode $\sigma$ in coordinates where $\al (s) =1$.) We then impose ${\delta\over \delta \Sigma}=0$ on the normal ordered vertex operator.

We have for the coincident two point functions:
\br    \label{Sig}
\lan Y ~Y\ran &~=&~ \Sigma \nonumber \\
\lan Y_n ~Y\ran &~=&~ {1\over 2}{\p \Sigma \over \p x_n}   \nonumber \\
\lan Y_n ~Y_m \ran &~=&~  {1\over 2}({\pp \Sigma \over \p x_n \p x_m} - {\p \Sigma \over \p x_{n+m}})
\er

Using this the normal ordering gives the following $\Sigma$ dependence:
\br   \label{LV}
\lpp &=& \gvk  \nonumber \\
&=& exp \{\ko ^2 \Sigma + \sum _{n >0} \kn .\ko  {\p \Sigma \over \p x_n} +  \nonumber \\
& & \sum _{n,m >0}\kn .\km {1\over 2}({\pp \Sigma \over \p x_n \p x_m} - {\p \Sigma \over \p x_{n+m}})\} \nonumber \\
& & :\gvk :
\er

\subsubsection{Two derivative EOM}

The important thing to note is that there are never more than two derivatives acting on $\Sigma$ and in fact only one when
it multiplies $\kn.\ko$. Thus on integrating by parts we never get more than two powers of $\ko$.

For instance:
\[
{\delta \over \delta \Sigma} [
\kn .\km {1\over 2}({\pp \Sigma \over \p x_n \p x_m} - {\p \Sigma \over \p x_{n+m}})
] :e^{i\ko .Y}:
= :({1\over 2}i\kom i\kon Y_n^\mu Y_m ^\nu + i\kom  Y_{n+m}^\mu ) e^{i\ko .Y}:
\]

This solves the problem of getting an EOM that has only two derivatives. 
\subsubsection{Gauge transformations}

Actually introduction of $\al(t)$ also solves the problem of gauge invariance. For, in order to be allowed to integrate by parts one must integrate over the $\xn$. Thus we effectively integrate over
$\al (n)$: $\int {\cal D} \al(t)  \equiv \prod _n d\xn$. But in that case we can always multiply $\al(t)$ by $\la (t) \equiv e^{\sum _{n\ge0}y_n t^{-n}}$. This just translates $\xn \to \xn+y_n$ under which the measure is invariant. But this is the same as \eqref{GT}
\[
k_\mu(t)\to \la(t) k_\mu(t)
\]

This should be a symmetry. We will show that this is in fact a gauge symmetry of the theory, that includes and generalizes the $U(1)$ gauge symmetry of electromagnetism described in \eqref{u1}.

Thus we expand $\la (t)$ in inverse powers of $t$, with  $\la _0 =1$.
\[
\la (t) = \sum _n \la _n t^{-n}
\]
Then we can write (\ref{GT}) as 
\be    \label{GT1}
\kn \to \sum _{m=0}^{n} \la _m k_{n-m}
\ee

In order to interpret these equations in terms of space-time fields 
we need a generalization of (\eqref{psi2}). They have to be extended to include $\la$ as in \eqref{psi1}. Thus we
assume that the string wave-functional is also a functional of $\la (t)$: $\Psi[\ko,\ki,...,\kn,...,\la_1,\la_2,...,\la_n,...]$
 
Thus we let terms involving one $\la$ to be equal to gauge parameters:
\br   \label{La}
\lan \la _1 \ran &~=&~ \Lambda _1 (\ko )\nonumber \\
\lan \la _1 \kim \ran &~=&~ \Lambda _{11\mu} (\ko ) \nonumber \\
\lan \la _2 \ran &~=&~ \Lambda _2 (\ko )
\er
The gauge transformations (\eqref{GT}) thus become, after mapping to
space time fields by evaluating
$\lan .. \ran$:

\br
A_\mu (\ko )~&\to &~ A_\mu (\ko ) + \kom \Lambda _1 (\ko ) \nonumber \\
S_{2\mu} (\ko )~&\to &~ S_{2\mu } (\ko ) +\kom \Lambda _2 (\ko ) + \Lambda _{11\mu }(\ko) \nonumber \\
S_{11\mu \nu}(\ko) ~&\to &~ S_{11\mu \nu}(\ko) + k_{(\mu0} \Lambda _{11\nu )}
\er  

\subsubsection{Explanation of gauge invariance of EOM and tracelessness of gauge parameters}

The  mechanism for gauge invariance can be understood as follows:
As we have seen, a gauge transformation, which is a translation of $\xn$,  changes the
normal ordered loop variable by a total derivative in $\xn$ 
which doesn't affect the equation of motion.
More precisely the gauge variation of the loop variable is a term of the form
${d\over d\xn } [A(\Sigma ) B]$, where $B$ doesn't depend
on $\Sigma$. The coefficient of $\delta \Sigma$ is obtained as
\[
\int ~~\delta ({d\over d\xn } [A(\Sigma ) B]) =
\int ~~ ({d\over d\xn}( {\delta A\over \delta \Sigma } \delta \Sigma ) B +
  {\delta A\over \delta \Sigma } \delta \Sigma {dB\over d\xn })
\]
\[
=\int ~~[ - {\delta A\over \delta \Sigma } {dB\over d\xn } +
 {\delta A\over \delta \Sigma } {dB\over d\xn }]\delta \Sigma =0
\]
Here we have used an integration by parts.

When we actually do the calculation, the above argument does not quite work. This is because we
have simplified some of the terms using \eqref{prop1}. Let us look at the terms up to level 3 - the pattern easily generalizes.
The loop variable including $\Sigma$ dependence is:
\br   \label{LV3}
\lpp &=& \gvk  \nonumber \\
&=& exp \{\ko ^2 \Sigma +  \ki .\ko  {\p \Sigma \over \p x_1} +  
\ki .\ki {1\over 2}({\pp \Sigma \over \p x_1^2 } - {\p \Sigma \over \p x_{2}})
 + \kt.\ko {\p \Sigma \over \p x_{2}} +\nonumber \\
&& \ki .\kt ({\pp \Sigma \over \p x_1\p x_2 } - {\p \Sigma \over \p x_{3}})
 + k_3.\ko {\p \Sigma \over \p x_{3}}\}\nonumber \\
 &&:e^{i(\ko.Y +\ki.Y_1 +\kt Y_2 + k_3.Y_3)}: 
\er
Let us extract the terms involving $\la_1$ after performing a gauge transformation. One finds in the exponent:
\[
\la_1[\ko^2 {\p \Sigma \over \p x_1}+\ki.\ko({\pp \Sigma \over \p x_1^2 } - {\p \Sigma \over \p x_{2}})+\ki.\ko  {\p \Sigma \over \p x_{2}}]
\]
\[
+\la_1[(\ki.\ki+ \ko.\kt)({\pp \Sigma \over \p x_1\p x_2 } - {\p \Sigma \over \p x_{3}})+\kt.\ko {\p \Sigma \over \p x_{3}}]
\]
\[
=\la_1{\p \over \p x_1}[\ko^2 \Sigma + \ki.\ko {\p \Sigma \over \p x_1}+\kt.\ko{\p \Sigma \over \p x_{2}}]+\la_1\ki.\ki({\pp \Sigma \over \p x_1\p x_2 } - {\p \Sigma \over \p x_{3}})
\]
If we set $\la_1 \ki.\ki=0$ then the change in the loop variable is a total derivative in $x_1$. This will give a gauge invariant contribution to the EOM by the arguments given in the previous paragraph. Since it is in the exponent, the condition $\la_1 \ki.\ki=0$ actually means $\la_1 \ki.\ki (...)=0$ where the three dots indicate any other combinations of loop momenta $\kn$.  For the gauge parameters, this is a tracelessness condition, which is known to be required in higher spin theories. At higher levels this becomes $\la_n \km.k_p(....)=0$ for $m,p >0$.

\subsubsection{Spin 2 EOM}

Let us apply ${\delta \over \delta \Sigma} |_{\Sigma =0}=0$ to \eqref{LV3}. One gets very easily:
\[
[i\ko^2 \kt.Y_2 -i \ki.\ko \ki.Y_2- i\kt.\ko \ko.Y_2+ i\ki.\ki \ko.Y_2]\e 
\]
\be
+[\ki.\ko \ko.\yi \ki.\yi + \hf \ki.\ki (\ko.\yi)^2 -\hf \ko^2 (\ki.\yi)^2]\e=0
\ee
We have separated by square brackets the two vertex operators at level 2. It is very easy to check that the equations are invariant
under
\be	\label{gt1}
\ki \to \ki + \la_1 \ko~~~;~~~\kt\to \kt + \la_1 \ki + \la_2 \ko
\ee
Converting to space time fields (using $\lan ...\ran$ \eqref{psi2}) one obtains:
\br
\Box S_{2\mu} - \p^\nu S_{\nu\mu} - \p_\mu \p^\nu S_{2\nu} + \p_\mu S^\nu_\nu&=&0 \\
\hf \Box S_{\mu\nu} + \hf \p_\mu\p_\nu S^\rho_\rho - \p_\nu \p^\rho S_{\rho\mu}&=&0
\er

The gauge transformations are
\br
S_{2\mu}&\to& S_{2\mu}+ \kom \Lambda_2 + \Lambda_{11\mu} \nonumber \\
S_{\mu\nu}&\to& S_{\mu\nu}+ k_{0(\mu}\Lambda_{11\nu)}
\er

\subsubsection{Spin 3 EOM}
By the same procedure one finds the EOM for spin 3:
\br  
& &[-i\ko^2{(\ki.\yi)^3\over 3!} + i \ki.\ko \ko.\yi {(\ki.\yi)^2\over 2!} -{i\over 2}\ki.\ki (\ko.\yi)^2(\ki.\yi)]+\nonumber \\&&
[i \ko^2 k_3.Y_3 - i\ki.\ko \kt.Y_3 -i\kt.\ko \ki.Y_3 +i\ki.\ki \ki.Y_3 - ik_3.\ko \ko.Y_3 + 2i \kt.\ki \ko.Y_3]+\nonumber \\
&&[-\ko^2 \ki.\yi \kt.Y_2 + \ki.\ko \ko.\yi \kt.Y_2 +\ki.\ko\ki.Y_2 \ki.\yi+ \kt.\ko \ko.Y_2 \ki.\yi \nonumber \\
&& - \ki.\ki \ko.Y_2 \ki.\yi - \ki.\ki \ko.\yi \ki.Y_2 - \kt.\ki \ko.Y_2 \ko.\yi]=0
\er
We have collected in square brackets, vertex operators of a given type. Thus these are three separate equations.
They are invariant under \eqref{gt1} along with 
\[
k_3\to k_3+\la_1\kt + \la_2 \ki +\la_3 \ko
\]

Space time fields for Spin 3 are defined later.
\section{ Dimensional Reduction}
\setcounter{equation}{0}

The equations obtained above are gauge invariant and describe massless fields. The second equation for the spin 2 level is in fact  the free equation for a massless spin 2 tensor such as the graviton. However we are trying to describe the massive spin 2 and the field $S_{2\mu}$ is required for this. We do not have a mass term in any of the equations. 
This is because what we have done so far is not strictly a beta function calculation. We have not included the $\sigma$ dependence associated with the engineering (classical) dimensions of the vertex operators - only the anomalous dimension. Including the engineering dimension is difficult to do because we are working with $\Sigma$, which is a linear combination of $\sigma$ and its derivatives.

 Now
in the BRST formalism it was shown that the extra auxiliary and Stueckelberg fields necessary for a gauge invariant description of string theory are obtained from the oscillators of a bosonized ghost coordinate, minus the first oscillator mode. Motivated by this let us assume that
one of the coordinates $Y^D$ is this extra coordinate. Let us call it $\theta$ and the corresponding momentum $q(t)$. We then perform
a Kaluza-Klein dimensional reduction and let $\qo$, the internal momentum be the mass of the field. To match with string theory $\qo^2$ will have to be set equal to the engineering dimension of the vertex operator. \footnote{ This means that it is not a simple $S^1$ compactification, which would have required $\qo$ to be an integer. We need $\qo^2$ to be an integer to match the string spectrum.}So (in some appropriate units) we can set $\qo^2=1$. Thus the engineering dimension arises as the anomalous dimension associated with the extra dimension. 

In the BRST formalism some field redefinitions have to be performed to get the fields into a form that is obtainable by dimensional reduction of a massless higher dimensional theory. This redefinition works only in the critical dimension and with the correct string spectrum. Here we automatically get it in that form directly. Furthermore gauge invariance is not affected by dimensional reduction, so $\qo^2$ can have any value.  We can choose it to match the string spectrum. But we will keep it as $\qo^2$ till we are forced to some specific value. There is also no critical dimension so far. However it was shown in \cite{BSOC} that if one wants the gauge transformations and constraints to be in the standard form of string theory, field redefinitions need to be done, and can be done only in the critical dimension and with the right spectrum.

With these comments out of the way, let us proceed with the dimensional reduction.

\subsection{Dimensional Reduction of the free EOM: Spin 2, Spin 3}

We first give explicitly the details regarding dimensional reduction of the Spin 2 and Spin 3 system described above.

\subsubsection{Spin 2}

The {\em naive} field content is 
\br
\lan \kim \kin \ran &\equiv& S_{11,\mu\nu}\\
\lan \kim \qi \ran &\equiv& S_{11,\mu}\\
\lan \qi \qi \ran &\equiv& S_{11}\\
\lan \ktm \ran&\equiv& S_{2,\mu}\\
\lan q_2 \ran&\equiv& S_{2}
\er
The gauge parameters are then
\br
\lan \la_2 \ran &\equiv& \Lambda_2\\
\lan \la_1\kim \ran &\equiv& \Lambda_{11,\mu}\\
\lan \la_1\qi \ran &\equiv& \Lambda_{11}\\
\er

The field transformations are
\br
S_{2,\mu}&\to& S_{2\mu}+ \kom \Lambda_2 + \Lambda_{11\mu}  \\
S_{2}&\to& S_{2}+ \qo \Lambda_2 + \Lambda_{11} \\
S_{11,\mu\nu}&\to& S_{11,\mu\nu}+ k_{0(\mu}\Lambda_{11\nu)}\\
S_{11,\mu}&\to& S_{11,\mu }+ k_{0\mu}\Lambda_{11}+\qo \Lambda_{11\mu}\\
S_{11}&\to& S_{11}+ 2q_0\Lambda_{11}\\
\er

Thus $S_{11,\mu\nu},S_{11,\mu},S_{11}$ form a gauge invariant massive spin 2 and $S_{2\mu},S_2$ form a massive spin 1. One can gauge away $S_{11,\mu},S_{11}, S_2$ using $\Lambda_{11\mu},\Lambda_{11},\Lambda_2$ and get a gauge fixed covariant description of a massive spin 2 and massive spin 1. However this is not the correct field content for open strings at the first massive level.

{\bf Comparison with String Spectrum:}

 For a string in $D$ dimensions, the physical states (defined by light cone oscillators $\al _{-2}^i, \al_{-1}^i\al_{-1}^j$) are $O(D-2)$ tensors given by:
 \ydiagram{1} (4),  ~~\ydiagram{2} (10)
 
 They are combined in the $O(D-1)$ symmetric traceless tensor:
 \ydiagram{2} (15-1=14)
 
 These are the transverse components of a massive $O(D)$ tensor (for which $O(D-1)$ is the little group). Although $D=26$ for the bosonic string we use a smaller number for $D$, say $D=6$ to specify the size of the rep. Thus the numbers in brackets are the dimensions of the reps for $D=6$.  For a covariant description we keep the trace. For a gauge invariant description an additional vector and a scalar are also needed as we have seen above.
 
 \subsubsection{Q-rules for level  2}
 
 The resolution is to use the identifications (called Q-rules):
\br
\qi \qi &=& q_2 \qo \\ q_1\kim &=&q_0 \ktm \\ \li q_1 &=& \lt q_0 
\er
These can be used to get rid of $\qi$ from the equations. This is motivated by the BRST prescription \cite{SZ} of eliminating the first oscillator in the bosonized ghost field.
These identifications are made by requiring that the LHS and RHS transform the same way under gauge transformation. This ensures that
gauge invariance is maintained during the truncation of the field content.
Then $S_{2\mu}$ which becomes identified with $S_{11\mu}$ up to a factor of $\qo \neq 0$ gets gauged away when we gauge away $S_{11\mu5}$.  Also $S_{11}$ and $S_2$ get identified and can be gauged away. Thus we are left with just $S_{11\mu\nu}$ - which gives the field content for a covariant description of a massive symmetric tensor and reproduces the open string spectrum.

 One important point to note is that these Q-rules are consistent
with the idea of a higher dimensional origin. Namely, if we let $\mu=5$ in the second equation, we do get the first equation. For higher levels these $q$ rules are quite non trivial but, very surprisingly, they continue to be consistent with dimensional reduction.  

{\bf Massive spin 2 equation:}

We give the resulting equation for massive spin 2 field. 
\be
\hf (k_0^2 + q_0^2) \kim\kin - \hf \ko .\ki k_{1(\mu} k_{0\nu)} + \hf \kom \kon \ki .\ki -\hf q_0^2 k_{2(\mu}k_{0\nu )} + \hf \kom \kon q_2q_0=0
\ee

\subsubsection{Spin  3}

\[ \lan \kim \kin \kir \ran \equiv S_{111}^{\mu \nu \rho};~~~\lan \kim \kin q_1 \ran \equiv S_{111}^{\mu \nu};~~~\lan\kim q_1q_1 \ran \equiv S_{111}^{\mu};~~~~\lan q_1 q_1 q_1 \ran \equiv S_{111}\]
\[\lan  \ktm \kin \ran \equiv S_{21}^{\mu \nu};~~~\lan \kim q_2 \ran \equiv S_{12}^{\mu};~~~\lan  \ktn q_1\ran \equiv S_{21}^{ \nu};\]
\[~~~\lan k_3^\mu \ran \equiv S_3^\mu;~~~\lan q_3\ran \equiv S_3\]

The gauge transformations again follow an obvious pattern:
\[\delta S_{111}^{\mu \nu \rho}=\lan \li (k_0^{(\mu}\kin k_1^{\rho )}\ran \equiv k_0^{(\mu} \Lambda_{111}^{\nu \rho )}\] 
\[\delta S_{111}^{\mu \nu}= \lan \li (q_0 \kim \kin +q_1 k_0^{(\mu}k_1^{\nu )})\ran\equiv q_0 \Lambda_{111}^{\mu \nu}+k_0 ^{(\mu}\Lambda_{111}^{\nu )}\]
\[\delta S_{111}^\mu=\lan \li ( 2 q_1 q_0 \kim + q_1^2\kom)\ran \equiv 2q_0\Lambda_{111}^\mu+ k_0^\mu \Lambda_{111}\]
\[\delta S_{111}=3\lan \li q_1^2q_0\ran \equiv 3 q_0 \Lambda_{111}\]
\[ \delta S_{21}^{\mu \nu}=\lan \li ( \kon \ktm + \kim \kin ) + \lt \kom \kin \ran \equiv k_0^\nu \Lambda_{21}^\mu +\Lambda_{111}^{\mu \nu} + k_0^\mu \Lambda _{12}^\nu \]
\[ \delta S_{12}^\mu =\lan \li (\kom q_2 + q_1 \kim )+ \lt \kim q_0\ran \equiv k_0^\mu \Lambda_{21}+ \Lambda_{111}^\mu + \Lambda_{12}^\mu q_0\]
\[\delta S_{21}^\mu=\lan \li (\ktm q_0 + q_1\kim) + \lt \kom q_1 \ran\equiv \Lambda_{21}^\mu q_0 + \Lambda_{111}^\mu + k_0^\mu \Lambda _{12}\]
\[\delta S_3^\mu =\lan \la_3 \kom + \lt \kim + \li \ktm \ran \equiv k_0^\mu \Lambda_3 + \Lambda_{12}^\mu + \Lambda_{21}^\mu\]  

These describe a massive spin 3, spin 2 and spin 1, as obtained by dimensional reduction of a massless theory in one higher dimension.

Once again this is not the open string spectrum.

{\bf Comparison with string spectrum:}

 The physical states (light cone oscillators $\al_{-3}^i, \al_{-2}^i\al_{-1}^j,\al_{-1}^i\al_{-1}^j\al_{-1}^k$)  are \ydiagram{1} (4), ~~\ydiagram{1}$\otimes$\ydiagram{1} (= \ydiagram{2} (10)$\oplus$ \ydiagram{1,1}(6)) and \ydiagram{3} (20) for a total of 40 states.

 These can be combined into a symmetric traceless 3-tensor and an antisymmetric 2-tensor of $O(D-1)$  (the little group for a massive particle):
 \ydiagram{3} (35-5=30) , \ydiagram{1,1} (10)

Fully covariant description of a three tensor requires a traceless three tensor, a vector (which can be taken to be the trace of a traceful three tensor) and a scalar. Clearly a truncation is required. Once again we 
 specify the Q-rules at this level.
 
 \subsubsection{Q-rules for level 3}
 
 \br	\label{Q3}
 \lan q_1 \kim\kin\ran &=& \hf \lan k_2^{(\mu}k_1^{\nu)} q_0\ran = \hf S_{21}^{(\mu \nu)}q_0 \nonumber \\
 \lan q_1 q_1 \kim \ran &=& \lan k_3^\mu q_0^2 \ran = S_3^\mu q_0^2  \nonumber \\
\lan q_1 \ktm \ran &=& \lan 2 k_3^\mu q_0 - q_2 \kim \ran = 2S_3^\mu q_0 - S^\mu_{12} \nonumber \\
 \lan q_1q_2 q_0\ran &=&\lan q_3 q_0^2\ran =\lan q_1^3\ran = S_3q_0^2  \nonumber \\
\lan \li q_1 \kim \ran &=& \lan \hf \lt \kim q_0+ \hf \li \ktm q_0\ran = \hf(\Lambda_{12}^\mu + \Lambda_{21}^\mu )q_0  \nonumber \\
\lan \lt  q_1 \ran&=&  \lan 2 \la _3 q_0 -\li q_2 \ran = 2 \Lambda _3 q_0 - \Lambda _{21} \nonumber \\
 \lan \li q_1 q_1\ran &=& \lan \la _3 q_0^2\ran = \Lambda _3 q_0^2
 \er

 This results in some modifications in gauge transformations.
 \be   \label{GT3}
\delta (q_2 \kim)= ({3\over 2} \lt \kim + \hf \li \ktm )q_0 + \li q_2 \kom ~~,~~~\delta q_3 = 3 \la _3 q_0
\ee

The resulting truncated set of fields are given below:
 \br   \label{spin3truncated}
 \lan \kim \kin \kir \ran &=& S_{111\mu\nu\rho}\nonumber \\
 \lan \hf k_{2[\mu}k_{1\nu]} \ran &= &A_{21\mu\nu} \nonumber \\
 \lan \hf k_{2(\mu} k_{1\nu)}\qo = \kim \kin q_1 \ran &=& S_{21\mu\nu}\qo\nonumber \\
 \lan k_{3\mu} q_0^2= \kim q_1^2 =\hf (\kim q_2 + \ktm q_1)q_0 \ran &=& S_{3\mu}q_0^2\nonumber \\
 \lan \kim q_2 \ran &=& S_{12\mu}\nonumber \\
 \lan q_3 q_0^2=q_2 q_1 q_0 = q_1^3 \ran &=& S_3 q_0^2
 \er
 Note that symmetrization and anti-symmetrization are not normalized to 1. Hence the factor of $\hf$ in the definition of the  index tensors.
  Their gauge parameters and transformations are:
 
 \br
 \lan \li q_1 q_1 = \hf(\lt q_1 +\li q_2)q_0  = \lambda _3 q_0^2\ran &=& \Lambda _3 q_0^2\nonumber \\
 \lan \hf(\li q_2 - \lt q_1) \ran &=& \Lambda _Aq_0\nonumber \\
 \lan \li q_1 \kim = \hf(\lt \kim +\li \ktm ) q_0 \ran &=&\hf (\Lambda _{12\mu} + \Lambda _{21\mu})q_0 = q_0\Lambda _S\nonumber \\
 \lan \hf(\lt \kim -\li \ktm )\ran &=& \hf (\Lambda _{12\mu} - \Lambda _{21\mu})=\Lambda _A\nonumber \\
 \lan \li \kim \kin \ran &=& \Lambda _{111\mu\nu}
 \er
 \br
 \delta S_3 &=&3 \Lambda _3 \qo\nonumber \\
 \delta S_{3\mu} &=& 2 \Lambda _{S\mu} + \kom \Lambda _3\nonumber \\
 \delta (S_{3\mu} q_0 - S_{12\mu})\equiv \delta S_{A\mu}\qo &=& \Lambda _{A\mu}q_0 + \kom \Lambda _{A\mu} \qo\nonumber \\
 \delta S_{\mu\nu} &=& \Lambda _{111\mu\nu} + k_{0(\mu } \Lambda _{S\nu)}\nonumber \\
 \delta A_{\mu\nu}&=& k_{0[\mu}\Lambda _{A\nu]}\nonumber \\
 \delta S_{111\mu\nu\rho} &=& k_{0(\mu }\Lambda_{111\nu\rho)}
 \er

The tracelessness condition on the gauge parameter is $\la _1 \ki.\ki +\la_1 q_1q_1=0$. Thus 
\be
\Lambda_{111~\mu}^{~~~\mu}+\Lambda_3 \qo^2=0
\ee

This is the field content for a covariant and gauge invariant equation.
All fields other than the three index symmetric tensor and the antisymmetric two tensor can be set to zero by a choice of gauge. This thus agrees with the open string spectrum at level 3. The set of constraints, gauge transformation and connection with the old covariant (OC) formalism is described in Section 8.

{\bf Massive spin 3 EOM:}

We take the massless equation (3.2.36) and perform the dimensional reduction and apply the $q$-rules given above.
As an illustration we give below the equation corresponding to the vertex operator $Y_1^\mu Y_1^\nu Y_1^\rho$. (In our notation, symmetrization, indicated by curved brackets, is not normalized to 1 - just the sum of permutations.)

\be
-{1\over 3!}(\ko^2+\qo^2)\kim\kin\kir - {1\over 3!}(\ki.\ko k_{1(\mu}k_{1\nu}k_{1\rho )} + k_{2(\mu}\kin k_{0\rho)}\qo^2) -{1\over 12} (\ki.\ki k_{1(\mu}\kor k_{0\nu)} + \qo^2 k_{3(\mu} \kon k_{0\rho)})=0
\ee

\subsection{Q-rules for higher levels}

The procedure for obtaining the massless equations and subsequent naive dimensional reduction is the same at higher levels. The new ingredient that crops up at each level is the consistent truncation of the spectrum to match with open strings. This is done by getting rid
of the states that involve $\qi$. This is done in  a manner that preserves gauge invariance by working out the analog of the $q$-rules at higher levels. Though we have not attempted to find a systematic procedure that works at all levels, we have explicitly worked out the rules for level 4 and 5. It turns out to involve solving a number of sets of  {\em overdetermined} linear equations for some unknown coefficients. It is interesting that they have a consistent solution. Even more interesting is that they can be chosen to be consistent with the idea of dimensional reduction - a requirement that on the face of it seems completely independent. This is also an overdetermined set which turns out to have a solution. We give the results below. It would be interesting to explore this further and find some underlying pattern. 

\subsubsection{Q-rules for level 4}

The basic procedure in obtaining the Q-rules is as follows.  Start with the highest spin field at that level, where the Q-rule
is uniquely fixed by the symmetry of the indices. This, in turn, also implies a corresponding Q-rule for the gauge parameter.
For example, at level 4 we have $q_1 \kim \kin \kir $. The only possibility is to set it equal to $1/3 k_{2(\mu}\kin k_{1\rho)}$. The factor $1/3$ compensates for the three permutations. Quite generally we  choose the sum of the coefficients on the RHS to be 1.

Now consider the gauge transformation of the LHS: $q_1 \li k_{0(\mu}\kin k_{1\rho)} + \li q_0 \kim \kin \kir$. 
Matching the coefficient of $\kor$ on both sides gives immediately:

\[ {\cal Q}: q_1 \li \kim \kin \rightarrow 1/3(\lt \kim \kin + \li \ktm \kin + \li \kim \ktn )\] 

This is a general pattern. Once we write down a Q-rule for the fields with $n$ indices, this fixes some Q-rules for gauge parameters with $n-1$ indices. Thus for instance the Q-rule for the two index field \footnote{Note that the antisymmetric combination, $q_1 k_{2[\mu} k_{1\nu]}$, is uniquely fixed to be $q_0k_{3[\mu}k_{1\nu]}$}:
\[ {\cal Q}: q_1 k_{2(\mu} k_{1\nu)} \rightarrow {A\over 2} q_0 k_{3(\mu}k_{1\nu)} + B q_2 \kim \kin + C \ktm \ktn \]

We immediately get constraints on $A,B,C$ matching the gauge parameters on both sides: $A=6-4C, B=3C-4$. Also by comparing coefficients of $\kom$ we get Q-rules for gauge parameters (with one index, such as $ q_1 \lt \kim, q_1 \li \ktm$) in terms of $A,B,C$.  

This continues till we have the full set of Q-rules for level 4. The results for the remaining fields are given below:

\begin{eqnarray}
q_1 \ktm \kin & = & \hf \Big({A\over 2} q_0 k_{3(\mu}k_{1\nu)} + B q_2 \kim \kin + C q_0 \ktm \ktn + q_0k_{3[\mu}k_{1\nu]}\Big) \nonumber \\ \nonumber
q_1 ^2\kim \kin & = & q_0 \Big({A_2\over 2} q_0 k_{3(\mu}k_{1\nu)} + B_2 q_2 \kim \kin + C_2 q_0 \ktm \ktn \Big)\\ \nonumber
q_1 k_{3\mu}  & = &  \Big(A_1 q_0 k_{4\mu} + B_1 q_2 \ktm  + C_1 q_3 \kim  \Big)\\ \nonumber
q_1^2 k_{2\mu}  & = & q_0 \Big(A_3 q_0 k_{4\mu} + B_3 q_2 \ktm  + C_3 q_3 \kim  \Big)\\ \nonumber
q_1^3 k_{1\mu}  & = & q_0^2 \Big(A_4 q_0 k_{4\mu} + B_4 q_2 \ktm  + C_4 q_3 \kim  \Big)\\ \nonumber
q_1q_2 k_{1\mu}  & = & q_0 \Big(A_5 q_0 k_{4\mu} + B_5 q_2 \ktm  + C_5 q_3 \kim  \Big)\\ \nonumber
q_1q_3   & = & a_1q_4 q_0 + b_1 q_2^2\\ \nonumber
q_1^2q_2   & = & a_2q_4 q_0^2 + b_2q_0 q_2^2\\ 
q_1^4   & = & a_3q_4 q_0^3 + b_3q_0^2 q_2^2
\end{eqnarray}

The corresponding Q-rules for gauge transformations are: 
\begin{eqnarray}
q_1 \lt \kin & = & \hf[ (1+ {A\over 2})q_0  \la _3 \kin  + ({A\over 2}-1) q_0 \li k_{3\nu}  + C q_0\lt \ktn + B q_2 \li \kin ] \nonumber \\ \nonumber
q_1 \li \ktn & = & \hf[ (-1+ {A\over 2}) q_0\la _3 \kin  + ({A\over 2}+1)q_0 \li k_{3\nu}  + C q_0\lt \ktn + B q_2 \li \kin ]\\ \nonumber
q_1^2 \li \kin & = &   {A_2\over 2} q_0^2 (\la _3 \kin  +  \li k_3^\nu ) + C_2 q_0^2\lt \ktn + B_2 q_0q_2 \li \kin \\ \nonumber
q_1 \la_3 &=& A_1 q_0\la _4 + B_1 \lt q_2 + C_1 \li q_3\\ \nonumber
q_1^2 \la_2 &=& q_0(A_3 q_0\la _4 + B_3 \lt q_2 + C_3 \li q_3)\\ \nonumber
q_1^3 \li &=& q_0^2(A_4 q_0\la _4 + B_4 \lt q_2 + C_4 \li q_3)\\ 
q_1 q_2 \li &=& q_0(A_5 q_0\la _4 + B_5 \lt q_2 + C_5 \li q_3)
\end{eqnarray}

All the parameters turn out to be  fixed in terms of two, (which we take to be $C,B_2$) when we require consistency with gauge transformations.

The general two parameter solution is given below:
\[\{A=6-4C,~~~ B=3C-4\}\]
\[ \{A_1= {3(C-2)\over 2-3C},~~~B_1={6-5C\over 2-3C},~~~C_1={2-C\over 2-3C}\}\]\[\{A_2= {2-4B_2\over 3},~~~C_2= {1+B_2\over 3}\}\]
\[\{A_3= {3C+2B_2-6\over2-3C},~~~B_3={2(10-2B_2-9C)\over3(2-3C)},~~~C_3={2(2-B_2)\over 3(2-3C)}  \}\]\[\{A_4={2+6B_2-3C\over 2-3C} ,~~~B_4=-{4B_2\over 2-3C},~~~C_4=-{2B_2\over 2-3C}  \}\]
\[\{A_5=-{2B_2\over 2-3C},~~B_5={2-4B_2-3C\over3(2-3C)},~~C_5={2(2-B_2-3C)\over3(2-3C)}\}\]
\[\{a_1={3(2-C)\over2(1+B_2)},~~b_1={2B_2+3C-4\over 2(1+B_2)}\}\]
\[\{a_2={1-2B_2\over1+B_2},~~b_2={3B_2\over 1+B_2}\}\]
\be\{a_3={3C-5B_2-2\over1+B_2},~~b_3={3(1+2B_2-C)\over 1+B_2}\}\ee

{\bf Consistency with Dimensional Reduction}

We explain the issue of this consistency with an example. Consider the term $q_1 \kim \kin \kir$. According to the Q-rules this is equal to
\[{\cal Q}:q_1 \kim \kin \kir\rightarrow {1\over 3} (\ktm \kin \kir + k_{2\rho} \kim \kin + \ktn \kir \kim)\]  Now dimensionally reduce both terms, choosing $\rho$ to be $\theta$. If this dimensional reduction commutes with the Q-rule it should be true that
\[{\cal Q}: q_1^2 \kim \kin ={\cal Q}: {1\over 3} ( q_1 \ktm \kin + q_1 \ktn \kim + q_2 \kim \kin)\]
The two parameter family of Q-rules in fact gives:
\[q_1^2 \kim \kin =  q_0 \Big({A_2\over 2} q_0 k_3^{(\mu}k_1^{\nu)} + B_2 q_2 \kim \kin + C_2 \ktm \ktn \Big)\] with $A_2={2-4B_2\over 3}, C_2 = {1+B_2\over 3}$. Similarly
\[ q_1 \ktm \kin + q_1 \ktn \kim= \Big({A\over 2} q_0 k_3^{(\mu}k_1^{\nu)} + B q_2 \kim \kin + C \ktm \ktn \Big)\] with
$A=6-4C, B=-4+3C$. Requiring agreement fixes $C=1+B_2$, thus fixing one parameter.  Continuing this process
one more step by setting $\nu=\theta$ gives one more constraint and fixes $C=1, B_2=0$. Interestingly, all other constraints for all other terms are satisfied with this choice.

We give the final solution below:
\[
\{A=2,~~~B=-1,~~~C=1\}\]\[ \{A_1=3,~~~B_1=-1,~~C_1=-1\}\]
\[\{A_2={2\over 3},~~~B_2=0,~~~C_2={1\over 3}\}\]\[\{A_3=3,~~~B_3=-{2\over 3},~~~C_3=-{4\over 3}\}
\]
\[
\{A_4=1,~~~B_4=0,~~~C_4=0\}\]\[\{A_5=0,~~~B_5={1\over 3},~~~C_5={2\over 3}\}
\]
\[
\{a_1={3\over 2},~~b_1=-{1\over 2}\}\]\[ \{a_2=1,~~b_2=0\}\]\be\{a_3=1,~~b_3=0\}
\ee

\subsubsection{Q-rules for Level 5}

Q rules for Level 5 are quite complicated and are given in the Appendix C \eqref{appenq}. Once again one obtains a highly overdetermined
set of linear equations, which turn out to have a four parameter set of solutions. Requiring consistency with dimensional reduction gives another overdetermined set of equations for these parameters, which again turn out to have a unique solution. It would be interesting to understand the underlying pattern that guarantees the existence of such solutions.

To summarize:
using the information presented thus far it is easy to write down gauge invariant {\em free} equations for all the higher spin modes of string theory.
In order to truncate the spectrum consistently one needs the Q-rules which have been given here up to level 5. For higher levels these have to be worked out. The gauge transformation has a simple form - it looks like a scale transformation. 

We now need to describe interactions. For this purpose we will generalize the scale (Weyl) invariance of the free theory to the exact
renormalization group (ERG) symmetry. This will automatically give us the interactions.

\section{Exact Renormalization Group in String Theory}
\setcounter{equation}{0}

In this section we review some basic facts about the ERG, its connection with the continuum $\beta$-function, and how this applies
to the world sheet description of string theory. 

\subsection{Generalities about the RG}

We recall the arguments of \cite{WK,W,P,W2} as applied to string theory. 
The world sheet action for the bosonic string in general open string backgrounds has the generic form:

\br    \label{1.1.1}
S~ &=&~ {1\over2}\int _\Gamma d^2\sigma \{\partial ^\alpha X^\mu 
 \partial _\alpha X_\mu \}~+~
\int _{ \partial \Gamma} dt L_1 \nonumber \\    
L_1~&=&~ \sum _i g^i M_i + \sum _i w^i W_{i} + \sum _i \mu ^i R_i
\er 

$\mu$ runs from $0~~ -~~ D-1$. $D$ is 26 for the bosonic string. $d^2 \sigma$ is the area
element in real coordinates and $dt$ the line element. 
Here $\Gamma$ denotes the (Euclidean) world sheet. Thus at tree level
$\Gamma$ is a disc (or upper half plane). $\partial \Gamma$ denotes the boundary
of $\Gamma$. Thus $d^2 \sigma = dxdy$ and $dt=dx$ for the upper half plane.

$L_1$ corresponds to the boundary action corresponding
to  condensation of open string modes.  (For concreteness
we restrict ourselves to open string backgrounds, which are boundary terms.)
We denote by $M_i ~,~ W_i$, and $ R_i $,  marginal, irrelevant and relevant 
operators
respectively. $ g^i ~, w^i ~, \mu ^i$, are the corresponding coupling constants.
In string theory, on shell vertex operators are marginal. If $k$ is the {\em space time} momentum of a string mode with mass $m$, the length dimension
of the corresponding vertex operator increases as $k^2$, and when $k^2=m^2$ it is marginal.

The world sheet theory is defined with an  ultraviolet cutoff, $\Lambda$. Thus the partition 
function
is 
\be
\int _{|p| < \Lambda} [dX(p)]exp \{-S[X(p), g_i,w_i,\mu _i]\} 
\ee

 It is more convenient  to discuss a finite RG ``blocking'' transformation
that takes the cutoff $\Lambda$ to $\Lambda \over 2$, rather than 
making an infinitesimal change. Denote it by $\cal R$. 
Thus $\cal R$ is to be implemented as follows:\newline
1. Perform the integral
 $\int _{{\Lambda \over 2}<|p|<\Lambda}[dX(p)]exp\{-S[X(p)]\}$.
\newline
2. Rescale momenta: Let $p'= 2p$. Now the range of $p'$ is again $0-\Lambda$.
\newline
3. Rescale the surviving $X(p), ~ 0<|p|<{\Lambda\over2}$. Let
$X(p) = Z X'(p')$. Choose $Z$ so that the kinetic term $p^2 X(p)X(-p)$ has the
same normalization as before.

There is a subtlety here that needs to be pointed out. The scaling of $X$ in general changes the integration measure
$[dX(p)]$. This is related to the trace anomaly in a field theory. In string theory this is related to the $\beta$ function associated with a closed string mode - the dilaton. We will postpone this point to a later section.

As a result of all the above we get  for the partition function:
\be
\int _{|p'| < \Lambda} [dX'(p')]exp \{-S[X'(p'), g'_i,w'_i,\mu ' _i]\} 
\ee

which is exactly the same as before except that the coupling constants have
different values. Thus effectively
\be
{\cal R }~:~ (g,w,\mu )  \longrightarrow (g',w',\mu ')
\ee
defines the discrete renormalization group transformation.

We can then define a recursion relation between the coupling constants:

\be    \label{RG}
{\cal R }~:~ (g_l,w_l,\mu _l )  \longrightarrow (g_{l+1},w_{l+1},\mu _{l+1})
\ee

At a fixed point, the couplings satisfy: 
\be
{\cal R }~:~ (g^*,w^*,\mu ^*  )  \longrightarrow (g ^*,w ^*,\mu ^*)
\ee

In string theory there are an infinite number of vertex operators, but to keep the discussion
simple we keep one of each type.

The recursion relation  thus takes the form
\br   \label{RR}
\mu _{l+1} &=& 4 \mu _l + N_\mu [\mu _l,g_l,w_l] \nonumber \\
g_{l+1}&=& g_l + N_g [\mu _l,g_l,w_l] \nonumber \\
w_{l+1}&=&{1\over 4}w_{l} + N_w [\mu _l,g_l,w_l]
\er

where the factor 4 characterizes a dimension-2 relevant operator
 (eg. a mass term, $X^2$)
 and the factor
1/4 characterizes a dimension-4 irrelevant operator, say of the form
 $(\p X \p X)^2$. $N_\mu , N_g $ and $N_w$ correspond to higher order
corrections which we take to be small in perturbation theory.

At a fixed point one has

\br     \label{beta}
g_l&=&g_{l+1} \nonumber \\ w_l&=&w_{l+1} \nonumber \\ \mu_l&=&\mu_{l+1}
\er
which means that doing
a block transformation does not change anything. 
This can only be true if there are no dimensionful physical quantities with which to compare the 
cutoff $\Lambda$. Otherwise we would be able to tell the difference between $\Lambda$ and $\Lambda/2$. Thus
the theory has an overall scale - the cutoff, $\Lambda$, and 
no other scale. So correlation lengths are either infinite or zero.

It is important to note that $w^* \ne 0$ in general. In the context of string theory at low (space time) energies massive modes are irrelevant 
couplings. Thus it is not true
in general that the massive modes have zero vev's.
Nevertheless we can eliminate the massive modes using their equations of motion and obtain
equations of motion for the massless modes - which at low energies correspond to
marginal operators. We can understand this in terms of the recursion relations.
We follow the discussion of Wilson \cite{W}.

Let $0\le l \le L$ be the range of the index $l$, with $\mu _0, g_0, w_0$
being the parameters of the action at high energies. From eqn. (\eqref{RR})
we see that to lowest order $\mu _L \approx 4^L \mu _0$.  This blows up
rapidly with $L$. This is the famous
``fine-tuning'' problem:  $\mu _0$ has to be tuned very accurately for $\mu_L$, the low energy
effective parameter to have some observed value.  Thus it is better to  use $\mu _L$ as
our input.  The irrelevant $w_0$, on the other hand keeps getting smaller and can be used
as an input parameter. This way it can be seen easily that $\mu _l, w_l, g_l$ 
rapidly lose their dependence on $w_0$: this is the statement of universality.
The marginal coupling, $g_l$ is important for all values of $l$. 

Now let us iterate the equations (\ref{RR}) a number of times. The equations are:
\be
\mu _l = 4^{l-L} \mu _L - \sum _{n=l}^{L-1} 4^{l-(n+1)}N_\mu [\mu _n, g_n, w_n]
\ee

\be
w_l=4^{-l} w_0 + \sum _{n=0}^{l-1} 4^{n+1-l}N_w[\mu _n, g_n, w_n] 
\ee

\be
g_l=g_{l_0} + \sum _{n=l_0}^{l-1}N_g[\mu _n, g_n, w_n]
\ee

We can solve these equation iteratively with the following starting inputs
obtained by neglecting the non-linear corrections:
\br
\mu _l &=& 4^{l-L}\mu _L \nonumber \\
w_l &=& 4^{-l} w_0 \nonumber \\
g_l &=& g_{l_0}
\er

The solution in general has the form
\[
g_l= V_g(g_{l_0},\mu _L,w_0,l,l_0,L)
\]
and similarly for $\mu _l,w_l$.

The solution simplifies 
when $l>>0$ and  $l<<L$.
 Namely the dependence on $\mu_L$ and $w_0$
of $g_l$ is so weak ($O(4^{l-L}$ and $4^{-l}$) that we can
set $\mu_L = w_0=0$ with negligible error. Furthermore
the summations can be extended to $+\infty$ for
$\mu _l$ and $-\infty$ for $w_l$.  The resulting equations
have a translational invariance in $l$ and $l_0$.
Thus 
\be   \label{CSGL}
g_l= V_g[l-l_0,g_{l_0}]
\ee

$l-l_0$ is the log of the ratio of the scales and $g$ is dimensionless to begin with.
There are no absolute scales in this equation. Furthermore
in this region the recursion relation can be approximated  by a differential equation, 
 the usual
Gell-Mann - Low, Callan-Symanzik $\beta$-function involving just the marginal coupling. The solution of this gives
us $g^*$. In string theory this corresponds to EOM for the massless fields, in which the massive fields have been integrated out.
Classically, this means solving their EOM.
One can also solve for the fixed point value  $w^*$, which is not zero. Thus the massive mode expectation values
are not zero. What is nice is that we don't need them to solve for $g^*$.

If we perform an infinitesimal RG transformation the equations (i.e. $N_w,N_g,N_\mu$) are polynomial in the couplings, while the $\beta$ functions are non polynomial. This is  exactly analogous to the relation between string field theory, which is polynomial in the fields and the low energy 
EOM, which are not. BRST open string field theory has a cubic action and a quadratic EOM. The ERG is also quadratic in the fields.

There is one more important point: This is the fact that the ERG is obviously written with a finite cutoff. One can perform an infinite number of 
RG iterations and reach the continuum. All theories on the same RG trajectory describe the same physics. The information about the cutoff scale is contained in the values of the couplings.  However, if the couplings are tuned to the values at the fixed point, 
 the value of the cutoff does not matter. It can be kept finite or taken to infinity or to zero. This can be illustrated by
just considering the normal ordering corrections of a vertex operator. ${e^{ik.X}\over a}=e^{(k^2-1) ln(a)\over 2}:e^{ik.X}:$.
Thus  when $k^2=2$, the dependence on the cutoff disappears. At this level of approximation this is the fixed point.
Conversely, off shell (read away from a fixed point) one needs to keep a finite cutoff. In string field theory the role of the cutoff is played by the string length, which is finite.

To summarize: the exact RG in this approach gives equation analogous to those of string field theory. It can be taken
to be an alternative way of writing down string field theory equations. What we will see in the following is that these equations can be made gauge invariant, as in BRST string field theory. Not only that, the world sheet formalism is manifestly background independent, so we obtain a background independent formalism.

\subsection{Exact RG in Position Space}

The discussion here follows \cite{WK} and \cite{P}. The exact RG \cite{WK} originally was written in momentum space. We work in position space because it is more convenient for two dimensional theories. We start with a quantum mechanical version and then generalize in a straightforward way to field theory.

\subsubsection{Quantum Mechanics}

Consider the Schroedinger equation
\be 
i\frac{\partial \psi}{\p t} =-\hf\frac{\pp \psi }{\p y^2}
\ee  
for which the Green's function is $\frac{1}{\sqrt{2\pi (t_2-t_1)}}e^{i\frac{(y_2-y_1)^2}{2(t_2-t_1)}}$. Let us
 change variables :$y = xe^\tau , it = e^{2\tau}$ and $ \psi ' = e^\tau \psi$. 
 We get
 \be
\frac{\p \psi '}{\p \tau} = \hf\frac{\p}{\p x }(\frac{\p}{\p x} + x ) \psi ' 
\ee
The Green's function is:
\be
G(x_2, \tau _2; x_1 , 0) = \frac{1}{\sqrt{2\pi (1- e^ {-2\tau _2})}}
e ^{- \frac{(x_2 - x_1 e^{-\tau _2)^2}} {2(1-e^{-2\tau _2 )}}}
\ee

Thus as $\tau _2 \rightarrow \infty $ it goes over to $\frac{1}{\sqrt{2\pi}}e^{-\hf x_2^2}$. As $\tau _2 \rightarrow 0$ it goes to $\delta (x_1-x_2 )$. 
\[ \psi (x_2, \tau _2 ) = \int dx_1 G (x_2, \tau _2; x_1,0 )\psi (x_1, 0)
\]

So 
\[
\psi (x_2, 0) =\psi(x_2)
\] which has all the information about the original function. 
 \[
 \psi (x_2,\infty)=\frac{1}{\sqrt{2\pi}}e^{-\hf x_2^2} \int dx_1 \psi (x_1)
 \]
where $x_1$ is integrated over fully and all information about the original function is lost.     
Thus consider
\be
 \frac{\p}{\p\tau } \psi (x_2, \tau) =  \hf\frac{\p}{\p x_2}(\frac{\p}{\p x_2} + x_2) \psi (x_2, \tau )
\ee
with initial condition $\psi (x, 0)$
Thus  we can define $Z(\tau ) = \int d x_2 \psi (x_2, \tau )$, where $\psi $ obeys the above equation, we see that $\frac{d}{d\tau} Z =0$. 
Although $Z$ is independent of $\tau$, the integrand changes from being the original function, to a Gaussian times the original function integrated over. 

Now take the initial wave function as $e^{\frac{i}{\hbar}S[x ]}$ where
$x$ denotes the space-time coordinates. Then for $\tau = \infty$ $\psi  \approx \int {\cal D }x e^{iS[x]}$ is the integrated partition function. At $\tau =0$ it is the unintegrated $e^{iS[x]}$. $\tau$ can parametrize a cutoff so that the high momentum modes get integrated out first. 

Following \cite{P}, we shall also split the action into a kinetic term and interaction term.
Thus  we write $\psi = e ^{-\hf x^2  f(\tau ) +L(x)}$  in the quantum mechanical case discussed above.

By choosing $a,b,B$ suitably  ( $b=2af , B = \frac{\dot f}{bf} $) in 
\[ \frac{\p \psi}{\p \tau} = B \frac{\p}{\p x} ( a \frac{\p}{\p x} + bx) \psi (x,\tau )
\] we get 
\be
\frac{\p L}{\p \tau} = - \frac{\dot f}{2f}+ \frac{\dot f}{2 f^2}[\frac{\pp L}{\p x^2} + (\frac{\p L}{\p x})^2 ]
\ee
Note that if $f = G^{-1}$ ($G$ can be thought of as the  propagator) then $\frac{\dot f}{f^2} = -\dot G$.
The first term on the RHS is a field independent constant term, which is normally neglected. The effect of neglecting this term
is that the partition function satisfies
\[
{dZ\over d\tau}= {\dot f\over 2f}Z
\]
or
\[
{d (ln~Z)\over d\tau}= {\dot f\over 2f}
\]
If we add to $L$ a field independent but $\tau$ dependent term, we can get rid of this. This is equivalent to adding a ``counterterm"
for the cosmological constant. In field theory however, if the background is curved,  this term produces a non local metric dependent
term which cannot be canceled by a local counterterm.  This is the trace anomaly. 

In most discussions of the exact RG in flat space this field independent term is neglected.

\subsection{Field Theory}
We now apply this to a Euclidean field theory.

\be    \label{FT}
\psi = e^{-\hf \int dz \int dz' X(z) G^{-1} (z,z') X(z') + \int dz L[X(z),X'(z)]}
\ee
Here $X'(z)=\p _z X(z)$. There is a straightforward generalization to the case  where higher derivatives $X''(z),X'''(z)...$ involving many coordinates. This will be required when we work with loop variables where there are an infinite number of $\xn$.

Generalizing the quantum mechanical case we apply 
\be   \label{27}
\int dz \int dz' B(z,z') \frac{\delta}{\delta X(z')} [\frac{\delta}{\delta X(z)} + \int b(z,z'')X(z'') ] 
\ee
to $\psi$ and require that this should be equal to $\frac{\p \psi}{\p \tau}$. 

\[
\ddXz \psi= [-\int du'~G^{-1}(z,u')X(u') + \ddXz \int du ~L[X(u),X'(u)]] \psi
\]
\[
{\delta^2\over \delta X(z)\delta X(z')} \psi = [-G^{-1}(z,z') + {\delta^2\over \delta X(z)\delta X(z')}\int du~L[X(u),X'(u)]]\psi +
\]
\[
 [-\int du'~G^{-1}(z',u')X(u') + \ddXzp \int du' ~L[X(u'),X'(u')]]\times
 \]
 \[
  [-\int du~G^{-1}(z,u)X(u) + \ddXz \int du ~L[X(u),X'(u)]]\psi
\]
The operator \eqref{27} thus becomes

\[ 
[-G^{-1}(z,z') + {\delta^2\over \delta X(z)\delta X(z')}\int du~L[X(u),X'(u)]]\psi +
\]
\[
 [-\int du'~G^{-1}(z',u')X(u') + \ddXzp \int du' ~L[X(u'),X'(u')]]\times
  \]
 \[
  [-\int du~G^{-1}(z,u)X(u) + \ddXz \int du ~L[X(u),X'(u)]]\psi+
\]
\[
b(z,z')\psi + \int dz''~b(z,z'')X(z'') [-\int du'~G^{-1}(z',u')X(u') + \ddXzp \int du' ~L[X(u'),X'(u')]]\psi
\]
Now choose $b(z,z')=2G^{-1}(z,z')$ and $B(z,z')= -\hf \dot G(z,z')$. The final equation simplifies to
\[
\int du~ {\p L\over \p \tau} \psi = \int dzdz'~\{-\hf \underbrace{\dot G(z,z') G^{-1}(z,z')}_{field~independent}- 
\]
\be    \label{ERG1}
 \hf \dot G(z,z')[{\delta^2\over \delta X(z)\delta X(z')}\int du~L[X(u),X'(u)]] + \ddXz \int du~L[X(u)]\ddXzp \int du'~L[X(u')]\}\psi =0
\ee
The field independent first term is usually dropped in discussions of the ERG but here we keep it. It has information about the trace anomaly which is proportional to the central charge. It is also supposed to contribute to the dilaton equation. In this subsection we focus on the free EOM which is obtained from the first term.

 We now take $\tau \approx ln~a$.  This comes from the scale dependence of the determinant (diagrammatically, the vacuum bubble).  
 \[
 {d\over d\tau}\hf Tr ~ln~G= \hf Tr[\dot G G^{-1}]
 \]
 
 The remaining terms which are the ones usually considered in RG studies, are diagrammatically easy to understand also  \cite{P}:  the first term  in the RHS, which is linear in $L$, represents
contractions of fields at the same point - self contractions within an operator. These can be understood as a pre-factor multiplying normal ordered vertex operators. The second term represents contractions between fields at two different points - between two different operators. For string theory purposes, the first term gives the free equations of motion and the second gives the interactions.

\subsection{ERG and Loop Variables: Open Strings}

The loop variable represents an infinite collection of all possible vertex operators of the open string. Thus the boundary term in
\eqref{1.1.1}, which is written as $\int dz ~L[X(z),X'(z)...]$ in \eqref{FT} is just:
\[
\int [dz]~{\cal L}(z)= \int dz ~{\cal D} \al (t) \lpp 
\]

 \be   \label{GVO}
 = \int \underbrace {[dz dx_1 dx_2...dx_n...]}_{[dz]}~~\gvk = \int [dz]~~{\cal L}[Y(z,\xn), \frac{\p Y}{\p x_1}, \frac{\p Y}{\p x_2},...,\frac{\p Y}{\p \xn}]
 \ee
From here on the variable $z$ will stand for $(z,x_1,x_2,...,x_n,...)$. And $z,z'$, will stand for
the sets of variables: 
\[(z_A, x_{1A},x_{2A},....,x_{nA},...),(z_B, x_{1B},x_{2B},....,x_{nA},...)
\] 
The integrals $\int dz$ in the ERG will be replaced
by $\int ... \int dz dx_{1A}dx_{2A}..dx_{nA}...$. Variables $\xn$ without a subscript will stand for the integration variable $u$ in the loop variable. Thus we will be allowed to integrate by parts on $u$, i.e.  $\xn$'s. This
is responsible for gauge invariance.  The free gauge invariant equations obtained in Section 3 will be reproduced.
  The interaction equation requires a further modification described later below.

\subsubsection{Free Equations}

\[
 \int du~ {\delta \over \delta Y^\mu(z')} {\cal L}(u)=\int du~\Big\{ {\p {\cal L} [Y(u),Y_{n}(u)]\over \p Y^\mu(u)} \delta (u-z')+{\p {\cal L} [Y(u),Y_{n}(u)]\over \p Y_1^\mu(u)} \p_{x_1}\delta (u-z')
\]
\be	\label{FEO}
 +{\p {\cal L} [Y(u),Y_{n}(u)]\over \p Y_2^\mu(u)} \p_{x_2}\delta (u-z')+...+{\p {\cal L} [Y(u),Y_{n}(u)]\over \p Y_n^\mu(u)} \p_{x_n}\delta (u-z')\Big\}
 \ee
 We can integrate by parts to get:
 \[
 = \int du~\Big\{ {\p {\cal L} [Y(u),Y_{n}(u)]\over \p Y^\mu(u)} \delta (u-z')-[\p_{x_1}{\p {\cal L} [Y(u),Y_{n}(u)]\over \p Y_1^\mu(u)}] \delta (u-z')
\]
 \be  \label{FnlDerI}
-[\p_{x_2}{\p {\cal L} [Y(u),Y_{n}(u)]\over \p Y_2^\mu(u)}] \delta (u-z')+...- [\p_{x_n}{\p {\cal L} [Y(u),Y_{n}(u)]\over \p Y_n^\mu(u)}]\delta (u-z')
  \Big\}
 \ee
This second form is convenient for the interacting term which is a product of first derivatives. For the free case, the first version is better.

\[
{\delta^2\over \delta Y^\mu (z) \delta Y^\mu(z')}\int du~{\cal L}(u)=
\]
\[
{\delta\over \delta Y^\mu(z)} \int du~ {\delta \over \delta Y^\mu(z')} {\cal L}(u)={\delta\over \delta Y^\mu(z)}\int du~\Big\{ {\p {\cal L} [Y(u),Y_{n}(u)]\over \p Y^\mu(u)} \delta (u-z')+{\p {\cal L} [Y(u),Y_{n}(u)]\over \p Y_1^\mu(u)} \p_{x_1}\delta (u-z')
\]
\be
+{\p {\cal L} [Y(u),Y_{n}(u)]\over \p Y_2^\mu(u)} \p_{x_2}\delta (u-z')+...+{\p {\cal L} [Y(u),Y_{n}(u)]\over \p Y_n^\mu(u)} \p_{x_n}\delta (u-z')\Big\}
\ee	\label{FnlDII}
We give the action on the $n$th term:
\[
{\delta\over \delta Y^\mu(z)}\int du~[{\p {\cal L} [Y(u),Y_{n}(u)]\over \p Y_n^\mu(u)} \p_{x_n}\delta (u-z')]=\sum_m
\int du~ {\pp {\cal L} [Y(u),Y_{n}(u)]\over \p Y_{m\mu}(u)\p Y_n^\mu(u)}[\p_{x_n}\delta (u-z')][\p_{x_m}\delta (u-z)]
 \]
Let us include the $z,z'$ integrals:
\[
\int dz~\int dz'~ \dot G(z,z')\sum _n \sum_m
\int du~ {\pp {\cal L} [Y(u),Y_{n}(u)]\over \p Y_{m\mu}(u)\p Y_n^\mu(u)}[\p_{x_n}\delta (u-z')][\p_{x_m}\delta (u-z)]
\]

Now we show that it can be written in the form given in Section 3 so that the rest of the derivation goes through as before.
We let the derivative on the delta function act on $z,z'$ instead of $u$ and integrate by parts to get
\[
\int dz~\int dz'~[\p_{x_{Bn}}\p_{x_{Am}} \dot G(z,z')]\sum _n \sum_m \int du~
 {\pp {\cal L} [Y(u),Y_{n}(u)]\over \p Y_{m\mu}(u)\p Y_n^\mu(u)}[\delta (u-z')][\p_{x_m}\delta (u-z)]
\]
\[
=\int dz~\int dz'~[\p_{x_{Bn}}\p_{x_{Am}} \dot G(z,z')]\sum _n \sum_m
 {\pp {\cal L} [Y(z),Y_{n}(z)]\over \p Y_{m\mu}(z)\p Y_n^\mu(z)}\delta (z-z')
\]
\[
=\int dz~~[\lan Y_n(z)Y_m(z)\ran]\sum _n \sum_m
 {\pp {\cal L} [Y(z),Y_{n}(z)]\over \p Y_{m\mu}(z)\p Y_n^\mu(z)}
 \]
 \be    \label{freestandardform}
 =
 -\sum_{n,m}\kn.\km \hf({\pp\Sigma\over \p \xn \p \xm}-{\p\Sigma\over x_{n+m}})
\ee
Thus we have derived the free equations using the ERG.

\subsubsection{Interacting Equations}
Two factors of the the functional derivative \eqref{FnlDerI} given below is what the interaction term is made of.
 \[
{\delta \over \delta Y^\mu(z')}\int du~  {\cal L}(u) = \int du~\Big\{ {\p {\cal L} [Y(u),Y_{n}(u)]\over \p Y^\mu(u)} \delta (u-z')-[\p_{x_1}{\p {\cal L} [Y(u),Y_{n}(u)]\over \p Y_1^\mu(u)}] \delta (u-z')
\]
 \be  
-[\p_{x_2}{\p {\cal L} [Y(u),Y_{n}(u)]\over \p Y_2^\mu(u)}] \delta (u-z')+...- [\p_{x_n}{\p {\cal L} [Y(u),Y_{n}(u)]\over \p Y_n^\mu(u)}]\delta (u-z')
  \Big\}
 \ee
 It can easily be checked that it is {\em not} gauge invariant. 
 
 The resolution of this problem, described in \cite{BSERGopen1,BSERGopen2}, is to introduced all derivatives in the loop variable.
 The basic idea is that the gauge variation of vertex operators of a given level should be of the form  $\la _n {\p \over \p \xn}$ of lower order vertex operators. This ensures gauge invariance.
 
 We give the results for Spin 1, 2  and then the general result.

{\bf Spin 1} 
 \[ i\kim {\p Y^\mu \over \p x_1} \rightarrow \li {\p \over \p x_1} (i\kom Y^\mu) \]
\[
{\delta \over \delta Y^\mu(z)} \int du~  {\cal L}(u)= \frac{\p {\cal L}}{\p Y^\mu(z)} - \p _{x_1} \frac{\p  {\cal L}}{\p \yim (z)}
\]
\[=
(i\kom {\cal L} - i\kim i\kon \yin {\cal L})|_{level~1}= -[\kom \kin -\kim \kon ]\yin \e
\]
This is the gauge invariant Maxwell field strength.

 {\bf Spin 2}
 \be
{\cal L}=[ iK_{11\mu} \frac{\pp Y^\mu}{\p x_1^2} + iK_{2\mu} \frac{\p Y^\mu}{\p x_2} - \hf \kim \kin \yim \yin]\e
\ee

We have introduced $K_{11\mu}, K_{2\mu}$ in place of $\ktm$ of the free theory.
We require 
\be	\label{gt3}
\delta K_{2\mu} = \lt \kom ~~;~\delta K_{11\mu} = \li \kim
\ee
This ensures that $\delta ( iK_{2\mu} \frac{\p Y^\mu}{\p x_2})= \lt {\p\over \p x_2}(i\ko.Y)$ and 
$\delta ( iK_{11\mu} \frac{\pp Y^\mu}{\p x_1^2})=\li {\p \over \p x_1} (i\ki.\yi)$
 \be	\label{K}
 K_{2\mu} \equiv (\bar q_2- {\bar q_1^2\over 2}) \kom   ~~;~~K_{11\mu} \equiv \ktm - K_{2\mu}
 \ee
 
 where $\bar q_n \equiv {q_n\over q_0}$, satisfies the required transformation property.  It is important
 to note that if we use $\frac{\pp Y^\mu}{\p x_1^2}= \frac{\p Y^\mu}{\p x_2}$, the vertex operators add up to 
 $\ktm \ytm$. Also note that dimensional reduction with mass is required for this construction. It has $\qo$ in the 
 denominator. $\qo$ is the mass, which therefore has to be non zero. \eqref{K} however is correct even for $\mu=\theta$
 when $\ktm = \qt$ etc.

 The quadratic term in the ERG is a product of 
 \[
{\delta \over \delta Y^\mu(z)} \int du~  {\cal L}(u)= \frac{\p {\cal L}}{\p Y^\mu(z)} - \p _{x_1} \frac{\p  {\cal L}}{\p \yim (z)} + \p_{x_1}^2\frac{\p  {\cal L}}{\p Y_{11}^\mu(z)}- \p _{x_2} \frac{\p  {\cal L}}{\p Y_2^\mu (z)}
 \]
at two points $z_A$ and $z_B$. Let us evaluate this for the  modified Lagrangian: 
\be
{\cal L}=[ iK_{11\mu} \frac{\pp Y^\mu}{\p x_1^2} + iK_{2\mu} \frac{\p Y^\mu}{\p x_2} - \hf \kim \kin \yim \yin]\e
\ee

\[
\frac{\p {\cal L}}{\p Y^\mu} = [i\kom  iK_{11\nu} \frac{\pp Y^\nu}{\p x_1^2} + i\kom iK_{2\nu} \frac{\p Y^\nu}{\p x_2}-i\kom \hf \kir \kin \yin Y_1^\rho]\e
\]
\[
\p _{x_1} \frac{\p {\cal L}}{\p Y_1^\mu} = -\kim k_1.Y_2 \e - \kim k_1.Y_1 i \ko .Y_1 \e
\]
\[
\p _{x_2} \frac{\p{\cal L}}{\p Y_2^\mu}=iK_{2\mu}  i \ko .Y_2 \e
\]
\[
\p _{x_1}^2 \frac{\p {\cal L}}{\p (\p _{x_1}^2Y^\mu)}= iK_{11\mu}(i \ko .Y_2 + (i\ko .Y_1)^2)\e
\]
Thus 
\[
\frac{\p {\cal L}}{\p X(z)} - \p _z \frac{\p{\cal L}}{\p X'(z)} + \p_z^2\frac{\p{\cal L}}{\p X''(z)}=\Big(i\kom  iK_{11\nu} \frac{\pp Y^\nu}{\p x_1^2} + i\kom i K_{2\nu} \frac{\p Y^\nu}{\p x_2}-i\kom \hf \kin \kir \yin Y_1^\rho\Big) \e 
\]
\[+\Big(\kim k_1.Y_2 \e + \kim k_1.Y_1 i \ko .Y_1 \e \Big) -iK_{2\mu} i \ko .Y_2 \e\]
\be   \label{erg2}
+iK_{11\mu}(i \ko .Y_2 + (i\ko .Y_1)^2)\e
\ee
We now replace $ \frac{\pp Y^\nu}{\p x_1^2}$ by $\frac{\p Y^\nu}{\p x_2}$ and simplify:

The coefficient of $Y_2^\nu$ is:
\be
V_{2\mu\nu}\equiv \Big(-\kom  K_{11\nu}- \kom  K_{2\nu}+\kim \kin+ K_{2\mu}  \kon- K_{11\mu}  \kon \Big)\e =\Big(-\kom  K_{11\nu}+\kim \kin- K_{11\mu}  \kon \Big)\e
\ee

The coefficient of $\yin Y_1^\rho$
is
\be
V_{11\mu\nu\rho}\equiv \Big(-i\kom \hf \kin \kir +i\hf \kim (\kin  \kor+ \kir \kon)-iK_{11\mu}  \kon  \kor\Big)\e
\ee 

Using (\eqref{gt1}\eqref{gt3}) we  see that they are invariant.

The components in the $\theta$ directions can be obtained from the above. For instance $V_{2\mu \theta}$ is
\be
V_{2\mu\theta} = \Big(-\kom  K_{11\theta}- \kom  K_{2\theta}+\kim q_{1\theta}+ K_{2\mu}  q_{0\theta}- K_{11\mu}  q_{0\theta} \Big)\e 
\ee

\subsubsection{General Construction of $K$'s}

Let us first introduce the following notation to generalize the construction of the spin 2 and spin 3 cases. Define
\[ K_{m\mu} : \delta K_{m\mu} =\la _m \kom;~~~ K_{mn\mu}  : \delta K_{mn\mu} =\la _m K_{n\mu} + \la _n K_{m\mu} , ~m\neq n \]
\be \label{Kmnp}
~~~K_{mnp\mu}: \delta K_{mnp\mu} =\la _m K_{np\mu}+\la_n K_{mp\mu} +\la_pK_{mn\mu} ,~~m\neq n\neq p\ee
and so on. 
For repeated indices
\[ K_{mm\mu} : \delta K_{mm\mu}  = \la _m K_{m\mu};~~~K_{mmm\mu} : \delta K_{mmm\mu} = \la _m K_{mm\mu} \]
Also
\[ K_{mmp\mu}: \delta K_{mmp\mu} = \la _m K_{mp\mu}+\la _p K_{mm\mu} \]
and so on.

The general rule is that if $[n]_i$ defines a particular partition of the level $N$, at which we are working, then 
\be	\label{Genrule}
\delta K_{[n]_i\mu} = \sum_{m\in [n]_i}\la_mK_{[n]_i/m\mu}
 \ee
where $[n]_i/m$ denotes the partition with $m$ removed, and the sum is over {\em distinct $m$'s}. (So even if $m$ occurs
more than once in the partition, the coefficient of $\la_mK_{[n]_i/m\mu}$ is still 1.)

Define 
\be	\label{qbar}
\bar q(t) \equiv {1\over q_0}q(t) = 1+ {\bar q_1\over t}+ {\bar q_2\over t^2}+...+{\bar q_n\over t^n}+... 
\ee
\[= e^{\sum _n y^n t^{-n}}= 1+ {y_1\over t} + {y_2+{y_1^2\over 2}\over t^2}+{y_3+y_1y_2+{y_1^3\over 6}\over t^3}+....\]
If we solve for $y_n$ in terms of $q_m$ we get
\[ \bar q_1 = y_1;~~~\bar q_2= y_2+{y_1^2\over 2} \implies y_2= \bar q_2- {\bar q_1^2\over 2};\] Similarly \[y_3=\bar q_3 - \bar q_2\bar q_1 + {\bar q_1^3\over 3}\] In general
$\sum_{n=0}^\infty {y_n \over t^n}= ln~ (\bar q(t))$.

Similarly define 
\[ \la (t) = 1 + {\li\over t} +...{\la_n \over t^n}+...=e^{\sum_0^\infty z_n t^{-n}}\]
The gauge transformation $ \bar q(t)\rightarrow \la(t)\bar q(t)$ is represented as $y_n\rightarrow y_n+z_n$. Since we are only interested in the lowest order in $\la$ we can set $z_n=\la_n$. Thus we have \be \label{yn} \delta y_n = \la_n\ee.

Let us now construct the $K_{mnp..\mu}$:
Let us start by defining $K_{0\mu} \equiv \kom$. Then $K_{1\mu} = \kim$, because $\delta K_{0\mu} = \li K_{0\mu}$.

{\bf Level 2:} 

Let 
\be
 \label {K2} K_{2\mu} = y_2 \kom \ee
 Using (\ref{yn}), it  clearly satisfies the requirement (\ref {Kmnp}), that $\delta K_{2\mu} = \lt K_{0\mu}$. Using $y_2=\bar q_2- {\bar q_1^2\over 2}$,
 \[K_{2\mu} = (\bar q_2 - {\bar q_1^2\over 2})\kom\]
Then we let
 \be \label{K11} K_{11\mu} = \ktm - K_{2\mu}
 \ee
  It is easy to check that $\delta K_{11\mu} = \li K_{1\mu}$.

We can now generalize this construction:

{\boldmath$K_n,~n\geq 2$:}

 Consider $K_{n\mu}$. Since we want $\delta K_{n\mu} = \la_nK_{0\mu}$, a correct choice is \be \label{Kn}
K_{n\mu} = y_n\kom
\ee

{\boldmath $K_{n1\mu},~n\geq 2:$}

We need $\delta K_{n1\mu} = \li K_{n\mu} + \la_n K_{1\mu}$. An obvious trial is to set 
\be \label{Kn1}
K_{n1\mu} = y_n K_{1\mu} = y_n \kim
\ee
Using (\ref{Kn},\ref{yn}) we see that it is correct.

{\boldmath $K_{mn\mu},~m\neq n;n,m\geq 2:$}

It is easy to check that
\be \label{Kmn}
K_{mn\mu} =y_ny_m \kom
\ee

satisfies $\delta K_{nm\mu} = \la_n y_m\kom + \la_m y_n\kom = \la_n K_{m\mu} + \la_m K_{n\mu} $ as required.

{\boldmath $K_{mm..\mu},~m\geq 2:$}

For repeated indices we see that what works is:
\be
K_{mm\mu}  = {y_m^2\over 2}\kom;~~~K_{mmm}^\mu = {y_m^3\over 3!}\kom;
\ee
 The pattern is easily generalized.

{\boldmath $K_{mn1\mu} ,~m\neq n;~m,n\geq 2:$}
\be \label{Kmn1}
K_{mn1\mu}  = y_ny_mK_{1\mu}
\ee
Satisfies \[\delta K_{mn1\mu}  = \la_n y_m K_{1\mu} + \la _m y_n K_{1\mu} + \li y_ny_m\kom= \la_n K_{m1\mu}+\la _m K_{n1\mu} + \li K_{mn\mu}\]
as required.

Again for repeated indices:
\be
K_{mm1\mu} = {y_m^2\over 2}K_{1\mu}
\ee

{\boldmath $K_{n11\mu},~n\geq 2:$}

We try 
\be
K_{n11\mu} = y_n K_{11\mu}
\ee

$\delta K_{n11\mu} = \la_n K_{11\mu} + \li y_n K_{1\mu} = \la_n K_{11\mu} + \li K_{n1\mu}$ as required.

The pattern is now clear: when all the $m,n,..\geq 2$ we just get 

{\boldmath $K_{mn...\mu},~m\neq n;~n\geq 2:$}
\be K_{mn..\mu} =y_my_n...\kom\ee

{\boldmath $K_{mn...1\mu},~n\geq 2:$}

When one of the indices is 1, we get \be K_{mn..1\mu}= y_my_n...\kim\ee 

{\boldmath $K_{mn...11\mu},~n\geq 2:$}
Similarly if two of the indices are
1 we get \be K_{mn..11\mu} = y_my_n...K_{11\mu} \ee

{\boldmath $K_{m.....11\mu},~n\geq 2:$}
\be K_{m\mu\underbrace{1111..1}_{n}}=y_mK_{\underbrace{1111..1\mu}_{n}}
\ee
For other repeated indices again the pattern is  obvious. Thus

\be
K_{mm\underbrace{111...\mu}_{n }} = {y_m^2\over 2}K_{\underbrace{111..\mu}_{n}}
\ee

{\boldmath $K_{\underbrace{1.....11\mu}_{n}}:$}

To complete the construction we need $K_{111..1\mu}$. For the second level we had $K_{11\mu}=\ktm - K_{2\mu}$.
Similarly one can check that \[ K_{111\mu} = k_{3\mu} - K_{21\mu}- K_{3\mu}\]
$\delta K_{111\mu} = \la_3 \kom + \lt \kim + \li \ktm - \lt K_{1\mu} - \li K_{2\mu} - \la_3 \kom= \li (\ktm -K_{2\mu}) = \li K_{11\mu}$ as required.

It is natural to try 
\be
K_{\underbrace {1....1\mu}_{n}} = k_{n\mu} - \sum_{[n]_i\in [n]' }K_{[n]_i\mu}
\ee
where $[n]'$ indicates all the partitions of $n$ {\em except} $\underbrace{1...1}_{n}$.

We prove this by recursion: 
\be  \label{Kn}
K_{[n]\mu}\equiv \sum _{[n]_i\in [n]} K_{[n]_i\mu} = k_{n\mu}
\ee 

{\bf Proof:}

Let us assume that the above is true for $n$. Consider $K_{[n+1]'_i\mu}$.
We have
\[
\delta K_{[n+1]'_i\mu}=\sum _{m\in [n+1]'_i} \la_m K_{[n+1]'_i/m\mu}
\]
The sum, as always, is over distinct $m$'s.
This is true because such $K$'s have all been explicitly constructed for all $n$.

For e.g. let us explicitly write out the coefficient of $\lt$ in the above equation - it is $\lt K_{[n+1]'_i/2}$.
Thus we can write
\[
 \delta K_{[n+1]'_i\mu}=\li K_{[n+1]'_i/1\mu}+\lt K_{[n+1]'_i/2\mu}+\la_3 K_{[n+1]'_i/3\mu}+...
\]
Note that $[n+1]'_i/2$ is a partition of $n+1$ with one 2 removed. If we sum over all $i$ this gives all the partitions of
$n+1$ with one 2 removed, i.e. {\em all partitions of $n-1$} i.e.  $ [n-1]$. Similarly $[n+1]'_i/3$ summed over all $i$ gives all partitions
of $n-2$, i.e. $[n-2]$. However $[n+1]'_i/1$ gives all partitions of $n$ {\em except for the one with all one's},i.e. it gives $[n]'$.
Now sum over $i$ and note that the LHS is  $\sum _iK_{[n+1]'_i}$ and has all the $K$'s at this level except for $K_{\underbrace{1...1}_{n+1}}$.
 
 Thus
\[
 \sum _i \delta K_{[n+1]'_i\mu}=\li K_{[n]'\mu}+\lt K_{[n-1]\mu}+\la_3 K_{[n-2]\mu}+..\la_mK_{[n+1-m]\mu}.
\]

Using (\ref{Kn}) we see that this becomes
\[
 \sum _i \delta K_{[n+1]'_i\mu}=\li K_{[n]'\mu}+\lt k_{(n-1)\mu}+\la_3 k_{(n-2)\mu}+..\la_mk_{(n+1-m)\mu} +....
\]
\[
=(\li (K_{[n]^\mu} - K_{\underbrace{1...1}_{n}\mu})+\lt k_{n-1\mu}+\la_3 k_{n-2\mu}+..\la_mk_{n+1-m\mu} +....
\]
\[
=(\li k_{n\mu} - \delta K_{\underbrace{1...1\mu}_{n+1}}+\lt k_{(n-1)\mu}+\la_3 k_{(n-2)\mu}+..\la_mk_{(n+1-m)\mu} +....
\]
So
\[
\sum _i \delta K_{[n+1]_i\mu}= \delta k_{(n+1)\mu}
\]Thus 
\[
K_{[n+1]\mu} = k_{n\mu}
\]
Since it is true for $n=2,3$ this completes the proof. 

This completes the construction of $K$'s for all levels.   There is one more step to be taken before we can use these to write down
an interaction term in string theory. This is to perform a truncation of the spectrum using the Q-rules which have been worked out up to level 5.
This can be implemented on all the terms in the EOM. This is a fairly mechanical step and we do not do it in this review.

{\bf Level 3}

We can now construct the level 3 gauge invariant field strength using these formulae:
 \[K_{3\mu} = y_3\kom =( \bar q_3 - \bar q_2\bar q_1 + {\bar q_1^3\over 3})\kom \]
\[K_{21\mu} = y_2\kim= (\bar q_2 - {\bar q_1^2\over 2})\kim\]
\[K_{111\mu} = k_{3\mu} - K_{21\mu} -K_{3\mu}\]

The level three part of the exponent of $\cal L$ is
\[K_{3\mu} {\p Y^\mu\over \p x_3} + K_{21\mu} {\pp Y^\mu \over \p x_2 \p x_1}+K_{111\mu} {\p^3 Y^\mu\over \p x_1^3}\]

and the full Lagrangian at Level 3 is ($Y^\mu_n \equiv {\p Y^\mu\over \p x_n}$):
\[
{\cal L}=[i K_{3\mu} {Y^\mu_3} + iK_{21\mu} {\pp Y^\mu \over \p x_2 \p x_1}+iK_{111\mu} {\p^3 Y^\mu\over \p x_1^3} - K_{2\mu} K_{1\nu} \ytm \yin 
\]
\be-K_{11\mu} K_{1\nu} {\pp Y^\mu\over \p x_1^2} \yin -i {\kim \kin \kir\over 3!}\yim \yin Y_1^\rho]\e
\ee
We evaluate the functional derivative
\[
{\delta \over \delta Y^\mu(z')}\int du~  {\cal L}(u)
\]
to obtain:
\[ L_{3\mu(z)}\equiv \Big[V_{3\mu \nu}Y_3^\nu(z) + V_{21\mu \rho \sigma} Y_2^\rho(z) Y_1^\sigma(z) + V_{111\mu \lambda \rho \sigma} Y_1^\lambda(z) Y_1^\rho (z)Y_1^\sigma (z)\Big] e^{i\ko.Y(z)}
\]
where
\br
V_{3\mu \rho}&=&-\kom[K_{3\rho} + K_{21\rho} + K_{111\rho}] + \kim [K_{11\rho} + K_{2\rho}] + K_{2\mu} \kir - K_{11\mu} \kir - K_{21\mu} \kor + K_{111\mu} \kor + K_{3\mu} \kor \nonumber \\
V_{21\mu \rho\sigma} &=& i\Big[ -\kom K_{11\rho} \kir +  \kim K_{11\rho} \kos + \kim K_{2\rho} \kos + \kim \kir \kis \nonumber
\\& &-2 K_{11\mu} \kir \kos -K_{11\mu} \kor \kis - K_{21\mu}\kor \kos + 3 K_{111\mu} \kor \kos \Big] \nonumber \\
V_{111}
^{\mu \la \rho \sigma}&= &
{1\over 3!} 
\kom k_{1\la} \kir \kis -
 {1\over 3!}\kim k_{1(\lambda} \kir \kos) + 
{1\over 3} K_{11\mu} k_{1(\lambda} \kor \kos) - K_{111\mu} k_{0\lambda} \kor \kos
\er
$L_{3\mu}(z)$ is a gauge invariant field strength for the massive level 3 fields. Note once again that the non-zero mass ($q_0$) is crucial for being able to construct such an object.  

\subsubsection{Interacting Equations of Motion}

The interacting part of ERG can be written as a sum of terms of the form given below. It is of the form:
Field strength $\times $ Field strength. 
The fact that the equations of motion are just quadratic is not surprising given that the basic vertex involves splitting and joining of strings.

However an interesting point is that the field strengths are gauge invariant under the {\em same}
transformations as that of the free theory. This is unlike non Abelian gauge theories, where the transformation law is modified
when the coupling constant is non zero. In this sense the equations are Abelian. Note also that if Chan Paton factors are included
to make the low energy sector Yang Mills instead of electromagnetism, then the gauge transformation laws would be modified even in the loop variable formalism.

For example an interaction involving two $V_3$'s given above is:
\[
\int dz ~\int dz' \dot G(z,z')\eta^{\mu\nu} V_{3\mu \rho}Y_3^\rho (z)e^{ik_{0A}.Y(z)} V_{3\nu \sigma} Y_3^\sigma (z')e^{ik_{0B}.Y(z')}
\]

In all these equations we have not made any restriction on the range of $\mu$. The construction superficially works for $\mu=\theta$ also.
However we need some restriction on the Green function $G^{\theta \theta}(z,z')$ if we are to reproduce string theory S-matrix. We certainly do not want the structure of the Veneziano amplitude to be modified. Thus the simplest solution is to set $G^{\theta \theta}(z,z')=0$
when $z\neq z'$ and leave it unchanged for $z=z'$ because the free equations require that. As a constructive algorith this is fine. Whether we can deduce this from first principles is an open problem.

In this BRST string field theory superficially looks non Abelian even when the gauge group for the massless gauge field is $U(1)$. The gauge transformation law of the massless photon has extra pieces due to interactions. However it is possible that there may be field redefinitions that get rid of these. There is some hint about this possibility in \cite{Justin}.

\section{Curved Space Time}
\setcounter{equation}{0}
\subsection{ Map to curved space time: problem of gauge invariance}
         
         Thus far we have been working in flat space time. The loop variable expressions are mapped in a straightforward way to space time fields and their derivatives. Expressions that are gauge invariant when written in terms of loop variables, continue to be gauge invariant  after the map to space time fields.  If our aim is a manifestly background independent formalism then we must learn how to apply all this in curved space time. This issue will become more acute when we deal with closed string theory in the next section. So it is appropriate to sort this issue out first. Using Riemann Normal Coordinates (RNC) is the first step towards solving this problem.
         
         The loop variable method gives us equations in momentum space. Thus writing  

\be	\label{exp}
f(x)=\int dk~e^{ikx}f(k)=\int dk~(1+ik.x + {(ik.x)^2\over 2!}+...)f(k)=f(0)+ x\p f(0) + {x^2\over 2!}\pp f(0)+...
\ee
we see that powers of $k$ correspond to terms in the Taylor expansion. Thus we need to do a Taylor expansion in curved space time. This is best done using Riemann Normal Coordinates \cite{AGFM,Pet,Eis}. Appendix A \eqref{appena} contains a summary of all the aspects of the Riemann Normal Coordinate system that are needed for this paper. Note that dimensional reduction has to be done before we talk about curved space time. So $\mu$ ranges from 0 to D-1.  
          
Consider a loop variable expression $\kom  \kir \kis$. In curved space time we work in the Riemann Normal Coordinate (RNC) system and interpret $i \kom \approx {\p\over \p \bar Y^\mu}$ where $\bar Y^\mu$ are RNC's.  
Then the map to space time fields  involves integrating over $\Psi[\ko,\ki,..\li,..]$ and is: 
\[
\lan \kir \kis \ran = S_{11\rho \s}(\ko)  ~~~;\int d\ko e^{i\ko.\bar Y(z)} S_{11\rho \s}(\ko)=S_{11\rho \s}(\bar Y(z))
\]
Thus
\be	\label{exp2}
\int d\ko [1+ i\kom \bar Y^\mu(z) +...] S_{11\rho \s}(\ko) = S_{11\rho \s}(0)+ \p_\mu S_{11\rho \s}(0) \bar Y^\mu (z)+...
\ee
Thus we can conclude that the loop variable expression $\kom  \kir \kis$ gets mapped to  $\p_\mu S_{11\rho \s}(0)$ which is the first term
in the Taylor expansion of the vector. Similarly $\kom  \kon \kir \kis$ gets mapped to $\p_\mu \p_\nu S_{11\rho \s}(0)$, the second term in the Taylor expansion etc. 
 We also know from \eqref{Taylor} of the Appendix A \eqref{appena}, how to write these Taylor expansion coefficients in a manifestly covariant form.

 Thus for instance
 \[
 \lan\kom  \kir \kis\ran= \n_\mu S_{11\rho\s}(0)
 \]
 and
 \[
\lan \kom \kon \kir \kis \ran = \{\n_\nu \n_\mu S_{\rho \s}(0) 
~-~{1\over 3}
 (R^\beta _{~\mu \rho \nu}(0) 
S_{ \beta \s}(0)+R^\beta _{~\mu \s \nu}(0) 
S_{ \beta \rho}(0))\}
\]
Note that the LHS is manifestly symmetric in $\mu,\nu$. Since covariant derivatives do not commute, the first term is not symmetric. The role of the second term is to compensate for this and make the RHS also symmetric in $\mu,\nu$.
Thus
 this method gives us a map from Loop Variable expressions to covariant expressions in curved space time. However as we will see below,  this map does not commute with gauge transformation.  Let $\cal L$ be the set of loop variable expressions that gets mapped to the set $\cal S$, of  expressions  involving space time fields. Let us call this map $\cal M$. 
 Thus 
 \be	\label{ex}
 {\cal M}: \kom \kir \kis \in {\cal L} \to \lan \kom \kir \kis\ran = \n_\mu S_{11\rho\s} \in {\cal S}
 \ee
 
 Now one can perform a gauge transformation $g$ on the loop variable expression and then map it to space time fields by $\cal M$. This can be compared with the gauge transformation of the expression involving space time fields. The question is whether the result is the same whether we go from ${\cal L}$ to ${\cal S}^g$ along either path.
 \br
{\cal M}: &{\cal L}& \longrightarrow {\cal S}\nonumber \\
&g \downarrow&~~~~g\downarrow\nonumber\\
{\cal M}:& {\cal L}^g& \longrightarrow {\cal S}^g
\er
 
 We know the gauge transformation law of $ S_{11\rho\s}$:
 \[
 g:  S_{11\rho\s}\to  S_{11\rho\s} + \n _{(\rho} \Lambda_{11\s)}
 \]
 Therefore
 \be    \label{mg}
 g:  \n_\mu S_{11\rho\s}\to \n_\mu  S_{11\rho\s} + \n_\mu \n _{(\rho} \Lambda_{11\s)}
\ee
 Simlarly
 \[
 g: \kom \kir \kis \to \kom \kir \kis+\kom (\li \kor \kis + \li \kis \kor)
 \]
 We now have to compare \eqref{mg}
 \be	\label{gm}
 {\cal M} :\kom (\li \kor \kis + \li \kis \kor) \to \{\n_{\mu} \n_{(\rho } \Lambda_{11\s)}(0)~-~{1\over 3}
 R^\beta _{~(\rho \s ) \mu}(0) 
\Lambda_{11 \beta }(0)\}
\ee
 
 We now have to compare \eqref{mg} with \eqref{gm} obtained by the other path. We see that they differ by terms proportional to the curvature tensor. Thus while in flat space they agree, in curved space they don't. The non commutativity of this process has the consequence that loop variable expressions that are gauge invariant  do not get mapped to gauge invariant expressions when mapped to space time fields.
 In a nutshell this is because $\kom \kon = \kon \kom$ but $\n_\mu \n_\nu \neq \n_\nu \n_\mu$.

\subsection{Prescription}

\subsubsection{Motivating the prescription}

We now give a well defined constructive algorithm for mapping to space time in such a way that gauge invariance is maintained. This was described in \cite{BSERGclosed3}.
 The prescription involves $q_n$ in a crucial way and so dimensional reduction with mass is important. In this section $\mu$ will range from 0 to D-1.  
 
We have seen the problem arises
when gauge transformation produces an extra derivative in the form $\kn \to \kn + \la_n \ko +....$. 
The problem is solved by rewriting the loop variable expression in such a way that no extra derivative terms appear in any gauge transformation.   All derivatives lurking in gauge transformations are made manifest right away. This can be implemented as follows:
Define
\be
 k_{n\mu} =\tilde k_{n\mu} + y_n\kom
\ee
where $y_n \to y_n + \la _n$ under a gauge transformation. $y_n$ have been defined earlier \cite{BSERGopen2}:
\be
\sum_{n=0} q_n t^{-n} = q_0 e^{\sum_{m=1} y_m t^{-m}}
\ee
Gauge transformation of $k_{n\mu}$ is given by
\be
k_{n\mu} \to k_{n\mu} + \la_1 k_{n-1\mu} + \la_2 k_{n-2\mu} +.....\la_{n-1}\kim + \la _n \kom
\ee 
Thus $\tilde k_{n\mu}$ satisfies a gauge transformation rule
\be
\tilde k_{n\mu} \to \tilde k_{n\mu} +   \la_1 k_{n-1\mu} + \la_2 k_{n-2\mu} +.....\la_{n-1}\kim
\ee
Once the loop variable expressions are written in terms of tilde variables, gauge transformations do not produce any new derivatives.
Thus all the required curvature couplings are present right in the beginning. Now we can map these to new space time fields (also with tildes). 
This ensures that the map to space time fields commutes
with gauge transformations. Expressions that were gauge invariant at the level of loop variables, continues to be gauge invariant in terms of the tilde space time fields.  The tilde fields are expressible in terms of the original fields - these are just field redefinitions. 
Once we have a covariant gauge invariant expression in terms of tilde space time fields, we can undo the field redefinitions. Field redefinitions done on the space time fields do not change any of the gauge transformation properties. 

\subsubsection{Illustration of Procedure}

Consider the level 2 field
\be
\lan  \kim \kin \ran = S_{11\mu \nu}
\ee 
We define tilde variables by 
\be
 \kim = \tilde k_{1\mu} + y_1 \kom
\ee
Then
\be
\kim \kin =   \tilde k_{1\mu}  \tilde k_{1\nu} +  \kom y_1 \tilde k_{1\nu} + \kon y_1  \tilde k_{1\mu} + y_1^2 \kom \kon
\ee
 

Define tilde space time fields by
\br   \label{tilde}
\lan  \tilde k_{1\mu}  \tilde k_{1\nu} \ran &=& \tilde S_{11\mu\nu}\nonumber \\
\lan y_1 \tilde k_{1\nu}\ran &=& {\tilde S_{11\nu}}\nonumber \\
\lan y_1^2\ran &=& {\tilde S_{11}}
\er
 The relations between the old space time fields and the tilde fields are given by:
 \be   \label{FD1}
 S_{11\mu\nu} = \tilde S_{11\mu\nu} + \nabla _{(\mu} {\tilde S_{11\nu)}} + \nabla_\mu\nabla_\nu {\tilde S_{11}}
\ee

\be   \label{FD2}
 \tilde S_{11\mu} ={S_{11\mu}\over \qo} -{1\over \qo^2} \nabla_\mu S_{11}~~~~\tilde S_{11}= {S_{11}\over \qo^2}
\ee  
where
\be
\lan q_1 \kim \ran = S_{11\mu} ;~~~~~\lan q_1^2\ran = S_{11}
\ee
and also $q_1=y_1 q_0$.

Let us turn to the gauge transformation laws for these fields: Using $\kim \to \kim + \li \kom$, $y_1 \to y_1+\li$, we obtain
\br  \label{GT1}
\delta \tilde S_{11\mu\nu} &=& 0 \nonumber \\
\delta \tilde S_{11\mu}&=& \lan \li \tilde k_{1\mu} \ran =\tilde \Lambda_{11\mu}\nonumber \\
\delta \tilde S_{11} &=& 2\lan y_1\li \ran = 2 \tilde \Lambda_{11}
\er
We see that there are no derivatives in the transformation laws.

If we define 
\be
\lan \li \kim \ran = \Lambda_{11\mu} = \lan \li \tilde k_{1\mu} + \li y_1 \kom \ran =\tilde \Lambda _{11\mu} + \nabla _\mu \tilde \Lambda_{11}
\ee
we see that combining \eqref{FD1}, \eqref{FD2} and \eqref {GT1}  gives:
\br   \label{GT2}
\delta S_{11\mu\nu} &=& \nabla_{(\mu}\Lambda_{11\nu)}\nonumber \\
\delta S_{11\mu}&=& \nabla _\mu \Lambda_{11} + \qo \Lambda_{11\mu}\nonumber \\
\delta S_{11} &=& 2 \Lambda_{11}\qo
\er 

This is of course the correct gauge transformation for the original space time fields. This illustrates the (obvious) fact that the above field redefinitions do not introduce any inconsistencies in gauge transformations. This has to do with the fact that gauge transformations of the tilde fields do not involve derivatives.

Let us now apply this prescription to the  loop variable expression considered earlier:
\be  \label{LV1}
\kom \kir \kis
\ee
In terms of tilde variables it is
\be  \label{LV2}
\kom (\tilde k_{1\rho} \tilde k_{1\sigma} + \kor y_1  \tilde k_{1\sigma}+\kos  y_1 \tilde k_{1\rho} + \kor \kos y_1^2)
\ee
Mapping to space-time fields in the usual way (described in the last subsection) we get
\be  \label{ST1}
\nabla_\mu \tilde S_{11\rho\sigma} + \nabla _\mu \nabla_\rho \tilde S_{11\sigma}+ \nabla _\mu \nabla_\sigma \tilde S_{11\rho} +{2\over 3}(R^\beta _{~\rho \mu \sigma} +R^\beta _{~\sigma \mu \rho}) \tilde S_{11\beta} + \nabla _\mu \nabla _\rho \nabla_\sigma \tilde S_{11} + {1\over 3}(R^\beta_{~\rho \mu \sigma} +R^\beta_{~\sigma \mu \rho})\nabla_\beta \tilde S_{11}
\ee

Since the gauge transformation of the loop variable expression \eqref{LV2} does not produce any extra $\kom$ its space-time map is guaranteed to coincide with the gauge transformation of \eqref{ST1}. Thus we have an internally self consistent prescription for mapping to space-time fields.

The last step of the prescription is to rewrite \eqref{ST1} in terms of our old space time fields. This does not modify the gauge transformation properties of this expression. We use the field redefinitions \eqref{FD1},\eqref{FD2}. The two and three derivative term cancel. This is clear because these terms do not involve the curvature tensor and one might as well have been in flat space. In flat space the original expression
had only one space time derivative. Performing some field redefinitions and then undoing them gets you back to the starting point.

\be \label{ST2}
\nabla_\mu S_{11\rho\sigma} + {2\over 3} (R^\beta_{~\rho \mu \sigma} +R^\beta_{~\sigma \mu \rho})[{S_{11\beta}\over \qo} - \hf \nabla _\beta {S_{11}\over \qo^2}]
\ee

Now we can compare the gauge transforms of \eqref{LV1} and \eqref{ST2} to verify that they agree. The gauge variation of \eqref{LV1} is
\be  \label{GT3}
\kom (\kor \li \kis + \kos \li \kir)
\ee
Mapping \eqref{GT3} to space-time fields gives
\be
\nabla_\mu (\nabla_\rho \Lambda_{11\sigma} +\nabla_\sigma \Lambda_{11\rho}) + {2\over 3}(R^\beta _{~\rho \mu \sigma} +R^\beta _{~\sigma \mu \rho})\Lambda_{11\beta}
\ee
The gauge variation of \eqref{ST2} is (using \eqref{GT2}) seen to be the same as above.

\subsubsection{Summary of prescription}

We summarize the prescription below:
\begin{enumerate}
\item Step 1: Define tilde variables by $k_{n\mu}= {\tilde k}_{n \mu} + y_n\kom$. Their gauge transformation law has no $\kom$.
Rewrite all loop variable expressions in terms of tilde variables.

\item Step 2: Define new space time fields by mapping the tilde variables to new (tilde) space time fields using the map $\cal M$. Obtain the relation between the old space time fields and the new ones.

\item Step 3: Map all loop variable expressions (now written in terms of tilde variables) to expressions involving the new space time tilde fields by the same map $\cal M$.  At this point a gauge invariant expression involving loop variables is mapped to a gauge invariant expression involving tilde space time fields. This expression may involve higher derivatives than the expression we started with.

\item Step 4: Rewrite the tilde  space time fields in terms of the old space time fields. All the higher derivative terms will cancel. We get
an expression involving space time fields which has the naive covariantization, plus some extra curvature couplings to Stuckelberg fields and derivatives thereof. 
\end{enumerate}
\subsection{Example: Open String Level 2}

The free equation of motion is \cite{BSLV,BSERGopen1,BSERGopen2}:
\[
-\ki.\ki i\ko Y_2 -\hf \ki.\ki (i\ko.Y_1)^2 -\ki.\ko (\ki.Y_1)(\ko.Y_1) 
\]
\be
+\hf \ko^2 (\ki.Y_1)^2 -\ko^2i\kt. Y_2 + i\ki.\ko(\ki.Y_2) +\kt.\ko i\ko.Y_2 =0
\ee
The coefficient of $\yim \yin$ is
\be	\label{EOMspin2}
\ko^2 \kim \kin - \ko .\ki k_{1(\mu }k_{\nu)0} + \ki.\ki \kom \kon =0
\ee 

This is  gauge invariant under $\kim \to \kim + \li \kom$. 
Dimensional reduction gives a massive field with EOM:
\be  \label{EOMspin2mass}
(\ko^2+\qo^2) \kim \kin -( \ko .\ki+ \qo q_1) k_{1(\mu }k_{\nu)0} + (\ki.\ki+q_1q_1) \kom \kon =0
\ee

This has to be mapped to space time fields.

Let us apply this procedure to a general loop variable expression 
\be  \label{genterm}
 \kom \kon \kir\kis
\ee
 All the terms in \eqref{EOMspin2mass} involving the tensor field can be obtained from this by contractions.
 
 {\bf Step 1}
 
 We let $\kim = \tilde k_{1\mu} + y_1 \kom$. Then \eqref{genterm} becomes
\be \label{step1}
\kom \kon \kir\kis=\kom \kon (\tkir \tkis + \tkir y_1 \kos + \kor y_1 \tkis +y_1^2 \kor \kos)
\ee

{\bf Step 2} has already been done - the required tilde space time fields were defined there in \eqref{tilde}.

{\bf Step3}

Let us consider each term in turn and map to space time fields, using the definitions \eqref{tilde} and the map {\cal M}.   We get
\br \label{step2}
\lan \kom \kon \tkir \tkis \ran &=& \hf[ \nabla _{(\mu }\nabla _{\nu)}\tilde S_{11\rho \sigma} + {1\over 3} R^\beta_{~ (\mu \nu) \rho}\tilde S_{11\beta \sigma} + {1\over 3}R^\beta_{~ (\mu \nu) \sigma}\tilde S_{11 \rho\beta}]\nonumber \\
\lan \kom \kon \kos \tkir y_1 \ran &=& {1\over 6}[\nabla_{(\mu}\nabla_\nu \nabla_{\sigma)} \tilde S_{11\rho}-R^\beta_{~(\mu |\rho|\nu } \nabla_{\rho)} \tilde S_{11\beta}- \hf \nabla_{(\sigma} R^\beta_{~\mu |\rho |\nu)}\tilde S_{11\beta}]\nonumber \\
\lan \kom \kon \kor \kos y_1^2\ran &=& {1\over 4!} \nabla_{(\mu} \nabla_\nu \nabla_\rho \nabla_{\sigma)} \tilde S_{11}
\er

The gauge transformations of the tilde variables are given in \eqref{GT1} and do not involve derivatives. Thus in \eqref{step2} the gauge transformations of the LHS and RHS are guaranteed to agree. Now we proceed to Step 4:

{\bf Step 4}

Now we can redefine fields in terms of the original fields using \eqref{FD1} and \eqref{FD2}:
\br   \label{step3}
\lan \kom \kon \tkir \tkis \ran &=& \hf \Big( \nabla _{(\mu }\nabla _{\nu)}[ S_{11\rho \sigma}- {\nabla_{(\rho} S_{11\sigma)}\over \qo^2}+ {\nabla_\rho \nabla_\sigma S_{11}\over \qo^2}] + \nonumber \\& &{1\over 3} R^\beta_{~ (\mu \nu) \rho}[ S_{11\beta \sigma}- {\nabla_{(\beta} S_{11\sigma)}\over \qo^2}+ {\nabla_\beta \nabla_\sigma S_{11}\over \qo^2}] + {1\over 3}R^\beta_{~ (\mu \nu) \sigma}[ S_{11\rho \beta}- {\nabla_{(\rho} S_{11\beta)}\over \qo^2}+ {\nabla_\rho \nabla_\beta S_{11}\over \qo^2}]\Big)\nonumber \\
\lan \kom \kon \kos \tkir y_1 \ran &=& {1\over 6}\Big(\nabla_{(\mu}\nabla_\nu \nabla_{\rho)} [ {S_{11\sigma}\over \qo} - {\nabla_\sigma S_{11}\over \qo^2}]-R^\beta_{~(\mu |\sigma|\nu } \nabla_{\rho)} [ {S_{11\beta}\over \qo} - {\nabla_\beta S_{11} \over \qo^2}]- \hf \nabla_{(\rho} R^\beta_{~\mu |\sigma |\nu)}[ {S_{11\beta}\over \qo} - {\nabla_\beta S_{11}\over \qo^2}]\Big)\nonumber \\
\lan \kom \kon \kor \kos y_1^2\ran &=& {1\over 4!} \nabla_{(\mu} \nabla_\nu \nabla_\rho \nabla_{\sigma)} { S_{11}\over \qo^2}
\er

We now
 substitute \eqref{step3} in the original loop variable expression written in terms of tilde variables \eqref{step1}. All the higher derivative terms  cancel and the final curvature independent term is  the expected covariantization of the flat space term: $\nabla_\mu \nabla_\nu S_{11\rho\sigma}$. There are also the terms associated with the naive covariantization. So the final expression starts off as
\be    \label{op-step3}
\underbrace{\nabla_\mu \nabla_\nu S_{11\rho\sigma} +{1\over 3}(R^\beta _{~\nu \mu \rho}+R^\beta _{~\rho \mu \nu})S_{\beta \sigma}+{1\over 3}(R^\beta _{~\nu \mu \sigma}+R^\beta _{~\sigma\mu \nu})S_{\rho \beta}}_{Naive~covariantization} + (curvature \times stuckelberg~fields)
\ee

The Stuckelberg fields $S_{11\mu} ,S_{11}$ are required for gauge invariance. The gauge transformations are given in \eqref{GT2}.  They can be set to zero by a gauge transformation.

The full answer is 
\br
\lan \kom \kon \kir \kis \ran &=& \hf [\nabla_{(\mu}\nabla_{\nu)} S_{\rho \sigma} -{1\over 3} R^\beta_{~(\mu |\rho|\nu)}S_{\beta \sigma}-{1\over 3}R^\beta_{~(\mu |\sigma|\nu)}S_{\rho \beta} ] + \nonumber \\ & &{1\over 6}[R^\beta_{~(\mu |\rho|\nu)}({\nabla_{(\beta}S_{\sigma)}\over \qo} -{\nabla_\beta \nabla_\sigma S\over \qo^2}) +R^\beta_{~(\mu |\sigma|\nu)}({\nabla_{(\beta}S_{\rho)}\over \qo} -{\nabla_\beta \nabla_\rho S\over \qo^2})]-\nonumber \\
& & {1\over 6}[R^\beta_{~(\mu |\sigma|\nu}\nabla_{\rho)}({S_\beta\over \qo} - {\nabla _\beta S\over \qo^2}) -\hf \nabla_{(\rho}R^\beta_{~\mu |\sigma|\nu)}({S_\beta\over \qo} - {\nabla _\beta S\over \qo^2})]\nonumber \\
& &-\hf[3 R^\beta_{~\sigma \rho \nu}\nabla_\mu S_\beta +3 R^\beta_{~\sigma \rho \mu}\nabla_\nu S_\beta + R^\beta_{~\nu \rho \mu}\nabla_\beta S_\sigma +R^\beta_{~\mu \rho \nu}\nabla_\beta S_\sigma + 3 R^\beta_{~ \rho \nu \mu}\nabla_\beta S_\sigma+3 R^\beta_{~ \sigma \nu \mu}\nabla_\rho S_\beta]\nonumber \\
& &+2(\nabla_\nu R^\beta_{~ \sigma \rho  \mu})S_\beta+2(\nabla_\mu R^\beta_{~ \sigma \rho  \nu})S_\beta]+\hf[R^\beta_{~ \sigma \nu  \mu}\nabla _\rho S_\beta +R^\beta_{~ \rho \nu  \mu}\nabla _\beta S_\sigma] \nonumber \\
& \equiv& F_{\mu\nu\rho\sigma}
\er

The EOM \eqref{EOMspin2mass} can now be written in terms of $F$ as 
\be
G^{\alpha \beta}[F_{\alpha \beta \mu \nu} - F_{\alpha(\mu|\beta|\nu)} + F_{\mu \nu \alpha \beta}] + \qo^2 S_{\mu\nu} -\qo \nabla_{(\mu}S_{\nu)}+\nabla_\mu \nabla_\nu S=0
\ee
where $G_{\mu\nu}$ is the space-time metric. This equation is both manifestly covariant and also gauge invariant by construction.
Note that the Stuckelberg fields $S_\mu,S$ can be set to zero by a gauge transformation \eqref{GT2}.

\subsection{Interacting Equation}

The interaction equation can be written in curved space using the same techniques. The main new complication is that the equation involves fields (or more precisely, gauge invariant field strengths) at two different space time points. Thus when we expand the exponential as in \eqref{exp} and \eqref{exp2}, it is about the origin of the RNC. For the interacting case we have two such expansions, both about the same origin.
The origin has been taken as $z=0$, with $\bar Y(0)=0$.  It is also possible to take one of the points to be $z=0$. In that case only one of the exponentials have to be expanded. 

A typical interaction terms is of the form

\be  \label{Int}
\int dz_1 dz_2 \dot G(z_1,z_2,a)K_{1\mu \nu \rho} [k_n]e^{i\ko.\bar Y(z_1)} \bar Y_1^\mu \bar Y_1^\nu \bar Y_1^\rho(z_1)K_{2\al \beta \gamma} [k'_n]e^{ik_0'.\bar Y(z_2)} \bar Y_1^\al \bar Y_1^\beta \bar Y_1^\gamma(z_2)
\ee

As mentioned above, the exponentials have to be understood as a power series that stands for the Taylor expansion described in \eqref{Taylor}. In mapping these expressions to space-time fields the same four step procedure can be followed.  The point of departure being that each interaction product will involve an infinite series of terms involving higher derivatives

The vertex operators can be taken to be normal ordered because the free equation takes into account all the contributions due to self contractions.

Thus for the OPE of normal ordered exponentials in flat space we have:
\br  \label{OPE}
: e^{i\ko.Y(z_1)}::e^{ip_0.Y(z_2)}: &=& e^{-\kom p_{0\nu} \lan Y^\mu(z_1) Y^\nu(z_2)\ran}:e^{i\ko.Y(z_1)+ip_0.Y(z_2)}:\nonumber \\
&=&e^{-\kom p_{0\nu} \lan Y^\mu(z_1) (Y^\nu(z_1)+ (z_2-z_1)\p_z Y^\nu(z_1) + {(z_1-z_2)^2\over 2!} \pp _zY^\nu(z_1)+...\ran}\nonumber \\
& &:e^{i\ko.Y(z_1)+ip_0.(Y^\nu(z_1)+ (z_2-z_1)\p_z Y^\nu(z_1) + {(z_1-z_2)^2\over 2!} \pp _zY^\nu(z_1)+...)}:
\er

Note that the contraction involves the green function $ \lan Y^\mu(z_1) Y^\nu(z_2)\ran$, which has been Taylor expanded. 
This can be done in curved space time also, provided we work in RNC. Each term in the expansion of the Green function is a {\em function} of $z_1-z_2$ and the cutoff $a$. The tensor structure in flat space is $\eta^{\mu\nu}$ which becomes $\bar g^{\mu\nu}(0)$ in the RNC and $g^{\mu\nu}(x_0)$ in a general coordinate system. The equations obtained are local ones, for fields at $x_0$, which is any arbitrary point on the manifold - so these equations are completely general.  

The expansion of $\bar Y(z)$ in the exponent generates higher dimensional vertex operators in flat space. Mapping this RNC expansion to curved space covariantly gives \eqref{VOP}. We see in general that the Riemann tensor appears at higher orders.

One can easily generalize \eqref{OPE} to do an OPE for arbitrary vertex operators and perform the same Taylor expansion using \eqref{Taylor} and \eqref{VOP} as done for example in \eqref{Vec} of the Appendix A \eqref{appena}.

We give a few examples below: (We let $X_o$ stand for the coordinates, in a general coordinate system $X$, of the point that is the origin for the RNC system. The coordinates in the RNC will be called $\bar Y$. The origin is  $\bar Y=0$. We also take for the $z$ dependnece $\bar Y(0)=0$. )

Let us take as an example:
\be
\lan  K_{\alpha \beta \gamma} [p_0,p_n]\ran =   F_{\alpha \beta \gamma} (\bar Y=0)\to  F_{\alpha \beta \gamma} (X_o)\\
\ee
Then
\br  \label{example}
\lan p_{0\mu} K_{\alpha \beta \gamma} [p_0,p_n]\ran &= & \bar \nabla _\mu F_{\alpha \beta \gamma} (\bar Y=0)\to \nabla _\mu F_{\alpha \beta \gamma} (X_o)\\
\lan p_{0\mu}p_{0\nu} K_{\alpha \beta \gamma} [p_0,p_n]\ran &= & \bar \nabla _\mu \bar \nabla _\nu F_{\alpha \beta \gamma} (\bar Y=0)+{1\over 3}\Big((\bar R^\lambda_{~\alpha \mu \nu}+\bar R^\lambda_{~\nu \mu \alpha})F_{\lambda \beta \gamma}(0) +\nonumber \\ & &(\bar R^\lambda_{~\beta \mu \nu}+\bar R^\lambda_{~\nu \mu \beta})F_{\alpha \lambda \gamma}(0)+(\bar R^\lambda_{~\gamma \mu \nu}+\bar R^\lambda_{~\nu \mu \gamma})F_{\alpha \beta \lambda}(0)\Big)\nonumber \\&\to& \nabla _\mu \nabla _\nu F_{\alpha \beta \gamma} (Y_o)+{1\over 3}\Big((R^\lambda_{~\alpha \mu \nu}+R^\lambda_{~\nu \mu \alpha})F_{\lambda \beta \gamma}(X_0) +\nonumber \\ & &(R^\lambda_{~\beta \mu \nu}+R^\lambda_{~\nu \mu \beta})F_{\alpha \lambda \gamma}(X_0)+(R^\lambda_{~\gamma \mu \nu}+R^\lambda_{~\nu \mu \gamma})F_{\alpha \beta \lambda}(X_0)\Big)
\er
All these expressions are tensors at the origin $X_0$ of the RNC.
Also for the contractions one needs Taylor expansions of the Green function, for example
\be \label{contr}
\lan \bar Y^\mu (z_1)\bar Y^\nu(z_1)\ran = \eta^{\mu\nu} G(z_1,z_1;a) ~~~;~~~\lan \bar Y^\mu (z_1)\p_z \bar Y^\nu(z_1)\ran = \eta^{\mu\nu} \p_{z_2}G(z_1,z_2;a)|_{z_2=z_1}
\ee
In a general coordinate system we simply replace $\eta ^{\mu \nu}$ by $g^{\mu \nu}(X_0)$ in the above equation.

Putting all this together one obtains on expanding the exponentials in \eqref{OPE}  
\[
\lan -\int \int~dz_1 dz_2~[\dot G(z_1,z_1;a)+ (z_1-z_2)\dot G'(z_1,z_1;a)+...][ 1-\kom p_{0\nu} g^{\mu\nu}(Y_0)G(z_1,z_1;a)+...]
\]
\[
K_{\lambda\sigma\rho}[\ko,\kn]\p_{z_1}Y^\lambda (z_1)\p_{z_1}Y^{\sigma}(z_1)\p_{z_1}Y^\rho(z_1)
K_{\alpha \beta \gamma}[p_0,p_n]\p_{z_1}Y^\alpha(z_1)\p_{z_1}Y^\beta(z_1)\p_{z_1}Y^\gamma(z_1)
\]
\[
e^{i(\ko+p_0).Y_0} (1+ (z_2-z_1)p_{0}.\p_{z_1}Y(z_1)+...)\ran
\]
\[
=-\int \int~dz_1 dz_2~[\dot G(z_1,z_1;a)+ (z_2-z_1)\dot G'(z_1,z_1;a)+...]\]\[[F_{\lambda \sigma\rho}(X_0) F_{\alpha \beta \gamma}(X_0)- g^{\mu\nu}(X_0) \nabla_\mu F_{\lambda \sigma\rho}(X_0)\nabla_\nu F_{\alpha \beta \gamma}(X_0)G(z_1,z_1;a)+...]
\]\be
\p_{z_1}Y^\lambda (z_1)\p_{z_1}Y^{\sigma}(z_1)\p_{z_1}Y^\rho(z_1)
\p_{z_1}Y^\alpha(z_1)\p_{z_1}Y^\beta(z_1)\p_{z_1}Y^\gamma(z_1)+...
\ee

where we have kept a sample term at level 6. This generates terms at lower levels from contractions as in:
\[
\p_{z_1}Y^\lambda (z_1)\p_{z_1}Y^{\sigma}(z_1)\p_{z_1}Y^\rho(z_1)
\p_{z_1}Y^\alpha(z_1)\p_{z_1}Y^\beta(z_1)\p_{z_1}Y^\gamma(z_1)=
\]
\[:\p_{z_1}Y^\lambda (z_1)\p_{z_1}Y^{\sigma}(z_1)\p_{z_1}Y^\rho(z_1)
\p_{z_1}Y^\alpha(z_1)\p_{z_1}Y^\beta(z_1)\p_{z_1}Y^\gamma(z_1): +\
\]
\[
\p_{z_1}\p_{z_2}\lan Y^{\lambda}(z_1)Y^{\alpha}(z_2)\ran |_{z_1=z_2}
:\p_{z_1}Y^{\sigma}(z_1)\p_{z_1}Y^\rho(z_1)\p_{z_1}Y^\beta(z_1)\p_{z_1}Y^\gamma(z_1):+...
\]

 and in addition there are also terms at higher levels coming from expanding the exponential.

\section{Closed String Theory}
\setcounter{equation}{0}

In \cite{BSCS} gauge invariant free equations for closed string was written down. The open string loop variable is extended to closed strings by adding the anti-holomorphic counterparts. Thus the starting point is
 \be \label{CLVnaive}
e^{i\ko . X (z) + \oint_c dt~ k(t) \al(t) \p_z X(z+t) + \oint_c d\bar t~ \bar k(\bar t) \bar \al(\bar t) \p_\zb X(\zb+\bar t)}
\ee
Expanding, we get
\be   \label{naive}
 \int dz~e^{i(\ko .Y + \sum_{n,\nb=1,2,...} (\kn . \yn +  k_\nb . Y_{\nb}) )}  = \int dz~\Big( \e (1+ i \kn . \yn + i k_{\nb}.Y_{\nb} - k_{n\mu} k_{\mb \nu} \yn ^\mu Y_{{\mb}}^\nu +...\Big)
 \ee 
The holomorphic anti-holomorphic separation originates in the world sheet equation of motion for $X(z,\zb)$
\[
\p_z \p_\zb X=0
\]
This means that physical states are represented by vertex operators that are products of terms of the form $\p_z^n X$ and $\p_\zb^n X$ but  do not involve mixed derivatives $\p_z \p_\zb X$. The constraint $L_0=\bar L_0$ sets the dimensions of the holomorphic and anti holomorphic parts equal.  
In \cite{BSCS} Weyl invariance was used to obtain gauge invariant equations. The method is a simple extension of what is described in Section 3 of this review. One noteworthy feature of the method is that terms involving mixed derivatives $\p_z \p_\zb \Sigma$ have to be retained. When the Liouville field is varied,  on integration by parts, they contribute terms to the equation of motion. One way to understand the necessity of adding mixed derivative terms is the following \cite{BSERGclosed1}. While the unregulated Green function $G(z,\zb;0) = ln~(z\zb)$ obeys
$\p_z\p_\zb G(z,\zb;0)=0$, the regulated Green function, which can be taken to be
\[
 G(z,\zb;0) = ln~(z\zb+a^2e^{2\sigma})
 \]
does not. Thus as long as there is a cutoff in place, holomorphic factorization does not take place. Therefore one should introduce vertex
operators involving mixed derivatives. In the limit that $a\to 0$, which  can be taken for on-shell S-matrix element calculation, vertex operators that correspond to mixed derivatives will have zero correlators with other vertex operators. These states therefore will not contribute to the S-matrix. But for off shell calculations one should retain these states. 

There is also another argument for the introduction of mixed derivatives. Recall the argument for gauge invariance of the interacting term. It was observed that
the gauge variation of the Lagrangian at level $N$ has to be derivatives of  lower level terms in the Lagrangian. For closed strings we have $\la _n$ as well as $\la _\nb$. Thus if $L_N$ denotes the
 Lagrangian at level $N$ its gauge variation has to be of the form:
 \be \label{LN} \delta L_N =  \sum _{n,\nb =1,2,...}\lambda _n {\p L_{N-n}\over \p \xn} + \lambda _\nb {\p L_{N-\nb}\over \p \xnb}
 \ee
where for closed strings, the level involves two numbers: $N=n+\mb$. Since $L_{N-\nb}$ certainly has purely holomorphic derivative terms (and $L_{N-n}$ certainly has purely anti holomorphic derivative terms), it is clear that $L_N$ must have mixed derivative terms.

Motivated by these considerations we generalize our loop variable \cite{BSERGclosed1} to 
\[
Exp ~\Big(i\Big(\ko . X (z) + \oint_c dt~ k(t) \al(t) \p_z X(z+t) + \oint_c d\bar t~ \bar k(\bar t) \bar \al(\bar t) \p_\zb X(\zb+\bar t)+ 
\]
\be	\label{LVC}
+\oint_c dt\oint_c d\bar t~K(t,\bar t) \al(t)
\bar \al (\bar t) \p_z \p_\zb X(z+t,\zb+\tb)\Big)\Big)
\ee

Expansion for $k(t),\al(t)$ are as given earlier and $\bar k(\tb) ,\bar \al(\tb)$ are anti-holomorphic versions of the same.
The first three terms in the exponent are the terms given in (\ref{naive}). The fourth term involves $K(t,\tb)$ defined below:
\be	\label{Kt}
K(t,\tb)\equiv K_{0;0}+\sum _{\mb=1}^\infty K_{0;\mb}\tb^{-\bar m} + \sum _{n=1}^\infty K_{n;0}t^{-n} + \sum _{n=1,\mb=1}^\infty K_{n;\mb}t^{-n}\tb^{-\mb}
\ee
Expanding $X(z+t,\zb+\tb)$ gives
\be
\p_z\p_\zb X(z+t,\zb+\tb)=\p_z\p_\zb X+ t \p_z^2\p_\zb X+ \tb \p_z \p_\zb^2 X + t\tb \p_z^2\p_\zb^2X+ t^2 {\p_z^3\p_\zb X\over 2!}+t^2 {\p_z\p_\zb^3X\over 2!}+...
\ee
Plugging all this in (\ref{LVC}) gives:
\[ \ko \Big(X+ \al_1 \p_zX + \al_2 \p_z^2 X +{\al_3\p_z^3X\over 2!}+...+\bar \al_1 \p_zX + \bar \al_2 \p_\zb^2X+...+{\al_n\bar \al_m\p_z^n\p_\zb^mX\over (n-1)!(m-1)!}+..\Big)\]
\[+\underbrace{K_{1;0}}_{=k_1} \Big(  \p_zX + \al_1 \p_z^2X + {\al_2\p_z^3X\over 2!}+..+\bar \al_1\p_z\p_\zb X+
\bar \al_2\p_z\p_\zb^2X +\]\[...+\al_1\bar \al_1 \p_z^2\p_\zb X+{\al_2 \bar \al_1 \p_z^3\p_\zb X\over 2!}+...+\al_1\bar \al_2\p_z^2\p_\zb^2X+...+{\al_n \bar \al_m \p_z^{n+1}\p_\zb^m X \over n!(m-1)!}+...\Big)+
\]
\[...+K_{n;\mb}\Big({\p_z^n\p_\zb^mX\over (n-1)!(m-1)!}+ {\al_1\p_z^{n+1}\p_\zb^mX\over (n)!(m-1)!}+...+{\al_p\bar \al_q\p_z^{n+p}\p_\zb^{m+q}X\over (n+p-1)!(m+q-1)!}+...\Big)\]

If we define the coefficient of $\ko$ to be $Y$, (\ref{LVC}) can be compactly written as
\be  \label{GVC}
Exp\Big(i\Big( \ko.Y + K_{1;0}.{\p Y\over \p x_1}+K_{0;\bar 1}.{\p Y\over \p \bar x_1}+K_{1;\bar 1}.{\pp Y\over \p x_1\p \bar x_1}+...
+K_{n;\mb}.{\p^2Y\over \p \xn \p\bar x_m}+...\Big)\Big)
\ee

This is the generalization to closed strings of $\gvk$. This is correct for the free theory. For the interacting theory we need a finer resolution
of the $K's$. For instance although ${\pp Y\over \p \xn \xm \xnb \xmb} = {\pp Y\over \p x_{n+m} \p x_{\nb +\mb}}$ we need separate
terms for each. Thus $K_{n,\mb}$ has to be generalized to $K_{[n],[\mb]}$ where $[n],[\mb]$ denote all the  partitions of $n,\mb$ respectively.
 In the next section we give the construction of $K_{[n],[\mb]}$.  
\subsection{Construction of $K_{[n],[\mb]}$}

The starting point for closed strings, is to make the identification
\[ K_{\mu n,m...;0}=K_{\mu n,m,..}\] where the RHS are the $K$'s that we have just defined (for open strings). Similarly
$K_{\mu 0;\nb,\mb...}$ is given by the same expressions with bars i.e. $\nb$ instead of $n$, $k_{\bar 1\mu}$ instead of $\kim$ etc.

Now let us construct the mixed $K$'s:
\be 
K_{1;\bar 1 \mu} = \bar y_1 \kim + y_1 k_{\bar 1 \mu} -y_1 \bar y_1 \kom  = \bar y_1 K_{1;0\mu} + y_1 K_{0;\bar1\mu} - y_1 \bar y_1 \kom
\ee
It is easily verified that
 \[
  \delta K_{1;\bar 1\mu}= \la _1 k_{\bar 1\mu} + \bar \la _1 \kim 
  \]

Similarly
\[ 
K_{1,1;\bar 1\mu} = \bar y_1 (\ktm - y_2 \kom ) + {y_1^2\over 2} k_{\bar 1\mu} - {y_1^2\over 2} \bar y_1 \kom
 \]
This can be rewritten as 
\be 
K_{1,1;\bar 1\mu}=\bar y_1 K_{1,1;0\mu} +{y_1^2\over 2} K_{0;\bar1\mu} -{y_1^2\over 2} \bar y_1 \kom
\ee
One can easily check that
\[ \delta K_{1,1;\bar 1\mu}= \la _1 K_{1;\bar 1\mu} + \bar \la _1 K_{1,1;0\mu}\] as required.
The pattern is very clear:
\be
K_{\underbrace{1,1,...1}_{n};\underbrace{\bar 1,\bar 1,...\bar 1}_{m}\mu} = {\bar y_1^m\over m!}K_{\underbrace{1,1,...,1}_{n};0\mu} + { y_1^n\over n!}K_{0;\underbrace{\bar 1,\bar1,...,\bar1}_{m}\mu} - {y_1^n\over n!}{\bar y_1^m\over m!}\kom
\ee

Let us check the variation:
\br
\delta K_{\underbrace{1,1,...1}_{n};\underbrace{\bar 1,\bar 1,...\bar 1}_{m}\mu}&=& \bar \la _1\Big({\bar y_1^{m-1}\over (m-1)!}K_{\underbrace{1,1,...,1}_{n};0\mu} + { y_1^n\over n!}K_{0;\underbrace{\bar 1,\bar1,...,\bar1}_{m-1}\mu} -{y_1^n\over n!}{\bar y_1^{m-1}\over (m-1)!}\kom \Big) \nonumber \\
 & +&\la _1\Big({\bar y_1^m\over m!}K_{\underbrace{1,1,...,1}_{n-1};0\mu} +{ y_1^{n-1}\over (n-1)!}K_{0;\underbrace{\bar 1,\bar1,...,\bar1}_{m}\mu}-{y_1^{n-1}\over (n-1)!}{\bar y_1^m\over m!}\kom \Big)\nonumber \\
&=& \bar \la _1K_{\underbrace{1,1,...1}_{n};\underbrace{\bar 1,\bar 1,...\bar 1}_{m-1}\mu}+\la _1K_{\underbrace{1,1,...1}_{n-1};\underbrace{\bar 1,\bar 1,...\bar 1}_{m}\mu}
 \er

One can then see that 
\be K_{p_1,p_2,..,\underbrace{1,1,...1}_{n};\bar q_1,\bar q_2,..,\underbrace{\bar 1,\bar 1,...\bar 1}_{m}\mu} = y_{p_1}y_{p_2}... \bar y_{q_1}\bar y_{q_2}.. K_{\underbrace{1,1,...1}_{n};\underbrace{\bar 1,\bar 1,...\bar 1}_{m}\mu} \ee with \[~~p_1,p_2,...,q_1,q_2,; \geq 2,~~~p_1\neq p_2\neq...; \bar q_1 \neq \bar q_2 \neq ...
\]
has the right gauge transformation. If any of the $p$ are repeated $i$ times, then $y_p$ is replaced by $y_p^i\over i!$. Similarly for the $\bar y_q$.

This completes the construction of $K_{\mu [n];[\mb]}$.   Since the basic variables are $\kn, k_{\nb}, q_n, q_\nb$ it is clear that no new degrees of freedom have been added to that of the free theory. However,
in principle one could add to $K_{[n]_i;[\mb]_j \mu}$, new variables of the form $k_{[n]_i;[\mb]_j \mu}$ with transformation rule 
\be \label{knmb}
\delta k_{[n]_i;[\mb]_j \mu} = \la _p k_{[n]_i/p;[\mb]_j \mu} + \bar \la _p k_{[n]_i;[\mb]/\bar p \mu}
\ee 
where as earlier $[n]_i/p$ stands for the particular partition $[n]_i$ with the one $p$ removed. (If $[n]_i$ does not contain $p$, that term does not contribute to the gauge transformation, and can be set to zero.) This is also discussed in Appendix D \eqref{appenc}.

\subsubsection{An interesting relation}

The $K$'s obey a relation of the form:
\be	\label{intrel}
\tilde K_{n;\mb \mu} \equiv \sum _{i,j} K_{[n]_i;[\mb]_j \mu} = \bar q_n k_{\mb \mu} + \bar q_{\mb} k_{n\mu} - \bar q_n \bar q_{\mb} \kom
\ee
Here, as earlier $[n]_i$ denotes a particular partition of $n$ denoted by $i$ and $\bar q$ was defined in (\ref{qbar}). Thus for instance
\[
\tilde K_{2,\bar 1\mu} \equiv K_{2;\bar 1\mu} + K_{1,1;\bar 1\mu} = \bar q_2 k_{\bar 1\mu} + \bar q_{\bar 1} \ktm - \bar q_{\bar 1} \bar q_2\kom\] 

The gauge transformation of $\tilde K_{n;\mb \mu}$ under $\la _p$ is easily seen to be:
\be
\delta \tilde K_{n;\mb \mu} = \la _p \tilde K_{n-p;\mb \mu}
\ee
{\bf Proof:} The only partitions $[n]_i$ that contribute to the gauge transformation, are the ones that have at least one $p$. Take these partitions and remove one $p$. The remaining numbers are all possible ways of making $n-p$ - so we get all the partitions of $n-p$. The gauge transformation law then forces (\ref{intrel}) to be true.
This relation will be used in the construction of the free equations. 

For the free equations one has to keep
only single derivatives in the loop variable. Thus we write ${\p \over \p x_{n_1+n_2+..}}{\p\over \p x_{\mb_1+\mb_2+..}}Y$ for $\p_{n_1}\p_{n_2}... \p_{\mb_1} \p_{\mb_2}Y...$. Thus the coefficient of  $Y^\mu_{n_1+n_2+...;\mb_1+\mb_2+..}$ is $\tilde K_{n;\mb \mu}$

Of course one can still add some new variables $k_{[n]_i,[\mb]_j \mu}$ with the correct gauge transformation law (\ref{knmb}), as mentioned earlier and then this would contribute to $\tilde K_{n,\mb \mu}$ also. This is in fact done in Appendix C \eqref{appenc}.

\subsection{Free Equation}

As in the case of the open string, equations are given by the ERG \eqref{ERG1}.
\[
\int du~ {\p L\over \p \tau} \psi = \int dzdz'~\{-\hf \underbrace{\dot G(z,z') G^{-1}(z,z')}_{field~independent}- 
\]
\be  
 \hf \dot G(z,z')[{\delta^2\over \delta X(z)\delta X(z')}\int du~L[X(u),X'(u)]] + \ddXz \int du~L[X(u)]\ddXzp \int du'~L[X(u')]\}\psi
\ee

The second term, when $L[Y(u)]$ is taken to be the loop variable \eqref{GVO} was worked out in Sec 5.4. We need to generalize this to the 
loop variable for closed strings \eqref{GVC}. This is straightforward - we need to include derivatives w.r.t. the anti holomorphic variables
$Y_\nb\equiv{\p Y\over \p \xnb}$ and also mixed derivatives $Y_{n,\mb}\equiv {\p Y\over \p \xn \p \xmb}$.

Thus \eqref{FEO} is generalized to
\br
 \int du~ {\delta \over \delta Y^\mu(z')} {\cal L}(u)&=&\int du~\Big\{ {\p {\cal L} [Y(u),Y_{n,\mb}(u)]\over \p Y^\mu(u)} \delta (u-z')+{\p {\cal L} [Y(u),Y_{n,\mb}(u)]\over \p Y_1^\mu(u)} \p_{x_1}\delta (u-z')\nonumber \\&+&{\p {\cal L} [Y(u),Y_{n,\mb}(u)]\over \p Y_{\bar1}^\mu(u)} \p_{\bar x_1}\delta (u-z')
+{\p {\cal L} [Y(u),Y_{n,\mb}(u)]\over \p Y_{1;\bar 1}^\mu(u)} \p_{x_1}\p_{\bar x_1}\delta (u-z')\nonumber \\&+&
{\p {\cal L} [Y(u),Y_{n,\mb}(u)]\over \p Y_2^\mu(u)} \p_{x_2}\delta (u-z')+{\p {\cal L} [Y(u),Y_{n,\mb}(u)]\over \p Y_{\bar2}^\mu(u)} \p_{\bar x_2}\delta (u-z')\nonumber 
\er
\be	\label{FEC}
~~~~~~~~~~~~~~~~~~~~~~~~~~~~~~~~~~~~~~~~~~~ +{\p {\cal L} [Y(u),Y_{n,\mb}(u)]\over \p Y_{2;\bar 1}^\mu(u)} \p_{x_2}\p_{\bar x_1}\delta (u-z')+...+{\p {\cal L} [Y(u),Y_{n,\mb}(u)]\over \p Y_n^\mu(u)} \p_{x_n}\delta (u-z')+..\Big\}
 \ee
The variables $\xn, \xmb$ in the derivatives acting on delta functions, correspond to $u$, and we can integrate by parts on $u$.
Thus for the second derivative ${\delta ^2\over \delta X^\mu(z') \delta X_\mu(z'')}$ we get
\[
\eta^{\mu\nu}{\delta \over \delta X^\nu(z')}
\int du~\Big\{ {\p {\cal L} [Y(u),Y_{n,\mb}(u)]\over \p Y^\mu(u)} \delta (u-z'')-\p_{x_1}{\p {\cal L} [Y(u),Y_{n;\mb}(u)]\over \p Y_{1;0}^\mu(u)} \delta (u-z'')
\]
\be - \p_{\bar x _1} {\p {\cal L}[Y(u),Y_{n;\mb}(u)] \over \p Y_{0;\bar 1} ^\mu(u) }
\delta (u-z'') +  \p _{ x_{1}}\p_{\bar x_{1}}{\p {\cal L}[Y(u),Y_{n,\mb}(u)] \over \p Y_{1,\bar 1}^\mu (u) }
 \delta (u-z'') +...\Big\}
\ee
\[
=\eta^{\mu\nu}\int du~[{\p\over \p Y^\nu(u)} +{\p\over \p x_1}\delta(u-z'){\p \over \p Y^\nu_{1;0}(u)}+{\p\over \p \bar x_1}\delta(u-z'){\p \over \p Y^\nu_{0;\bar1}(u)}+{\p\over \p x_2}\delta(u-z'){\p \over \p Y^\nu_{2;0}(u)}+\]\[{\p\over \p \bar x_2}\delta(u-z'){\p \over \p Y^\nu_{0;\bar 2}(u)}+
 {\pp\over \p x_1\p \bar x_1}\delta(u-z'){\p \over \p Y^\nu_{1;\bar 1}(u)}+...]\]
 \[
 \Big\{ {\p {\cal L} [Y(u),Y_{n,\mb}(u)]\over \p Y^\mu(u)} \delta (u-z'')-\p_{x_1}{\p {\cal L} [Y(u),Y_{n;\mb}(u)]\over \p Y_1^\mu(u)} \delta (u-z'')
\]
\be - \p_{\bar x _1} {\p {\cal L}[Y(u),Y_{n;\mb}(u)] \over \p Y_{\bar 1} ^\mu(u) }
\delta (u-z'') +  \p _{ x_{1}}\p_{\bar x_{1}}{\p {\cal L}[Y(u),Y_{n,\mb}(u)] \over \p Y_{1,\bar 1}^\mu (u) }
 \delta (u-z'') +...\Big\}
\ee

The details of the algebra are worked out in Appendix B \eqref{appenb}. The free equation was derived using Weyl invariance in \cite{BSCS}. One difference between that derivation and this one is noteworthy. There it was assumed that $\p_z \p_\zb X(z,\zb)=0$. So there were no mixed derivative vertex operators. However $\p_z \p_\zb \sigma \neq0$. Thus in the loop variable generalization, this meant that $\p_{\xn} \p_{\xmb}\Sigma \neq 0$. Thus, if we assume that mixed derivatives are absent, then in terms of Green functions
\be   \label{nmd}
\p_{\xn} \p_{\xmb}\Sigma = \p_{\xn} \p_{\xmb}\lan Y(z,\zb)Y(z,\zb)\ran=2\lan \p_{\xn}Y(z,\zb) \p_{\xmb}Y(z,\zb)\ran
\ee
On the other hand in the present approach
\be   \label{md}
\p_{\xn} \p_{\xmb}\Sigma = \p_{\xn} \p_{\xmb}\lan Y(z,\zb)Y(z,\zb)\ran=2\lan \p_{\xn}Y(z,\zb) \p_{\xmb}Y(z,\zb)\ran+2\lan \p_{\xn}\p_{\xmb}Y(z,\zb) Y(z,\zb)\ran
\ee
In the Weyl invariance approach of \cite{BSCS} the coefficient of the term in \eqref{nmd} is $\kn.k_{\mb}$. In the present approach the two terms on the RHS of \eqref{md} come with {\it a priori} different coefficients: the first with $\kn.k_{\mb}$ and the second with $K_{n,\mb}.\ko$. Thus when we set these two equal, we get the same (correct) free equations. These constraints were called K-constraints in \cite{BSERGclosed1}
and are derived in Appendix D \eqref{appenc}.

We first derive the EOM in flat space time. Later we generalize to curved space time.

\subsubsection{Results: Graviton}
We get for level $(1,\bar 1)$:
\be   \label{grav}
[-\ko^2 \kim k_{\bar 1\nu} + \ko .\ki \kom k_{\bar 1\nu} + \ko. k_{\bar 1}\kon \kim - \ki.k_{\bar 1}\kom \kon]Y_{1;0}^\mu Y_{0;\bar 1}^\nu=0
\ee

Let us write this equation in terms of space time fields and analyze the gauge transformations:
Define \[
\lan \hf k_{1(\mu} k_{\bar 1\nu)}\ran = h_{\mu \nu}~~~
\] 
\be
\lan \hf k_{1[\mu} k_{\bar 1\nu]}\ran = B_{\mu \nu}
\ee
 Let
us also define 
\[
\lan \hf(\la _1 k_{\bar 1\mu} + \bar \la _1 k_{1\nu})\ran = \eps _\mu ~~
\] 
 \be
\lan \hf(\la _1 k_{\bar 1\mu} - \bar \la _1 k_{1\nu})\ran = \Lambda _\mu
\ee Then the gauge transformation laws are 
\[	
\delta_G h_{\mu \nu} = \p_{(\mu}\eps_{\nu)}  ~~~
\] 
 \be	\label{gauge}
\delta_G B_{\mu \nu}=
 \p_{[\mu}\Lambda_{\nu]}
 \ee

The equation splits into two parts:
\be
-\Box h_{\mu\nu} + \p^\rho \p_\mu h_{\rho \nu} + \p^\rho \p_\nu h_{\mu\rho} - \p_\mu\p_\nu h^\rho_{~\rho}=0
\ee
the linearized graviton equation about flat space time,
and
\be
\p^\rho[\p_\rho B_{\mu\nu} + \p_\nu B_{\rho\mu} + \p_\mu B_{\nu \rho}]=0
\ee
The quantity in square brackets is just the field strength ($H=dB$) for $B$:
\[
H_{\rho\mu\nu}= \p_\rho B_{\mu\nu} + \p_\nu B_{\rho\mu} + \p_\mu B_{\nu \rho}
\]
The graviton equation can also be written as
\be
2\p^\rho \Gamma_{\rho \mu \nu}-\p_\mu\p _\nu h^\rho_{~\rho}=0
\ee
where 
$\Gamma_{\rho \mu \nu}$ is the Christoffel connection
\[
\Gamma_{\rho \mu \nu}=\hf[\p_\mu h_{\rho\nu}+\p_\nu h_{\mu\rho}-\p_\rho h_{\mu\nu}]
\]

Note that at the linearized level $\eps _\mu$ does not have the interpretation of a coordinate transformation. At this level
it is a gauge parameter with no geometrical interpretation. When we include interactions we will be forced to the geometrical interpretation.
It is then tempting to speculate that the gauge parameter for the $B$ field also has such an interpretation. In that case space time would seem to be complex with $\eps_\mu$ and $\Lambda_\mu$ being the transformation of the real and imaginary parts respectively \cite{BSERGclosed1}.

\subsubsection{Level $(2,\bar 2)$: Spin 4}
\begin{enumerate}
\item {\bf Physical States}
The closed string physical states are direct products of the open string states.
We have seen that for open strings the states at level 2 come from a two index traceless symmetric tensor. But the covariant description requires the trace. Thus  we have the diagram   
  \vspace{1cm}

\begin{center}
 \ydiagram{2} 
 $~~~\otimes~$
 \ydiagram{2}
 ~~~=~~
 \ydiagram{4}
 $~~~\oplus$
 \ydiagram{3,1}
 $~~~~\oplus$
 \ydiagram{2,2}
 \end{center}
 \vspace{1cm}
 
The gauge invariant description requires many other tensor fields. In open strings $q_1$ was not allowed and had to be replaced. The corresponding rule for closed strings was discussed above, is that the number of $q_1$'s and $q_{\bar 1}$'s should be equal. 

\item{\bf Field Content and Gauge Transformation}

\begin{itemize}
\item {\bf Scalars}
The allowed combinations are:
\be
\delta S^{2\bar2}=\lan \delta (q_2 q_{\bar 2})\ran  = \lan 2\la _2 q_0 q_{\bar 2} + 2 \la _{\bar 2} q_{\bar 0} q_2\ran  = 2\Lambda^{2\bar2}\qo +2\Lambda^{\bar 2 2} q_{\bar 0}
\ee
Since $\qi^2 \qib^2 = \qt q_{\bar 2}\qo q_{\bar 0}$, $S^{11\bar 1\bar 1}$ is not an independent field. 
Here we have used the q-rules separately for the left and right modes separately:
\be 
q_1^2 = q_2 q_0;~~~\li q_1 = \lt q_0
\ee

\item{\bf Vectors}

\br
\delta S_\mu^{2\bar 2}=\lan \delta (\ktm \qtb)\ran &=&\lan  2 \ltb q_{\bar 0} \ktm + \li \kim \qtb + \kom \lt \qtb\ran = 2q_{\bar 0}\Lambda^{\bar 2 2}_{~\mu} + \kom \Lambda ^{2 \bar 2} \nonumber \\
\delta S_\rho^{\bar 2 2}=\lan \delta (\ktrb \qt) \ran &=& \lan 2 \lt q_0 \ktrb + \lib \kirb \qt + \kor \ltb \qt\ran = 2q_{ 0}\Lambda^{2\bar 2 }_{~\rho} + \kor \Lambda ^{ \bar 22}
\er

We have used a notation that the first number superscript refers to the level of the $k$'s and the second to the $q$'s. For the gauge parameters, the first index refers to the level of $\la$, the second to $k$ and the third to $q$. The space time index is directly below the corresponding $k$-level number.

We have two vectors and four vector gauge parameters. We can thus set 
\be 
\lan \li \kim \qtb\ran=0
\ee
 without any damage to our ability to gauge  away Stuckelberg fields.

\item{\bf 2-Tensors}

Q-rules relate $\qi \qib \kim \kirb$ to $\ktm \ktrb$ so it is not an independent field. Thus we have

\br
\delta S_{\mu\rho}^{2\bar 2}=\lan \delta(\ktm\ktrb )\ran &=&\lan  \li \kim \ktrb + \lib \kirb \ktm + \kom \lt \ktrb + \kor \ltb \ktm \ran =\Lambda^{11\bar 2}_{~\mu\rho}+\Lambda^{\bar 1\bar 1 2}_{~\rho\mu} + \kom \Lambda ^{2 \bar 2}_{~\rho} +\kor \Lambda ^{\bar 22}_{~\mu}
\nonumber \\
\delta S_{\mu\nu}^{11\bar 2}=\lan \delta (\kim \kin \qtb)\ran&=& 2\lan  \ltb q_{\bar 0} \kim \kin + \underbrace{k_{0(\mu} \li k_{1\nu)} \qtb}_{=0}\ran = 2 \Lambda ^{\bar 2 1 1}_{~\mu\nu} \nonumber \\
\delta S^{\bar 1\bar 1 2}_{\rho \sigma}=\lan \delta (\qt \kirb \kisb)\ran&=&\lan  2 \lt \qo \kirb\kisb +  \underbrace{k_{0(\rho} \lib k_{1\bar \sigma)} \qt}_{=0}\ran= 2 \Lambda ^{2\bar 1 \bar 1}_{~\rho \s}\nonumber
\er

Some of the gauge parameters that have already been set to zero earlier are shown here as being set to zero also.

\item{\bf 3-Tensor}

\be
\delta S_{\mu\nu\rho}^{11\bar 2}=\lan \delta(\kim\kin\ktrb)\ran = \lan \lib \kirb \kim\kin + k_{0(\mu}\li k_{1\nu)} \ktrb + \kor \ltb \kim\kin\ran =
\Lambda^{\bar 1 \bar 1 11}_{~\rho\mu\nu} + k_{0(\mu}\Lambda^{11\bar 2}_{~\nu)\rho} + \kor \Lambda ^{\bar 2 1 1}_{~\mu\nu}
\ee
There is also the conjugate with bars exchanged.

\item{\bf 4-Tensor}
\end{itemize}

We focus on the four index tensor, which contains all the physical states.  The four index tensor is also interesting because it includes as shown above, tensors with mixed symmetry.

The world sheet action has a term $(\ki . Y_1)^2 (k_{\bar 1} . Y_{\bar 1})^2$ corresponding to the 4-tensor:
\be		\label{S1111}
\lan \kim \kin k_{\bar 1 \rho} k_{\bar 1\sigma}\ran = S_{\mu \nu\rho \sigma}^{11\bar1\bar1}
\ee
We can define tensor irreps by writing (brackets denote symmetrization: $S_{(\mu \sigma)}=S_{\mu \sigma} +S_{\sigma \mu}$) the "resolution of unity":
\be
S_{\mu \nu \rho \sigma} = {1\over 24} \underbrace{S^S_{\mu \nu \rho \sigma}}_{ \ydiagram{4}}+{1\over 8}\underbrace{S^{(3,1)}_{\mu \nu (\rho \sigma)}}_{\ydiagram{3,1}}+{1\over 12}\underbrace{S^{(2,2)}_{\mu \nu \rho \sigma}}_{\ydiagram{2,2}}
\ee

 Hereafter, for simplicity we write $S_{\mu\nu\rho \sigma}$ instead of $S ^{11\bar1\bar1}_{\mu\nu\rho \sigma}$. Its gauge variation is
\br
\delta \lan \kim\kin \kirb \kisb \ran &=& \lan \li \kom \kin \kirb \kisb \ran  + \lan \li \kon \kim \kirb \kisb \ran	+\lan \lib \kor \kim \kin  \kisb \ran  +\lan \lib k_{0\sigma} \kim\kin \kirb  \ran \nonumber \\
\implies \delta S_{\mu \nu \rho \sigma}&=& \p_\mu \Lambda _{~~\nu \rho \sigma}^{11\bar 1\bar 1} +\p_\nu \Lambda _{~~\mu \rho \sigma}^{11\bar 1\bar 1}+\p_\rho \bar \Lambda _{~~\sigma \mu \nu }^{\bar 1\bar 111}+\p_\sigma \bar \Lambda _{~~\rho\mu \nu  }^{\bar 1\bar 111}
\er	  
For the gauge transformation parameter $\Lambda_{ \nu \rho \sigma} = \lan \li \kin \kirb \kisb\ran$  irreps are defined by the resolution of unity which reads as:
\be
\Lambda_{\nu \rho \sigma}={1\over 6}\underbrace{\Lambda^ S _{\nu \rho \sigma}}_{\ydiagram{3}} -{1\over 3}\underbrace{\Lambda^ I_{\sigma \rho \nu}}_{\ydiagram{2,1}}
\ee
In terms of these fields and gauge parameters one obtains:

\br
{1\over 24}\delta S^S_{i_1i_2i_3i_4}&=&{1\over 12}[\p_{i_1}\Lambda ^S_{i_3i_4i_2}+\p_{i_2}\Lambda^ S_{i_3i_4i_1}+\p_{i_3}\Lambda^ S_{i_4i_1i_2}+\p_{i_4}\Lambda^ S_{i_3i_1i_2}]\nonumber\\
{1\over 8}\delta S^{(3,1)}_{i_1i_2(i_3i_4)}&=&{1\over 12}[\p_{(i_1}\Lambda^ S_{|i_3i_4|i_2)}-\p_{(i_3}\Lambda^ S_{|i_1i_2|i_4)}]-{1\over 6}[\p_{(i_1}\Lambda^ I _{|i_3i_4|i_2)}  -  \p_{(i_3}\Lambda^ I _{|i_1i_2|i_4)}]   \nonumber \\
{1\over 12}\delta S^{(2,2)}_{i_1i_2i_3i_4}&=&-{1\over 6}[\p_{(i_1}\Lambda^I_{|i_3i_4|i_2)}+\p_{(i_3}\Lambda^I_{|i_1i_2|i_4)}]
\er
There is an identical complex conjugate equation involving $\bar \Lambda$ which we do not bother to write down.
 \item {\bf Free Equation}
   
The free equation of motion (EOM) can be written as:
\[
   -{1\over 4} \ko^2 (\ki .Y_1)^2 (k_{\bar 1} .Y_{\bar 1})^2 +\hf \ko.\ki (\ko .Y_1)(\ki .Y_1) (\kib .\yib)^2 + \hf \ko.\kib (\ko.\yib)(\kib .\yib)(\ki.\yi)^2 +
\]
\be
   -{1\over 4} \ki.\ki (\ko.\yi)^2(\kib.\yib)^2 -{1\over 4} \kib.\kib (\ko.\yib)^2(\ki.\yi)^2 - \ki.\kib (\ko.\yi)(\ko.\yib)(\ki.\yi)(\kib.\yib)
\ee
  It is gauge invariant under 
\[
  \kim \to \kim + \li \kom;~~~~\kimb \to \kimb + \la_{\bar1} \kom
\] 
  if we use the tracelessness condition on the gauge parameters: 
\[
   \li \ki.\kib \kimb= \li \kib.\kib \kim =0= \lib \ki.\kib \kim = \lib \ki.\ki \kimb
\] 
   
   Using \eqref{S1111} the EOM becomes:
   \[
	   -\pp S_{\mu\nu \rho\sigma} + \p^\la \p_{(\mu}S_{\nu)\la \rho \sigma} + \p^\la \p_{(\sigma} S_{|\mu \nu \la|\rho)}
   \]
   \be   \label{FEOM}
   -\p_\mu \p_\nu S^{\la}_{\la\rho\sigma}- \p_\rho \p_\sigma S_{\mu\nu ~\la}^{~~\la}-\p_{(\sigma}\p_{(\nu}S_{\mu)~\la|\rho)}^{~~\la}=0
   \ee
 \item{\bf Free Action} 
  
  It turns out that an action can also be written for this free theory:
  \[
S_{free}=  -\hf S^{abcd}\Box S_{abcd} - \p_a S^{aefg}\p^bS_{befg} - \p_a S^{efga}\p^bS_{efgb}
  \]
  \[
  -\p_a\p_b S^{abfg}S^c_{~cfg} -\p_a \p_b S^{fgab}S_{fg~c}^{~~c}-4\p_a\p_b S^{eafb}S_{e~fc}^{~c}
  \]
  \[
  +\hf\big(S^{c~fg}_{~c}\Box S^a_{~afg} +S^{fgc}_{~~~~c}\Box S_{fg~a}^{~~a} +4 S^{cf~g}_{~~c}\Box S^a_{~fag}\Big)
  \]
  \[
  +2\Big(S^{c~de}_{~c}\p_d\p^b S_{b~ae}^{~a} +  S^{dec}_{~~~~c}\p_e\p^a S_{d~ba}^{~b}\Big)
  \]
  \be
  -\hf\Big( S^{c~de}_{~c}\p_e\p^a S^b_{~bad} + S^{dec}_{~~~~c}\p_e\p^a S_{ad~b}^{~~b}\Big)
  \ee
 The EOM obtained from this action are linear combinations of \eqref{FEOM} and its traces. 
 
 The action is of the form $S_aM^{ab}S_b$, where $M$ is a symmetric in its indices. Its gauge variation is therefore
 $S_a M^{ab}\delta S_b$. Since $M^{ab} S_b$ is the EOM, which we know is gauge invariant, it must be true that
 $ \delta (M^{ab} S_b)=M^{ab}\delta S_b=0$. Thus it follows that the action is also gauge invariant. 
  
\end{enumerate}

\subsection{Interactions}

We now turn to the issue of closed string interactions. The interactions are given by the second term in \eqref{ERG1}. It involves 
gauge invariant expression that we called a field strength because for Maxwell theory it is indeed the field strength. 

\subsubsection{Gauge Invariant Field Strength}
We start with level 2. The interaction Lagrangian $L$ at level 2 is best obtained by starting with the generalized loop variable, which we denote by $\cal L$. 
\be
{\cal L} = e^{i\Big( \ko.Y + K_{1;0}Y_{1;0}+K_{0;\bar 1}Y_{0;\bar 1} + K_{1;\bar 1}Y_{1;\bar 1} + K_{2;0}Y_{2;0}+K_{0;\bar 2}Y_{0;\bar 2} + K_{1,1;0}Y_{1,1;0}+K_{0;\bar1,\bar1}Y_{0;\bar 1,\bar 1}+....\Big)}
\ee 

The field strength is given by: 
\[
{\delta \over \delta Y^\mu(z')} \int du~  {\cal L}(u)=\int du~\Big\{ {\p {\cal L} [Y(u),Y_{[n],[\mb]}(u)]\over \p Y^\mu(u)} \delta (u-z')+{\p {\cal L} [Y(u),Y_{n;\mb}(u)]\over \p Y_1^\mu(u)} \p_{x_1}\delta (u-z')
\]\[+ {\p {\cal L}[Y(u),Y_{[n],[\mb]}(u)] \over \p Y_{\bar 1} ^\mu(u) }
 \p_{\bar x _1}\delta (u-z') + {\p {\cal L}[Y(u),Y_{[n],[\mb]}(u)] \over \p Y_{1,\bar 1}^\mu (u) }
 \p _{ x_{1}}\p_{\bar x_{1}} \delta (u-z') \Big\}
 \]
 \[
 = \int du~\Big\{ {\p {\cal L} [Y(u),Y_{[n],[\mb]}(u)]\over \p Y^\mu(u)} \delta (u-z')-[\p_{x_1}{\p {\cal L} [Y(u),Y_{[n],[\mb]}(u)]\over \p Y_1^\mu(u)}] \delta (u-z')
\]
\[- [\p_{\bar x _1} {\p {\cal L}[Y(u),Y_{n;\mb}(u)] \over \p Y_{\bar 1} ^\mu(u) }]
\delta (u-z') +  [\p _{ x_{1}}\p_{\bar x_{1}}{\p {\cal L}[Y(u),Y_{[n],[\mb]}(u)] \over \p Y_{1,\bar 1}^\mu (u) }]
 \delta (u-z') +
\]
\[
  [\pp _{ x_{1}}{\p {\cal L}[Y(u),Y_{[n],[\mb]}(u)] \over \p Y_{1,1;0}^\mu (u) }]\delta(u-z')+ [\pp _{ \bar x_{1}}{\p {\cal L}[Y(u),Y_{[n],[\mb]}(u)] \over \p Y_{0;\bar1,\bar 1}^\mu (u) }]\delta(u-z')
  \]
  \[- [\pp _{ x_{1}}\p_{\bar x_ 1}{\p {\cal L}[Y(u),Y_{[n],[\mb]}(u)] \over \p Y_{1,1;\bar 1}^\mu (u) }]\delta(u-z')- [\pp _{ \bar x_{1}}\p_{x_1}{\p {\cal L}[Y(u),Y_{[n],[\mb]}(u)] \over \p Y_{1;\bar1,\bar 1}^\mu (u) }]\delta(u-z')
   \Big\}
 \]
 \be  \label{FnlDer}
 +
  [\pp _{ \bar x_{1}}\pp_{x_1}{\p {\cal L}[Y(u),Y_{[n],[\mb]}(u)] \over \p Y_{1,1;\bar1,\bar 1}^\mu (u) }]\delta(u-z')
\ee
\[   
=
 \Big\{i\kom{\cal L }(z') -i K_{1;0\mu} \p_{x'_1} {\cal L}(z')-i K_{0;\bar 1\mu} \p_{\bar x' _1}  {\cal L}+iK_{1;\bar 1\mu} \p _{ x'_{1}}\p_{\bar x'_{1}} {\cal L}(z')+
 \]
 \be \label{FldStr}
  iK_{1,1;0\mu} \pp_{x_1}{\cal L}+ iK_{0;\bar 1,\bar 1}\pp_{\bar x_1}{\cal L}-iK_{1,1;\bar 1}^\mu \pp_{x_1}\p_{\bar x_1}{\cal L}-iK_{1;\bar 1,\bar 1}\p_{x_1}\pp_{\bar x_1}{\cal L}+iK_{1,1;\bar 1,\bar 1\mu} \pp_{x_1}\pp_{\bar x_1}{\cal L}\Big\}
 \ee
We have kept only terms that contribute to level $(1;\bar 1)$ and $(1,1;\bar 1,\bar 1)$. 
 From the structure of ${\cal L}$ we can see that 
 \be \label{Lvar}
 \delta {\cal L} = \sum_{n,\nb =1,2,...} (\la _n {\p \over \p \xn}{\cal L}+\la _{\bar n} {\p \over \p \xnb}{\cal L})
\ee
Using (\ref{Lvar})  we can easily check that (\ref{FldStr}) is invariant under $\la_1,\la _{\bar 1}$ variations, and at level 2, is the gauge invariant field strength for closed strings. We write it explicitly below:

\[
-i\kom ( K_{1;0}.Y_{1;0})(K_{0;\bar 1}.Y_{0;\bar 1})\e -\kom K_{1;\bar 1}. Y_{1;\bar 1}\e\]
\[
i K_{1;0\mu}( \ko. Y_{1;0})( K_{0;\bar1}.Y_{0;\bar 1}) \e +K_{1;0\mu} K_{0;\bar1}Y_{1,\bar 1}\e\]
\[
iK_{0;\bar 1\mu} (K_{1;0}.Y_{1;0})(\ko . Y_{0;\bar1})\e +K_{0;\bar 1\mu} K_{1;0}Y_{1;\bar 1}\e
\]
\be
-i K_{1;\bar 1\mu} (\ko Y_{0;\bar1})(\ko.Y_{1;0})\e - K_{1;\bar 1\mu} \ko.Y_{1;\bar1}\e
\ee

 The coefficient of $Y_1^\mu Y_{\bar 1}^\nu$ can be seen to be
 \be	\label{fldstrlevel2}
 -\kor k_{1\mu} k_{\bar 1\nu} + k_{1\rho} \kom k_{\bar 1\nu} + k_{\bar 1\rho} k_{1\mu} \kon - K_{1;\bar 1\rho} \kom \kon 
 \ee 
In terms of space time fields this is 
\[
G_{\rho\mu\nu}\equiv\Big( -\p _\rho (h_{\mu \nu} + B_{\mu \nu})+ \p _\mu (h_{\rho \nu}+B_{\rho \nu})+ \p _\nu (h_{\mu \rho}+ B_{\mu \rho})\Big)- \p_\mu \p _\nu S_\rho
\] 
\be   \label{spin2}
=\Gamma_{\rho \mu \nu} + H_{\rho \mu \nu} - \p_\mu \p _\nu S_\rho
\ee

\subsubsection{Problems with massless spin 2 field strength}

We have defined a field $S_\mu = \lan K_{1;\bar 1\mu}\ran$. However, 
at level 2 the physical fields are the graviton, antisymmetric tensor and dilaton. In fact since $K_{1;\bar 1}$ involves $\bar q_1$, this field
strength is well defined only if the graviton and dilaton are massive and $q_0\neq0$.  The gauge 
transformation of $S_\mu$  is 
\[
\delta S_\mu =2 \eps_\mu
\]
This means that it is a Stuckelberg field and one can fix a gauge such that it is zero. This would mean that
 the graviton has extra degrees of freedom - corresponding to a massive spin 2 field. While this is 
internally consistent this cannot
describe the usual closed string states which are massless at this level. In the free equation of motion,
 only the combination $\kom   K_{1;\bar 1}^{\mu}$ is involved. This was replaced by $\ki.\kib$ and that solved 
the problem. For the interacting case we see that some drastic modification is called for. 

Actually there is another problem. In the case of the open string we have seen that the theory continues
 to look Abelian even at the interacting level. The gauge transformation law is not modified by interactions.
We have also pointed out that this may benot be unreasonable because we know that the gauge symmetry of the 
massless field is an Abelian U(1). For closed strings we know (from hindsight) that the massless spin 2
 describes a graviton, which is ``non Abelian'' in the sense that the gauge transformation is different for 
the interacting theory.

If we combine this observation with the earlier one, a logical possibility is that both problems are solved 
if we modify our symmetry transformation rule so that we do not need a new field $S_\mu$. This was done in 
\cite{BSERGclosed1}. 

\subsubsection{Solution: Introduction of ``Reference "(or ``Background") metric and modification of symmetry transformation}

The idea is to do two things: 1) Let us denote the loop variable gauge transformation for the massless
fields by $\delta _G$. Let us consider another transformation $\delta_T X^\mu = -\xi^\mu$. We attempt to make the action invariant under $T$. If $h_{\mu\nu}$
transforms as a 2 index symmetric tensor under this transformation then $h_{\mu\nu}\p_z X^\mu \p_\zb X^\nu$
is invariant.
 \[
\delta _Th_{\mu \nu}\equiv \xi ^\la h_{\mu \nu,\la}+\xi ^\la_{~,\mu}h_{\la \nu}+\xi^\la_{~,\nu}h_{\mu\la};~~~~
\delta_T X^\mu = -\xi ^\mu
\]

But $T$ is not a symmetry of the theory because the {\em kinetic} term $\eta_{\mu\nu}\p_z X^\mu \p_\zb X^\nu$
is not invariant under this transformation.
\be \label{kin}
\delta_T( \eta_{\mu \nu} \p_z X^\mu \p_\zb X^\nu) = -\bar \xi_{(\mu,\nu)}\p_z X^\mu \p_\zb X^\nu
\ee
Here $\bar \xi_\mu = \eta_{\mu\nu}\xi^\nu$.
 To remedy  this we modify the kinetic term to 
\be
(\eta _{\mu\nu}+ h_{\mu\nu}^R) \p_z X^\mu \p_\zb X^\nu
\ee
and assign to $h_{\mu\nu}^R$ the transformation law \footnote{If we define a fully covariant $\xi_\mu = g^R_{\mu\nu}\xi^\nu$ then the same transformation law can be written as  $\delta _T h^R_{\mu \nu}(X)= \nabla^R_{(\mu} \xi _{\nu)}$.}
\be
\delta _T h^R_{\mu \nu}(X)=\xi^\la h^R_{\mu\nu,\la}+ \xi ^\la_{~,\mu}h^R_{\la\nu}+\xi^\la_{~,\nu}h^R_{\mu\la} +
\bar \xi_{(\mu,\nu)};~~~~\delta_TX^\mu = -\xi^\mu(X)
\ee

Now the kinetic term is also invariant. However we have made the action depend on an arbitrary quantity
$h^R_{\mu\nu}$. So let us add the negative of this term to the interaction term, which becomes
\be
(h_{\mu\nu}-h^R_{\mu\nu}) \p_z X^\mu \p_\zb X^\nu
\ee
While this term is not invariant under $T$, it is invariant under the combined action of
$G$ and $T$- provided we identify $\eps_\mu = \bar \xi_\mu$!The kinetic term is also invariant under $T+G$ because $G$ does nothing to it and it was designed to be invariant under $T$. The transformation $T+G$ can be called a background
 general coordinate transformation. Under this both $h_{\mu\nu}$ and $h_{\mu\nu}^R$ transform as if they were
 metric fluctuations about $\eta_{\mu\nu}$ with the result that the difference $h-h^R$ transforms as an
 ordinary tensor.

What we have achieved is the following: By introducing a reference metric and including it in the action
we have made the action invariant under a {\em new} symmetry. This is very close to ordinary
 general coordinate transformation (GCT) but it is not the same because it also transforms the 
reference metric. This seems pointless, because we want invariance under GCT which should act only on 
physical fields, and not on auxiliary constructs. However what saves the situation is that the full action
 does not depend on $h^R_{\mu\nu}$ because we have added and subtracted it out. Since our equations treat the 
kinetic term and interaction term separately,  equations {\em will}
 depend on $h_{\mu\nu}^R$ at all intermediate stages of the calculation. However the final solution obtained after solving all the equations, should not 
depend on $h_{\mu\nu}^R$ because the starting point did not! Thus invariance under $G+T$ is equivalent to invariance under GCT - because the final answer depends only on the physical field $h_{\mu\nu}$ and on this
 they both have the same action.

 \subsubsection{Higher spin and Massive modes}
 
 In the above discussion we did not mention the massive modes. On the one hand they all have to be made covariant under $G+T$ . This means introducing background covariant derivatives depending on $h_{\mu\nu}^R$. On the other hand we want the action to not depend on $h_{\mu\nu}^R$. So one must subtract some terms. Note that if $h^R_{\mu\nu}$ were equal to $h_{\mu\nu}$ the dependence would be allowed. So what must be subtracted out is a function of $h_{\mu\nu}-h_{\mu\nu}^R$.  Thus for instance:
 \[
K_{n;\bar m \mu}{D^2 Y^\mu\over D \xn D {\bar x_m}}=K_{n;\bar m \mu}( {\pp Y^\mu\over \p \xn \p {\bar x_m}}+ \Gamma ^{R\mu}_{\rho \sigma}Y^\rho_n Y^\sigma_{\bar m})
\]
 
 We must subtract out the $h^R$ dependent term - but not all of it. Thus
 \[
 K_{n;\bar m \mu}( {\pp Y^\mu\over \p \xn \p {\bar x_m}}+ \Gamma ^{\mu}_{\rho \sigma}Y^\rho_n Y^\sigma_{\bar m})
\]
is the required  covariantization to be done when $h^R_{\mu\nu}=h_{\mu\nu}$. So we subtract \footnote{This differs from \cite{BSERGclosed1,BSERGclose d_{2}} where the subtracted term was just $\Gamma ^{R\mu}_{\rho \sigma}Y^\rho_n Y^\sigma_{\bar m}$ and not the difference. The present prescription is superior because the subtracted term is a tensor.}
\be	\label{subt}
K_{n;\bar m \mu}(\Gamma ^{R\mu}_{\rho \sigma}-\Gamma ^{\mu}_{\rho \sigma})Y^\rho_n Y^\sigma_{\bar m} 
\ee
from a higher massive mode vertex operator (which amounts to a field redefinition of the corresponding space time field):
\[
k_{n\rho} k_{\mb \s}Y^\rho_n Y^\sigma_{\bar m}\to [k_{n\rho} k_{\mb \s}-K_{n;\bar m \mu}(\Gamma ^{R\mu}_{\rho \sigma}-\Gamma ^{\mu}_{\rho \sigma})]  Y^\rho_n Y^\sigma_{\bar m}
\]
 Note also that \eqref{subt} is a tensor under $G+T$. Thus the tensorial property of the redefined field $\lan k_{n\rho} k_{\mb \s}\ran $ is not affected. 
 
 These subtractions and field redefinitions of course have to be done at all levels in a systematic way. The result is a theory that is background covariant, but for which the full action does not depend on the reference metric - even though the kinetic and interaction term
 separately do.
 
 \subsubsection{Choosing the reference metric equal to the physical metric}
 
 With this justification in mind, one can further simplify things by setting $h_{\mu\nu}^R=h_{\mu\nu}$. There are no subtractions to be made. Then the action depends only on $h_{\mu\nu}^R(=h_{\mu\nu})$ at all stages of the calculation, and the background symmetry $G+T$ which is now identified with GCT,  is  manifest at all stages of the calculation.

In non Abelian gauge theories background field methods have similar benefits. While gauge fixing makes the usual gauge symmetry non manifest, a new symmetry involving the background field is manifest. This restricts
the form of the effective action. The background, which is arbitrary can then be chosen to be equal to the physical field. So the background gauge symmetry is the physical symmetry. So at all stages of the calculation the gauge symmtery is manifest. The utility of this method is described in an Appendix of \cite{BSERGclosed1} for the classical case, following the original discussion in \cite{Abbott} for the more complicated quantum case.

\subsubsection{On $K_{1;\bar1\mu}$}

Because $\qo=0$ the expression for $K_{1;\bar 1\mu}$ does not make sense for the lowest level. So we will not use that expression at all.
The role of  $K_{1;\bar1\mu}$ is now played somehow by $h_{\mu\nu}^R$. We can try to understand this as follows. An integration by parts
can be done can be done for the following term in the interaction Lagrangian:
\be
\int d^2z~ K_{1;\bar1\mu} \p_z\p_\zb Y^\mu \e = -\hf i (\kon K_{1;\bar1\mu}+ \kom K_{1;\bar1\nu}) \p_z Y^\mu \p_\zb Y^\nu
\ee
Consider the gauge transformation $\delta K_{1;\bar1\mu} = 2\eps _\mu$. This gives for the interaction term a change:
\be
- i (\kon \eps_\mu + \kom \eps_\nu) \p_z Y^\mu \p_\zb Y^\nu\e \approx  -\p_{(\mu} \eps_{\nu )} \p_z Y^\mu \p_\zb Y^\nu
\ee
But after the identification  $\eps_\mu = \eta_{\mu\nu}\xi^\nu$, this is exactly the linearized transformation law for $-\int d^2z~h_{\mu\nu}^R \p_z Y^\mu \p_\zb Y^\nu$, the term that we added to the interaction Lagrangian. Thus the same role is played by a different term. The new field strength is gauge invariant : $\Gamma_{\rho\mu\nu}-\Gamma^R_{\rho\mu\nu}$. Thus we can set

\be	\label{htilde}
\hf\lan k_{1(\mu}k_{\bar 1\nu)}-(\kon K_{1;\bar1\mu}+ \kom K_{1;\bar1\nu})\ran=h_{\mu\nu}-h^R_{\mu\nu}\equiv\tilde h_{\mu\nu}
\ee
Let us use $\circ$ for the D+1 dimensional dot product and $.$ for the D dimensional one.
This combination is gauge invariant (under background transformations, where both $h$ and $h^R$ transform).  However $\mu$ runs
from 0 to D-1 in the above identification because the metric fluctuation $h_{\mu\nu}$ is not there in the D'th direction.  The K-constraint is
\[
\ko\circ K_{1;\bar 1}=\ki\circ \kib =  \ki.\ki + \qi\qib 
\]
\[ \implies \ko .K_{1;\bar 1}+ \qo Q_{1;\bar 1}-\ki.\kib  = \qi\qib
\]
$\lan \qi \qib \ran = \Phi_D$ is the dilaton. Also using \eqref{htilde} we get
\[
 -\tilde h^\mu_\mu +\qo Q_{1;\bar 1}=\Phi_D~;\mu=0,..,D-1
\]
Although $\qo$ is zero we have not set $ \qo Q_{1;\bar 1}$ to zero. In fact if we use the usual expression for $K_{1;1\mu}$ we
obtain that $ \qo Q_{1;\bar 1} = \qi \qib$. However as mentioned above, we do not wish to use that expression for the lowest level.
So we leave it arbitrary. In fact we can use the K constraint to define $\lan \qo Q_{1;\bar 1}\ran $ as $\tilde h^\mu_\mu + \Phi_D$.
This equation relates the trace of the metric fluctuation to the dilaton in a gauge invariant way. We remind the reader that in the gauge fixed old covariant formulation,
the trace of the graviton field is the dilaton.  

From equation \eqref{htilde} we can write
\[
\hf k_{1(\mu}k_{\bar 1\nu)}  = \tilde h^\mu_\mu + \hf k_{0(\mu}K_{1;\bar 1\nu)}
\]
Let us substitute this into the graviton free equation \eqref{grav}:
\be   
[-\ko\circ \ko \kim k_{\bar 1\nu} + \ko \circ \ki \kom k_{\bar 1\nu} + \ko\circ  k_{\bar 1}\kon \kim - \ki\circ k_{\bar 1}\kom \kon]=0
\ee
In this equation all dot products range from $0$ to $D$. Since $\qo=0$, in the first three terms we can drop $\qo$. The last term has a trace
that includes $\qi\qib$.
  We get (  $\mu =0,...,D-1$.)
\be
-\ko^2 \tilde h_{\mu\nu} + \kom k_0^\rho \tilde h_{\rho\nu} +\kon k_0^\rho \tilde h_{\rho \mu} - \kom\kon (\ki \circ \kib - \ko .K_{1;\bar 1})=0
\ee
From the K-constraint $\ki \circ \kib - \ko .K_{1;\bar 1}=\qo Q_{1;\bar 1}=\tilde h^\mu_\mu + \Phi_D$. Thus we get (we set $\Phi_D=0$ for convenience):
\be
-\ko^2 \tilde h_{\mu\nu} + \kom k_0^\rho \tilde h_{\rho\nu} +\kon k_0^\rho \tilde h_{\rho \mu} - \kom\kon \tilde h^\rho_{~\rho}  =0
\ee

Let us write this equation as
\be
k_0^\rho[- \kor \tilde h_{\mu\nu} + \kom  \tilde h_{\rho\nu} +\kon  \tilde h_{\rho \mu}] - \kom\kon \tilde h^\rho_{~\rho}  =0
\ee

Viewed as a linearized equation for a perturbation about flat space this equation can also be written as
\[
\p^\rho (\Gamma _{\rho \mu \nu}-\Gamma^R _{\rho \mu \nu}) -\p_\mu\p_\nu \tilde h^\rho_{~\rho}
\]

If we interpret this as an equation in the RNC, we can immediately covariantize it to \footnote{Note that while $\Gamma ^\rho_{~ \mu \nu}-\Gamma^{R\rho} _{~~ \mu \nu}$ is a tensor under background coordinate transformations, $\Gamma _{\rho \mu \nu}-\Gamma^R _{\rho \mu \nu}$ is not.}
\be
\nabla^{R\rho}[- \nabla^R_\rho \tilde h_{\mu\nu} + \nabla^R_\mu  \tilde h_{\rho\nu} +\nabla^R_\nu  \tilde h_{\rho \mu}] - \nabla^R_\mu \nabla^R_\nu \tilde h^\rho_{~\rho}  =0
\ee

This is to be compared with the linearized equation for the graviton fluctuation about a given background. We will check this in Section 7.6.3.

In addition to the gauge transformation, there is also the tensor rotation of $h_{\mu\nu}^R$ and $h_{\mu\nu}$. This is manifestly a symmetry
of the action when the indices are contracted.

\subsection{Covariant Description Summarized}

In the above discussion we started with flat space and explained the problem of gauge invariance and the role of $h^R_{\mu\nu}$ in solving this problem. We then showed that the symmetry $G+T$ also called background covariance, is present in the action. We now summarize this by explaining the covariance starting directly from the curved space viewpoint. Thus we show that the kinetic term and interaction term are separately invariant under $G+T$. We then
discuss how the ERG can be made manifestly invariant. The main tool is the Taylor expansion that can be done in an RNC and generalized,
for scalar objects, to other coordinate systems.

\subsubsection{Kinetic term}

\be
\int dz~(\eta_{\mu\nu}+h^R_{\mu\nu}(Y))Y_1^\mu Y_{\bar 1}^\nu =\int dz ~g^R_{\mu\nu}(Y(z))Y_1^\mu(z) Y_{\bar 1}^\nu(z)
\ee
This term is manifestly invariant under background general coordinate transformations where $g^R_{\mu\nu}(Y)$ transforms like a metric tensor. 

To do quantum calculations we need to Taylor expand  $g^R_{\mu\nu}(Y)$ about a fixed point, $Y_0$, which we take to be the origin of the RNC. We will refer to the RNC as $\bar Y^\mu$ and let  $\bar Y^\mu =0$ be the origin. We use \eqref{Taylor} to expand $\bar g^R_{\mu\nu}$
to get \cite{Pet}
(as usual the bar indicates that we are in the RNC):
\[
\bar g^R_{\al \beta}(\bar Y) 
= \bar g^R_{\al  \beta}(0) 
-{1\over 3}\bar R^R_{\al \mu \beta \la}(0)\bar Y^\mu \bar Y^\la-
{1\over 3!}\bar R^R_{\al \gamma \beta \la ,\mu}(0)\bar Y^\la \bar Y^\mu \bar Y^\gamma
\]
\be \label{Taylormetric}
+{1\over 5!}\{ -6 \bar R^R_{\al \delta \beta \gamma ,\la \mu}(0)+ {16\over 3} \bar R^{R~~~\rho}_{\la \beta \mu}(0)\bar R^R_{\gamma \al \delta \rho}(0)\}\bar Y^\la \bar Y^\mu \bar Y^\gamma \bar Y^\delta +...
\ee

In the above expansion $\bar g_{\al \beta}^R(0)$ can be taken to be $\eta_{\al \beta}$, but for the moment we leave it as it is.
The first derivative vanishes at the origin in the RNC. 
Thus the kinetic term in the RNC becomes 
\be   \label{kinetic}
\bar g^R_{\mu\nu}(\bar Y) \bar Y^\mu_1 (z)\bar Y^\nu_{\bar 1}(z)
\ee
where we substitute \eqref{Taylormetric} for $\bar g^R_{\mu\nu}(\bar Y)$. This term as it stands is manifestly a scalar at the point P of the manifold and is easily generalized to a general coordinate system as 
\be \label{kineticgen}
g^R_{\mu\nu}( Y) Y^\mu_1 (z) Y^\nu_{\bar 1}(z)
\ee
Here $Y^\mu_1 (z)$ is a vector at the point P (with coordinates $Y^\mu$) of the manifold as is clear from
\[
{\p Y'^\mu(z)\over \p z^\al}= {\p Y'^\mu(z)\over \p Y^\nu(z)}{\p Y^\nu(z)\over \p z^\al}
\]

Now for the quantum treatment it is necessary to expand the metric in a Taylor series \eqref{Taylormetric} so that we have an ordinary kinetic term that can be inverted to define green function, plus additional terms that will be interpreted as interactions.  Each term in the Taylor expansion is a tensor {\em at the origin} O. Thus only if we interpret $Y^\mu_1 (z)$ as a vector at the origin will it become a sum of scalars.
But this is true because (see Appendix A \eqref{appena}) the covariant derivative of a vector $V^i$ is defined as
\be   
D_\beta V^i(z) \equiv {\p V^i(z)\over \p z^\beta} + \Gamma^i_{ab}(X(z)){\p X^a(z)\over \p z^\beta}  V ^b(z)
\ee
In the RNC at the origin, this becomes 
\[
\bar D_\beta V^i(z) \equiv \p_\beta V^i(z)
\]
because 
$\bar \Gamma^i_{ab}(0)=0$. Thus $\bar Y^\mu_1(z) = {\p \bar Y^\mu \over \p x_1}$ can be thought of as a vector at the origin (note that $\bar Y^\mu$ is a vector at the origin). Thus in a general coordinate system we will write $y^\mu(O)$ instead of $\bar Y^\mu$ and $\bar Y^\mu_1(z)$ will be written as
\[
D_1 y^\mu (O)
\] The first term in the expansion of \eqref{kineticgen} becomes
\[
 \bar g^R_{\al  \beta}(0)\bar Y^\al_1 \bar Y^\beta _{\bar 1} = g^R_{\al \beta}(O) D_1 y^\al(O) D_{\bar 1}y^\beta(O)
 \]
which is manifestly a scalar at the origin O with coordinates $x_0$.

Similarly the second term becomes
\be	\label{Int}
-{1\over 3} \bar R^R_{\al \mu \beta \la}(0)\bar Y^\mu(0) \bar Y^\la(0) \bar Y_1^\al(0) \bar Y_{\bar 1}^\beta(0)=-{1\over 3} R^R_{\al \mu \beta \la}(O)y^\mu(O) y^\la(O)D_1 y^\al(O) D_{\bar 1}y^\beta(O)
\ee

which is a quartic interaction in the quantum world sheet theory. This will be included as part of $L$ in the ERG. 

Thus we have shown that the kinetic term can be written in a manifestly background covariant form. Let us now turn to the interaction term involving the massless graviton.

\subsubsection{Regularization and Higher Derivative Kinetic Term}

The Wilsonian ERG was originally formulated in momentum space. Regularization is achieved by keeping only low momentum modes in the theory. In the present formulation, position space is being used. Furthermore one has to be careful about not violating general coordinate invariance. 

In position space one obvious way to regulate the theory is to add higher (world sheet) derivative terms to the kinetic term. If one wants a finite cutoff in (world sheet) position space then one has to work with a non local action (eg a lattice).  In continuum language this becomes equivalent to adding arbitrarily high derivative terms to the kinetic term. This superficially seems to be in conflict with GCT because $\p_z^2X^\mu$
is not a vector unlike $\p_zX^\mu$. One solution to this is to write $(D^R_z)^2X^\mu$ where $D^R_z$ is the covariant derivative introduced in Appendix A \eqref{appena}. The superscript $R$ denotes that it is background covariant. In that case one can add terms of the form
\[
\Delta S_0 =\int d^2z~\sum _n c_n a^{2n}(\eta_{\mu\nu}+h_{\mu\nu}^R)(D^R_z)^n X^\mu (D^R_\zb) ^n X^\nu
\]
to the kinetic term action. The coefficients $c_n$ characterize the regulator. Powers of $a^n$ have been added to make the term dimensionless.
Note that this introduces an additional dependence on the arbitrary $h^R_{\mu\nu}$. 

In the loop variable formalism there is another option available. One can write
\[
\Delta S_0 =\int d^2z~\sum _{n,\nb} c_{n,\nb} a^{n+\nb}(\eta_{\mu\nu}+h_{\mu\nu}^R)Y_n^\mu Y_\nb^\nu
\]
Unlike ${\p^nY^\mu\over \p x_1^n}$, $Y_n^\mu \equiv {\p Y^\mu \over \p \xn}$ is a vector and the above term is background GCT invariant. 

Thus as the above preliminary analysis indicates, it is possible to regulate the theory while maintaining general coordinate invariance. In this review we assume this is possible. As a result there is a well defined regularized Green function, for which one can write a covariant Taylor expansion.

If $h_{\mu\nu}^R = h_{\mu\nu}$ then any dependence in the action on $h^R_{\mu\nu}$ is allowed. Otherwise we have to subtract terms in the interaction Lagrangian to cancel the unwanted dependence on $h^R$ introduced in the kinetic term. This involves adding terms of the form $\tilde h_{\mu\nu}Y_n^\mu Y_\nb^\nu$. Such terms are already present in the action so this merely results in  field redefinitions of massive fields.

In \cite{BSERGclosed1} some speculations were made on the possible new space time symmetries of string theory suggested by the presence of these terms. Since the ideas have not been developed sufficiently, we do not describe them in this review.

\subsubsection{Massless Interactions}

The graviton term in the world sheet theory can be written in terms of $\tilde h_{\mu\nu}\equiv h_{\mu\nu}-h^R_{\mu\nu}$ in the RNC as
\be  \label{gravVO}
\int dz~\tilde h_{\mu\nu}(\bar Y(z))\bar Y^\mu_1(z) \bar Y^\nu_{\bar 1}(z)=\int dz~\int d\ko ~\tilde h_{\mu\nu}(\ko)e^{i\ko .\bar Y(z)} \bar Y^\mu_1(z) \bar Y^\nu_{\bar 1}(z)
\ee
$\tilde h_{\mu\nu}(Y)$ is an ordinary tensor under background general coordinate transformation and thus this term is manifestly invariant.
As was done above, this can be expanded in a Taylor series using \eqref{Taylor}. This is equivalent to expanding the exponential in powers of $\ko$. Thus \eqref{gravVO} becomes (commas are covariant derivatives):

\[
[\bar{\tilde h}_{\mu\nu}(0)+ \bar{\tilde h}_{\mu\nu,\rho}(0) \bar Y^\rho(O) + {1\over 2!} \{\bar{\tilde h}_{\mu\nu,\rho \s}(0) - {1\over 3} (\bar R^{R\beta}_{\rho \mu \s }\bar{\tilde h}_{\beta\nu}(0)+\bar R^{R\beta}_{\rho \nu \s }\bar {\tilde h}_{\mu\beta}(0))\}\bar Y^\rho(O)\bar Y^\s(O)+...]\bar Y_1^\mu(O) \bar Y_{\bar 1}^\nu(O)
\]
\be
=[\tilde h_{\mu\nu}(0)+ \tilde h_{\mu\nu,\rho}(0) y^\rho(O) + {1\over 2!} \{\tilde h_{\mu\nu,\rho \s}(0) - {1\over 3} (R^{R\beta}_{\rho \mu \s }\tilde h_{\beta\nu}(0)+R^{R\beta}_{\rho \nu \s }\tilde h_{\mu\beta}(0))\}y^\rho(O)y^\s(O)+...]D_1y^\mu(O) D_{\bar 1}y^\nu(O)
\ee

Thus we have shown that the kinetic term and massless graviton vertex operator term in the world sheet action can be written
in a manifestly background covariant form (i.e. invariant under $G+T$). For the massive modes  Section 7.3.4 explains how manifest invariance under $G+T$ is achieved. 

This concludes our discussion of how the world sheet action can be made manifestly invariant, starting from an action written in RNC.

\subsection{ERG in curved space time}

We now turn to the ERG which is also initially written using RNC. We show that it can also be written in a manifestly background covariant form. If both the world sheet action, and the ERG acting on it are manifestly symmetric, the resulting equations will also have the symmetry.
In an actual calculation it is much easier to work in the RNC system. Thus the strategy will be to work in the RNC, obtain the equations, and then covariantize using the techniques of Section 6.  
\subsubsection{Covariantizing ERG}

We reproduce the ERG equation here. This is written in flat space time. We will replace $X^\mu$ by $\bar Y^\mu$ when we are in curved space time. 

\[
\int du~ {\p L\over \p \tau} \psi = \int dzdz'~\{-\hf \underbrace{\dot G(z,z') G^{-1}(z,z')}_{field~independent}- 
\]
\be    
 \hf \dot G^{\mu\nu}(z,z')[{\delta^2\over \delta X^\mu(z)\delta X^\nu (z')}\int du~L[X(u),X'(u)]] + {\delta\over \delta X^\mu(z)}\int du~L[X(u)]{\delta\over \delta X^\nu(z')} \int du'~L[X(u')]\}\psi =0
\ee

Here $G^{\mu\nu}(z,z')\equiv \lan X^\mu(z) X^\nu(z')\ran$. In curved space time $X^\mu$ is not a vector and the Green function does not have nice transformation properties. More precisely, the combination $X^\mu(z) {\delta\over \delta X^\mu(z)}$ is what occurs in the ERG
and this is not an invariant object. $\bar Y^\mu {\delta\over \delta \bar Y^\mu(z)}$ on the other hand is a well defined object. In Appendix A \eqref{appena}
equation \eqref{Vecfld} defines 
\[
y^\mu(P)={\p Y^\mu \over \p \bar Y^\nu}|_P\bar Y^\mu(\bar Y_P)
\]
a geometric object, namely the tangent vector to the geodesic at P. Thus $y^\mu(P) {\delta\over \delta  Y^\mu(z)}|_P$ is an invariant quantity.
Thus
\[
\lan y^\mu(P) y^\nu(P')\ran  {\delta\over \delta  Y^\mu(z)}|_P{\delta\over \delta  Y^\nu(z')}|_{P'}
\]

is a well defined scalar object. When $z=z'$ it is the first term in the ERG, acting on $L$. 
\[
\int dz\int dz'~\lan y^\mu(z) y^\nu(z)\ran {\delta ^2 \over \delta  Y^\mu(z)\delta Y^\nu (z)}\int du~ L[u]
\]

And when $z\neq z'$ it gives the second term
\[
\int dz \int dz'~\lan y^\mu(z)y^\nu(z')\ran {\delta\over \delta  Y^\mu(z)}\int du~ L[u]  {\delta\over \delta  Y^\nu(z')}\int du'~ L[u']
\]

\subsubsection{Interaction Term and Covariant OPE}

The second term involves vertex operators at different points. Typically one performs an OPE to rewrite it as a sum of operators. The coefficient of any particular operator is the interaction term of an equation of motion for a field dual to that operator. The coefficient of that operator in the first term of the ERG  provides the free part of the equation of motion for that field. It is thus essential to perform the OPE in a covariant way.

The issue of the covariant OPE was discussed in Section 6 in the context of open strings. The same issues are present for closed strings.
The only new ingredient is that one has to expand in $z$ and $\zb$. We repeat some of the points here for convenience. 
There are two ingredients in an OPE: a Taylor expansion and a contraction.  Let us illustrate this with the simplest example.
\footnote{Let us take ${1\over 2\al '}\int d^2x ~(\nabla X)^2 ={1\over \al '}\int d^2z ~\p_z X\p_\zb X$ as our action. Then
$\lan X(z) X(w)\ran = - {\al'\over 2\pi} ln |{z-w\over R}|$. We set $\al' =  4\pi$ for convenience.} 

\be
e^{ikX(z)}e^{ipX(0)} = e^{ikX(z) + ip X(0)}=e^{ik(X(0)+z\p_z X(0) + \zb \p_\zb X (0) + \hf z^2 \p_z^2 X(0)+ z \zb \p_z \p_\zb X(0)+ \hf \zb^2 \p_\zb^2X(0)+...) +ipX(0)}
\ee
In order to take care of self contractions we can introduce the normal ordered vertex operators by
\be
e^{ikX(z)} = e^{-{k^2\over 2} ln~a^2}:e^{ikX(z)}:
\ee
and
\[
e^{ikX(z) + ip X(0)}= e^{\hf \langle (ik.X(z)+ipX(0))(ik.X(z)+ipX(0))\rangle} :e^{ikX(z) + ip X(0)}:
\]
\[
= e^{-  {(k^2 +p^2)\over 2} ln~a^2 - k.p~ ln~(|z|^2+a^2)} :e^{ikX(z) + ip X(0)}:
\]
\be
=e^{-  {(k^2 +p^2)\over 2} ln~a^2 - k.p~ ln~(|z|^2+a^2)}:e^{ik(X(0)+z\p_z X(0) + \zb \p_\zb X (0)+ \hf z^2 \p_z^2 X(0)+ z \zb \p_z \p_\zb X(0)+ \hf \zb^2 \p_\zb^2X(0)..) +ipX(0)}:
\ee
We have used a choice of cutoff Green function $G(z,0;a)=  ln ~(|z|^2+a^2)$ for illustration. These equations are written in flat space.
In curved space time one has to be more careful. It has been argued in an earlier section that it is possible to regularize the theory on the world sheet while maintaining space time background general coordinate invariance. However non local expressions cannot easily be written in covariant form and the simplest way to get covariant expressions is to perform covariant Taylor expansions.
Thus to begin with the Green function needs to be Taylor expanded. Thus
$\lan y^\mu (z) y^\nu (0)\ran$ has to be expressed as a power series in $z,\zb$ and then each term has to be covariantized. The results are given in Appendix A \eqref{appena}.   
Thus for instance we write in the RNC (below symmetrization does not have  a normalization factor of $n!$ - so that is explicitly multiplied.)
\[
\bar Y^i (z) = \bar Y ^i (0) + z^\al  \bar Y_\al ^i (0) + {z^\al z^\beta\over 2!} \p_\al  \bar Y^i _\beta(0) + {z^\al z^\beta z^\gamma\over 3!} \p_\al \p_\beta \bar Y^i_\gamma (0)+  {z^\al z^\beta z^\gamma z^\delta \over 4!} \p_\al \p_\beta \p_\gamma  \bar Y^i _\delta(0) +...
\]
\[
= \bar Y ^i (0) + z^\al \bar Y ^i _\al(0) + {z^\al z^\beta\over 2!} D_\al  \bar Y^i _\beta(0) + {z^\al z^\beta z^\gamma\over 3!} D_\al D_\beta  \bar Y^i _\gamma(0)
\]
\be
+  {z^\al z^\beta z^\gamma z^\delta \over 4!}[ D_\al D_\beta D_\gamma \bar Y^i_\delta (0) + {1\over 48} (R^i_{~dac}(0)+ R^i_{~cad}(0))\bar Y^d_{(\delta} (0)\bar Y^c_\gamma (0)\bar D_\beta Y_{\al)}^a(0)]+...
\ee
which is a covariant expansion.  

Thus the Green function is expanded as
\be
G^{ij}(z,0;a)= G^{ij}(0,0;a) + z^\al (\p_\al G^{ij}(z,0;a))|_{z=0}+ {z^\al z^\beta\over 2!}(\p_\al \p_\beta G^{ij}(z,0;a))|_{z=0}+...
\ee
Note that every term is finite because of the presence of a cutoff. Each term involves the metric tensor, Riemann tensor and (covariant) derivatives thereof, all evaluated at one point, which can be taken to be the origin of the RNC.
Similar expansions have to be done for the terms in the world sheet action which are products of space time fields and vertex operators. These are given in Appendix A \eqref{appena}. The covariant looking expansions are done in RNC, but the full object (product of field and vertex operators occurring in the interaction term of the ERG) is a scalar and the expansion is therefore valid in any coordinate system. Thus as an example
\eqref{9} in Appendix A \eqref{appena} gives the following result, which can be used in an OPE:
\br
S_i (X(z))X_\al^i(z)&=& S_i (X(0))X_\al^i(0) + z^\beta [ \hf S_i(X(0))D_{(\beta}X_{\al)}^i(0) + \n_jS_i(X(0))X_\al^iX_\beta ^j(0)]\nonumber \\
& & {z^\beta z^\al\over 2!}[\hf \n_j S_i (X(0)) D_{(\beta}X_{\al)}^i + ( {\n_{(i}\n_{j)}S_i(X(0))\over 2} + {1\over 3} R^l_{~jki}(X(0))S_l(X(0)))X_\beta ^k X_\gamma ^j X_\al^i(0) + \nonumber \\
& &+ {D_{(\al}D_\beta X_{\gamma)}^i(0)S_i(X(0))\over 6} + (\n_jS_i(X(0)))D_{(\beta} X_{\al)})^i X_\gamma^j(0)]\nonumber
\er
\be		
+...~~~~~~~~~~~~~~~~~~~~~~~~~~~~~~~~~~~~~~~~~~~~~~~~~~~~~~~~~~~~~~~~~~~~~~~~~~~~~~~~~
\ee

Thus to conclude: We have all the necessary ingredients for a covariant equation. Both the ERG and the action have been written in manifestly covariant form. We also have a covariant OPE. The result of all this is therefore covariant. Having assured ourselves that the result is covariant, we are free  to work in the RNC where the equations are simpler, and covariantize everything in the end. This is computationally far simpler. In Section 6 we gave an algorithm for covariantizing the loop variable equation. This can be used.

Two points that need to be noted:

{\bf 1.} In Section 6 we had to perform some field redefinitions at the intermediate stages
of the calculation in order to preserve the gauge invariance of the equations in curved space time. When the world sheet action is written in covariant form this step has to be incorporated and the same field redefinitions have to be performed. We will not bother to do this because we never actually use the covariant form of the world sheet action. Instead, are going to be working in the RNC with loop variables and  the covariantization is only done at the last stage when we map to space time fields. 

{\bf 2.} The metric $g_{\mu\nu}^R$ is completely arbitrary. So it is consistent to set it equal to the physical metric. In that case the field
$\tilde h_{\mu\nu}$ that represents the graviton is actually zero. All interaction involving the gravitational field involves only the curvature tensor. Furthermore the symmetry $G+T$ which is manifest is the usual general coordinate invariance. For many purposes this choice is simpler and more convenient. 

While keeping in mind the option of setting $\tilde h_{\mu\nu}=0$, we now proceed to write the background covariant form of the equations involving the graviton $\tilde h_{\mu\nu}$.

\subsection{Background Covariant Equation for $\tilde h_{\mu\nu}$}

We first evaluate the field strength corresponding to $\tilde h_{\mu\nu}$ which involves acting with the functional derivative \eqref{FnlDer}
 once. We write $S_{int}= \int du L[u]$. We work in the RNC and covariantize at the end.

\subsubsection{Field Strength}

The functional derivative below gives the field strength:
\[	
{\delta S_{int} \over \delta \bar Y^\rho(z)} - {\p \over \p x_1}{\delta S_{int} \over \delta \bar Y_1^\rho(z)}- {\p \over \p \bar x_1}{\delta S_{int} \over \delta \bar Y_{\bar 1}^\rho(z)}=
\]
\[
  \hf[{\p \bar {\tilde h}_{\mu \nu}\over \p \bar Y^\rho}(\bar Y(z))  - {\p \bar {\tilde h}_{\rho \nu}\over \p \bar Y^\mu}(\bar Y(z))-{\p \bar {\tilde h}_{\mu \rho}\over \p \bar Y^\nu}(\bar Y(z))
+{2\over 3} \bar Y^\beta(z) (\bar R^R_{\rho \nu \mu \beta}(0)+\bar R^R_{\rho \mu \nu \beta}(0))]\bar Y^\mu_1(z) \bar Y^\nu _{\bar 1}(z)\]
\be	\label{GravFS}
+[ \bar {\tilde h}_{\rho \mu}(\bar Y(z))+{1\over 6} \bar Y^\alpha(z) \bar Y^\beta (z)(\bar R^R_{\rho \alpha \mu \beta}(0)+\bar R^R_{\mu \alpha \rho \beta}(0))]\bar Y^\mu_{1;\bar 1}(z)
\ee
where $\bar{\tilde h}_{\mu\nu}= \bar{h}_{\mu \nu}-\bar {h}^R_{\mu\nu}$ and all arguments of fields  have been displayed to avoid confusion. The bars on the metric fluctuation and curvature tensor are just to remind us that we are working in the RNC. The free indices are contracted with vertex operators and the second field strength in the interaction term. It then becomes a scalar. Anticipating this, to go to a general coordinate system we just remove the bars.
The field strength tensor is, to this order,
\be \label{CovGravFS2}
F_{\rho \mu \nu}(\bar Y)= \hf[{\p \bar {\tilde h}_{\mu \nu}\over \p \bar Y^\rho}(\bar Y(z))  - {\p \bar {\tilde h}_{\rho \nu}\over \p \bar Y^\mu}(\bar Y(z))-{\p \bar {\tilde h}_{\mu \rho}\over \p \bar Y^\nu}(\bar Y(z))
+{2\over 3} \bar Y^\beta(z) (\bar R^R_{\rho \nu \mu \beta}(0)+\bar R^R_{\rho \mu \nu \beta}(0))+...]
\ee

The field $\tilde h$, and the curvature tensor are gauge covariant and thus so is the field strength. 

The first term involving $\tilde h$ is at a general point $\bar Y$ and has to be Taylor expanded about the origin in powers of $\bar Y$. The term involving the curvature tensor is already at O($\bar Y$) and is non leading. (It contributes to leading order in the free equation derived below.)

 We can write  the leading term in a general coordinate system by  the usual procedure  of writing background covariant derivatives: 
\be  \label{CovGravFS}
(\Gamma_{\rho \mu \nu }-\Gamma^R_{\rho \mu \nu })\rightarrow  \hf(\nabla ^R _\mu \tilde h_{\rho \nu} + \nabla^R _\nu \tilde h_{\rho \mu}  - \nabla ^R _\rho \tilde h_{\mu\nu})\equiv\tilde \Gamma^R_{\rho \mu \nu} 
\ee

Similarly a non leading term (in powers of $\bar Y$) is 
\[
{2\over 3} \bar Y^\beta(z) (\bar R^R_{\rho \nu \mu \beta}(0)+\bar R^R_{\rho \mu \nu \beta}(0))
\]
which can be covariantized to 
\be	\label{CovGravFS2}
{2\over 3}  y^\beta(z) ( R^R_{\rho \nu \mu \beta}(0)+ R^R_{\rho \mu \nu \beta}(0))
\ee
where $y^\mu$ was defined in \eqref{Vecfld} of Appendix A \eqref{appena} and used in Sec 7.5 while covariantizing the ERG. 
In addition to this there are terms of order O($\bar Y$) coming from the Taylor expansion of $\tilde \Gamma _{\rho \mu \nu}$
\be	\label{CovGravFS3}
\bar Y^\sigma \bar \nabla^R \bar{ \tilde \Gamma }_{\rho \mu \nu}(0) \to y^\sigma \nabla ^R  \tilde \Gamma _{\rho \mu \nu}(0)
\ee
The full field strength is a sum of all these: \eqref{CovGravFS}, \eqref{CovGravFS2} and \eqref{CovGravFS3} and higher order terms.

If we choose $h_{\mu\nu}=h^R_{\mu\nu}$ then the only term in the field strength is \eqref{CovGravFS2} and higher derivatives of the curvature tensor. 
\be	
{2\over 3}  y^\beta(z) ( R_{\rho \nu \mu \beta}(0)+ R_{\rho \mu \nu \beta}(0))+...
\ee

\subsubsection{Free Equation}

The free graviton equation for the metric fluctuation about flat space was derived in Section 7.3 and is (with the dilaton field $\Phi_D=0$):
\be 	\label{GravFree}
\p^\rho (\Gamma_{\rho\mu\nu}-\Gamma^R_{\rho\mu\nu})-\p_\mu\p_\nu \tilde h^\rho_{~\rho}=0
\ee
 \eqref{GravFree} is the RNC version ( at the origin where $\Gamma^R(0)=0$) of 
the covariant equations in a general coordinate system:
\be   \label{GravFreeCov1}
\nabla^{R}_{\sigma}( g^{R\sigma \rho}\tilde \Gamma_{\rho\mu\nu}) - \hf\nabla^R_\mu \nabla^R_\nu \tilde h^\rho_{~\rho} =0
\ee
where 
\[
\tilde \Gamma_{\rho\mu\nu}=\hf[- \nabla^R_\rho \tilde h_{\mu\nu} + \nabla^R_\mu  \tilde h_{\rho\nu} +\nabla^R_\nu  \tilde h_{\rho \mu}]
\]
We  work out for completeness the contribution due to the rest of the terms involving the background curvature tensor.  The equation in loop variable notation is
 \be
\hf[ -\ko^2 \kim \kinb + \ko .\ki (\kom \kinb + \kon \kimb) - \kom \kon \ki .\kib] =0
 \ee
 The equation is being evaluated at the origin O, where $\bar Y^\mu=0$, so only the quadratic term contributes - the cubic and higher order terms do not contribute.

Therefore  we use 
 \[ 
 \kim \kinb = {1\over 6} \bar Y^\al \bar Y^\beta    ( \bar {R}^R_{\mu \al \nu \beta}(0) +\bar R^R_{\nu \al \mu \beta}(0))
 \]
 in the above and obtain
 \be
 - \bar R^R_{\mu \nu} \bar Y_1^\mu \bar Y_{\bar 1}^\nu
 \ee
 Thus the total for the graviton contribution  to the free graviton EOM is (dropping bars):
 \be \label{GravFreeCov}
(  R^R_{\mu \nu}+\nabla^{R}_{\sigma}( g^{R\sigma \rho}\tilde \Gamma_{\rho\mu\nu})-\hf\nabla^R_\mu \nabla^R_\nu \tilde h^\rho_{~\rho})  Y_1^\mu  Y_{\bar 1}^\nu
\ee

\subsubsection{Comparison with Einstein's Equation}

This free equation in the first case should be compared with what one expects for a graviton from Einstein's vacuum equation $R_{\mu\nu}=0$ expanded to linear order in $\tilde h$, 
about a background. One can expand as follows:
\be   \label{Eins}
R_{\mu \nu} = R^R_{\mu \nu} + \delta R_{\mu \nu}
\ee

To evaluate $\delta R$,
 go to an inertial frame with $\Gamma =0$ at the point under consideration,
\[R^\alpha _{~\mu \beta \nu} = \p_\beta \Gamma ^\alpha _{~\mu \nu} - \p_\nu \Gamma ^\alpha _{~\mu \beta}
\]
So
\[\delta R^\alpha _{~\mu \beta \nu} = \p_\beta \delta \Gamma ^\alpha _{~\mu \nu} - \p_\nu \delta \Gamma ^\alpha _{~\mu \beta}
\]
Now unlike $\Gamma ^\alpha _{~\mu \nu}$, $\delta \Gamma ^\alpha _{~\mu \nu}$ is a tensor, so the above equation, if written covariantly, is valid in all frames:
\be 
\delta R^\alpha _{~\mu \beta \nu} = \nabla_\beta \delta \Gamma ^\alpha _{~\mu \nu} - \nabla_\nu \delta \Gamma ^\alpha _{~\mu \beta}
\ee
So we get the Palatini equation:
\be 	\label{Palatini}
\delta R_{\mu \nu}=\delta R^\alpha _{~\mu \alpha \nu} = \nabla_\alpha \delta \Gamma ^\alpha _{~\mu \nu} - \nabla_\nu \delta \Gamma ^\alpha _{~\mu \alpha}
\ee
We now show that to linear order in $\tilde h$ (or $h$), 
\be   \label{Result}
\delta \Gamma ^\alpha _{~\mu \nu}\equiv\Gamma ^\alpha _{~\mu \nu}-\Gamma ^{R\alpha} _{~\mu \nu}=g^{R \alpha\rho}\tilde \Gamma _{\rho \mu \nu}
\ee
where
\[
 \hf(\nabla ^R _\mu \tilde h_{\rho \nu} + \nabla^R _\nu \tilde h_{\rho \mu}  - \nabla ^R _\rho \tilde h_{\mu\nu})\equiv\tilde \Gamma^R_{\rho \mu \nu} 
 \]
Writing $g^{\rho \sigma}= g^{R\rho \sigma}+\delta g^{\rho \sigma}$ we get 
\[
\Gamma ^\sigma _{~\mu \nu} -\Gamma ^{R\sigma} _{~\mu \nu}= g^{R\rho \sigma}(\Gamma _{\rho \mu \nu} -\Gamma^R _{\rho \mu \nu})+\delta g^{\rho \sigma} \Gamma _{\rho \mu \nu}
 \]
\be \label{LHS}
=g^{R\rho \sigma}(\Gamma _{\rho \mu \nu} -\Gamma^R _{\rho \mu \nu})+\delta g^{\rho \sigma} \Gamma^R _{\rho \mu \nu}
\ee
to linear order in $h$.

Now consider the RHS of \eqref{Result}. Expand the covariant derivatives:
\be \label{RHS}
g^{R\rho \sigma} [(\Gamma _{\rho \mu \nu} - \Gamma^R _{\rho \mu \nu})-\Gamma^{R\alpha}_{~\mu \nu} \tilde h _{\rho \alpha}]
\ee

If we now take into account the fact that $\delta g^{\rho \sigma}= -g^{R\rho \alpha}h_{\alpha \beta}  g^{R\beta \sigma}$
we see that \eqref{LHS} and \eqref{RHS} are equal and we have the result \eqref{Result}. Furthermore taking the trace of \eqref{Result} we get
\be \label{Trace}
\Gamma ^\alpha _{~\mu \alpha}-\Gamma ^{R\alpha}_{~~\mu \alpha} = g^{R\alpha \rho}\hf(\nabla _\mu^R\tilde h_{\rho \alpha}+\nabla^R_\alpha \tilde h_{\mu \rho} -\nabla ^R_\rho\tilde h_{\mu \alpha})=\hf \nabla _\mu^R \tilde h^\rho_{~\rho}
\ee

Inserting \eqref{Result} and \eqref{Trace} into \eqref{Palatini} we obtain the  equation for $\tilde h$ in the background metric (including for completeness the background contribution):
\be
R^R_{\mu \nu}+ \nabla^R_\sigma (g^{R\sigma \rho} \tilde \Gamma _{\rho \mu \nu})-\hf \nabla^R_\nu\nabla _\mu^R \tilde h^\rho_{~\rho} =0
 \ee
 which is the covariantized equation that we obtained in the loop variable approach in Section 7.6.2, \eqref{GravFreeCov}.

Finally if we set $h_{\mu\nu}=h^R_{\mu\nu}$ (so that $\tilde h_{\mu\nu}=0$)  then the free equation simply reduces to Einstein's equation
\[
R_{\mu\nu}=0
\]

\subsubsection{Interactions of the graviton}

The interaction terms will
involving OPE of the field strength with itself as well as with field strengths of other modes. The field strength was given to leading order in
Section 7.6.1 (\eqref{CovGravFS},\eqref{CovGravFS2},\eqref{CovGravFS3}). One has to expand the OPE covariantly. This has been described in Section 7.5.2 and can be applied here in a fairly obvious way. We will not describe this again.

In the last few  subsections we have described at some length the contribution of the graviton to the equation of motion. There was some subtlety in this because of the dual role of the graviton: On the one hand it is just another mode of the closed string and on the other, being massless one has to describe in a convenient way the gauge symmetry associated with it - which is general covariance. 

The massive modes are more straightforward from this point of view. Nevertheless one has to resolve the clash between general covariance of their EOM and the higher ``broken" gauge symmetries associated with them.  This clash, at the algebraic level, is the same
as the one we faced for the open string modes for which a solution was presented in Section 6. The solution was given there  as a four step algorithm and the same algorithm can be applied {\it mutatis mutandis} for closed string massive modes also.

\subsection{Example: Closed String - Level $(2,\bar 2)$}

In this section we  work out the result for the first massive level of closed strings. The free equation of motion and free action were worked out in Section 7.2.2. We give here the free equation in a curved space time background using the technique for mapping from loop variable expressions to fields in curved space time described in Section 6. The loop variable equation of motion is
\[
   -{1\over 4} \ko^2 (\ki .Y_1)^2 (k_{\bar 1} .Y_{\bar 1})^2 +\hf \ko.\ki (\ko .Y_1)(\ki .Y_1) (\kib .\yib)^2 + \hf \ko.\kib (\ko.\yib)(\kib .\yib)(\ki.\yi)^2 +
\]
\be    \label{Spin4}
   -{1\over 4} \ki.\ki (\ko.\yi)^2(\kib.\yib)^2 -{1\over 4} \kib.\kib (\ko.\yib)^2(\ki.\yi)^2 - \ki.\kib (\ko.\yi)(\ko.\yib)(\ki.\yi)(\kib.\yib) =0
\ee
  It is gauge invariant under 
\[
  \kim \to \kim + \li \kom;~~~~\kimb \to \kimb + \la_{\bar1} \kom
\] 
  if we use the tracelessness condition on the gauge parameters: 
\be   \label{const}
   \li \ki.\kib \kimb= \li \kib.\kib \kim =0= \lib \ki.\kib \kim = \lib \ki.\ki \kimb
\ee

The fields were also defined in Sec 7.2. We specialize to the four index tensor and define the dimensionally reduced fields:
\br
\lan \kim \kin \kirb \kisb \ran &=&S_{1\mu 1\nu \bar 1\rho \bar 1\sigma }\nonumber \\
\lan q_1  \kin \kirb \kisb \ran &=& S_{11\nu\bar 1\rho \bar 1\sigma }\qo \nonumber \\
\lan q_{\bar 1} \kim \kin \kirb \ran&=& S_{1\mu 1\nu \bar 1 \rho 1}\qo \nonumber \\
...& & \\
\er
and similarly for the remaining fields.  We hope the notation is clear to the reader.
Consider the first term in \eqref{Spin4} (the bar on $Y$ indicates RNC): 
\be
 \ko^2 (\ki .\bar Y_1)^2 (k_{\bar 1} .\bar Y_{\bar 1})^2 = \ko^2 \kim \kin k_{\bar 1 \rho} k_{\bar 1 \sigma} \bar Y_1^\mu \bar Y_1^\nu \bar Y_{\bar 1}^\rho \bar Y_{\bar 1}^\sigma
\ee
Thus we need to map $ \ko^2 \kim \kin \kirb \kisb$ to a space-time field using our modified prescription. 

The four index tensor equation map is quite tedious to work out. There is no new complication that arises except that we need the Taylor expansion in RNC \eqref{Taylor} to higher orders. So we will only give outlines. 

The constraints \eqref{const}  can be mapped directly to space-time field constraints. If it is zero in flat space, it continues to be zero even in curved space since the extra curvature couplings in curved space are also linear in the constraint.

{\bf Step 1}

Let $\kim = \tkim + y_1 \kom$ and $\kimb = \tilde k_{\bar 1\mu} + y_{\bar 1} \kom$. Then we obtain:
\be	\label{redef}
\ko^2 \kim \kin \kirb \kisb= \ko^2(\tkim +y_1\kom)(\tkin+ y_1\kon)(\tilde k_{\bar 1\rho} + y_{\bar 1} \kor)(\tilde k_{\bar 1\sigma} + y_{\bar 1} \kos)
\ee

This has to be done for each term in \eqref{Spin4}.

{\bf Step 2}

We define some tilde fields at the intermediate stage as:
\br
\lan \tkim \tkin \tilde k_{\bar 1\rho} \tilde k_{\bar 1\sigma} \ran &=&\tilde S_{1\mu 1\nu \bar 1\rho \bar 1\sigma}\nonumber \\
\lan y_1  \tkin   \tilde k_{\bar 1\rho} \tilde k_{\bar 1\sigma}\ran &=& \tilde S_{11\nu \bar 1\rho\bar 1 \sigma} \nonumber \\
\lan y_{\bar 1}\tkim  \tkin   \tilde k_{\bar 1\rho} \ran &=& \tilde S_{1\mu 1\nu \bar 1\bar 1\rho } \nonumber \\
\lan y_1 y_{\bar 1} \tkim \tilde k_{\bar 1\sigma}\ran &=& \tilde S_{11\mu \bar 1\bar 1 \sigma}\nonumber \\
\lan y_{\bar 1}^2 \tkim \tkin\ran &=& \tilde S_{1\mu 1\nu \bar 1 \bar 1} \nonumber \\
...& &\nonumber\\
\er
etc.

We need to work out the map between these sets of fields.
We have
\be
\lan \qi \qi \qib \qib \ran = S_{11\bar 1 \bar 1}= \qo^4 \lan y_1y_1y_{\bar 1}y_{\bar 1}\ran = \qo^4 \tilde S_{11\bar 1 \bar 1}
\ee

\br
\lan \qi \qi \qib \kirb \ran &=& \qo^3 \lan y_1 y_1 y_{\bar 1} (\tilde k_{\bar 1 \rho} + y_{\bar 1} \kor) \ran \nonumber \\
\implies S_{11\bar 1 \bar 1\rho} &=& \qo^3  \tilde S_{11\bar 1 \bar 1\rho} + \qo^3 \nabla _\rho \tilde S_{11\bar 1 \bar 1}
\er

\br
\lan \qi \qi \kirb \kisb\ran &=& \qo^2 y_1^2\lan \tilde k_{1\rho}\tilde k_{\bar1\sigma} + y_{\bar 1} \kor \tilde k_{\bar1\sigma} +y_{\bar 1} \kos \tilde k_{\bar1\rho} + y_{\bar 1}^2 \kor \kos \ran \nonumber \\
S_{11\bar 1 \rho \bar 1 \sigma} &=& \qo^2 \tilde S_{11\bar 1 \rho \bar 1 \sigma} +\qo^2 \nabla _\rho \tilde  S_{11\bar 1  \bar 1 \sigma}+\qo^2
\nabla _\sigma S_{11\bar 1  \bar 1 \rho}+ \qo^2 \nabla _\rho \nabla _\sigma \tilde S_{11\bar 1 \bar 1}
\er

Solving these equation for the tilde fields one obtains:

\br   \label{2tensor}
\tilde S_{11\bar 1 \bar 1}&=&{1\over \qo^4}S_{11\bar 1 \bar 1}\nonumber \\
\tilde S_{11\mu \bar 1 \bar 1} &=&  {S_{11\mu \bar 1 \bar 1}\over \qo^3} - {\nabla_\mu S_{11\bar 1 \bar 1}\over \qo^4}\nonumber \\
\tilde S_{11 \bar 1\rho \bar 1\sigma}&=& {S_{11 \bar 1\rho \bar 1\sigma}\over \qo^2} - { \nabla_{(\rho}S_{11 \bar 1 \bar 1\sigma)}\over \qo^3}
+{ \nabla_\rho\nabla_\sigma S_{11\bar 1 \bar 1}\over \qo^4}
\er
The above equations are essentially the same as was given in the last section for open strings. We further need expressions for the three and four index tensors.

After some straightforward algebra one finds the following relation for the three index tensor:
\br   \label{3tensor}
\tilde S_{11\nu \bar 1 \rho \bar 1 \sigma}&= & {S_{11\nu \bar 1 \rho \bar 1 \sigma}\over \qo} -{1\over \qo^2}[ \nabla_\nu S_{11\bar 1 \rho \bar 1 \sigma}+\nabla_\rho S_{11\nu \bar 1  \bar 1 \sigma}+\nabla_\sigma S_{11\nu \bar 1 \rho \bar 1 }]\nonumber \\ 
& &+ {1\over \qo^3}[\nabla_{\rho} \nabla_\nu S_{11 \bar 1 \bar 1\sigma} +\nabla_{\sigma} \nabla_\nu S_{11 \bar 1 \bar \rho1} +\nabla_{\sigma} \nabla_\rho S_{11\nu \bar 1 \bar 1}] + {1\over \qo^4}\nabla_\sigma \nabla _\nu \nabla_\rho  S_{11\bar 1 \bar 1}
\nonumber \\
& &+{2\over 3}(R^\lambda_{~\rho \nu \sigma}+R^\lambda_{~\sigma \nu \rho})[{S_{11 \bar 1 \bar 1\lambda}\over \qo^3}
-{\nabla_\lambda S_{11\bar 1 \bar 1}\over \qo^4}] +{1\over 3}(R^\lambda_{~\sigma\rho \nu }+R^\lambda_{~ \nu \rho \sigma})[{S_{11\lambda \bar 1 \bar 1}\over \qo^3}] 
\er

Finally using the same methods the four index tensor is seen to satisfy a relation of the form
\[
S_{1\mu 1\nu \bar 1 \rho \bar 1 \sigma} = \tilde S_{1\mu 1\nu \bar 1 \rho \bar 1 \sigma} + (lower~index ~tensors)
\]

Using \eqref{2tensor} and \eqref{3tensor}, one can solve for $ \tilde S_{1\mu 1\nu \bar 1 \rho \bar 1 \sigma}$ in terms of the ordinary fields.
We do not work it out here. Thus we have expressions for the tilde fields in terms of original fields.
This is the end of Step 2.

{\bf Step 3}

Using the results of \eqref{Taylor} we obtain for instance:
\br    \label{cl-step2}
\lan \ko^2 \tkim \tkin \tilde k_{\bar 1\rho}\tilde k_{\bar 1\sigma}\ran &=&\nabla^2 \tilde S_{11\bar 1\bar 1 \mu \nu \rho \sigma}-{1\over 3}(R^\lambda _{~\mu} \tilde S_{11\bar 1\bar 1 \lambda \nu \rho \sigma}+R^\lambda _{~\nu} \tilde S_{11\bar 1\bar 1 \mu \lambda  \rho \sigma}+R^\lambda _{~\rho} \tilde S_{11\bar 1\bar 1 \mu \nu \lambda \sigma}+R^\lambda _{~\sigma} \tilde S_{11\bar 1\bar 1 \mu \nu \rho \lambda})\nonumber \\
\lan\ko^2 y_1 \kom \tkin \tilde k_{\bar 1\rho}\tilde k_{\bar 1\sigma}\ran &=&\nabla_\mu \nabla^2 \tilde S_{11\bar 1\bar 1 \nu \rho \sigma}-(R^\lambda _{~\nu}\nabla_\mu \tilde S_{11\bar 1\bar 1 \lambda \rho \sigma}+R^\lambda _{~\rho}\nabla _\mu \tilde S_{11\bar 1\bar 1\nu  \lambda  \sigma}+R^\lambda _{~\sigma}\nabla _\mu \tilde S_{11\bar 1\bar 1 \nu \rho \lambda}) \nonumber \\ & &- \hf (\nabla _\mu R^\lambda _{~\nu} \tilde S_{11\bar 1\bar 1\lambda  \rho  \sigma}+\nabla _\mu R^\lambda _{~\rho} \tilde S_{11\bar 1\bar 1\nu  \lambda  \sigma}+\nabla _\mu R^\lambda _{~\sigma} \tilde S_{11\bar 1\bar 1 \nu \rho \lambda})
\er
We do not bother to write down the rest of the terms. As the number of derivatives increase the expressions become more complicated. Hopefully it is clear to the reader that given the taylor expansion \eqref{Taylor} the terms can easily be written down.
 
{\bf Step 4}

The last step is to plug in the results of Step 2 into \eqref{cl-step2} and obtain expressions involving the original fields
 and then insert these into \eqref{redef}. Since the map from loop variables to space time fields has been done in such a way that
 gauge transformations are well defined, the gauge invariance of the loop variable expression guarantees the gauge invariance of the field theory expression.
 
 This has to be done for each term in the equation of motion \eqref{Spin4}. We do not work out the details since the details are not very
illuminating and there are no further conceptual issues.

  It should also be clear that all higher derivative terms not involving the curvature tensor cancel and reproduce the flat space result. Then the curvature couplings to the four index tensor field give the naive covariantization just as in \eqref{op-step3}.  The remaining terms are curvature coupling to Stuckelberg fields. These involve higher derivatives and  are required for gauge invariance. However they can be set to zero by a choice of gauge and so propagation of  physical fields is described by a second order differential equation. Presumably this is sufficient for classical consistency of the theory.  

\subsection{Central Charge}

In the BRST formulation of string theory, which is a gauge invariant one, the extra dimension corresponds to a bosonized ghost field field
on the world sheet. Accordingly the central charge contribution of this is -26. This is cancelled by a contribution of +26 from 26 space time coordinates. This cancellation is crucial for gauge invariance. Furthermore although formally an extra dimension, in detail the ghost has different dynamics. Not only that, the equations of motion for the massive fields are not in a form that would be obtained by dimensional reduction of a massless theory in one higher dimension. In \cite{SZ} it was also shown that {\em in 26 dimensions} it is possible to do field redefinitions that bring the free equations of motion into standard form - i.e. of a dimensionally reduced massless theory.

In the loop variable approach the viewpoint is that space time gauge invariance is the primary requirement. The construction described in the earlier sections achieves this requirement. However the extra coordinate behaves more or less like an extra dimension. It plays the role of producing the effects of the naive world sheet scaling dimension in terms of an anomalous dimension $\qo^2$ where $\qo$ is the momentum in the extra dimension. But if this is not to affect the value of the poles of the scattering amplitudes it must be true that the correlation function of these field must vanish except at coincident points. Thus it must behave like a massive (on the world sheet) field. A priori it is not clear
what the connection is with the bosonized ghost field. Furthermore it is not clear why the central charge should be -26 and hence there is no critical dimension. It seems natural to conclude that only in 26 dimensions is this theory equivalent to critical string theory. 
In fact in \cite{BSOC} it was shown (and reviewed here in Section 8) that at least for level 2 and 3 open string modes, it is possible to redefine fields such that the physical state constraints and gauge transformations in the loop variable approach have exactly the same form as in string theory old covariant formalism
{\em provided the critical dimension is 26 and the masses of the fields have the string theory value.}

In other dimensions the theory can still be gauge invariant and have massless gauge fields, so it does not seem to have the properties usually ascribed to non critical strings. We do not have an answer to this question. Nevertheless there is a more limited sense in which the question can be asked. In the world sheet RG approach to string theory the critical dimension constraint arises as an equation of motion for the dilaton whose vertex operator is taken as  $\p_z \p_\zb \sigma$. The EOM for the dilaton thus has a term $D-26$ in it. Thus we can at least ask about the dilaton equation in the loop variable formalism and whether the central charge term arises. We give  a partial answer to this question.

In Section 5.3 we saw that the ERG has a field independent term $\hf Tr \dot G G^{-1}$. This gives the contribution to the trace anomaly from the determinant (or equivalently the measure in Fujikawa's interpretation). It has non universal divergent parts and a universal finite part proportional to $\p_z \p_\zb \sigma$ where $\sigma$ is the Liouville mode and gives the central charge. 

The connection to the dilaton comes from the observation \cite{CDMP, Polchinskitextbook}  that normal ordering operators such as $\p_z X \p_\zb X$ produces
$\p_z \p_\zb \sigma$ on a curved world sheet. This is the form of the trace of the kinetic term and the coefficient of this is the dilaton.
Alternatively in the loop variable approach one can take operators of the form $Q_{1;\bar 1}\p_z \p_\zb \theta e^{i\qo \theta}$ which on normal ordering produces
$\qo Q_{1;\bar 1}\p_z \p_\zb \sigma$. Thus we can expect $\qo Q_{1;\bar 1}$ to stand for the dilaton. In fact we have seen that it stands for
$\tilde h^\rho_\rho + \Phi_D$. So if we set $h_{\mu\nu}=h^R_{\mu\nu}$ (so that $\tilde h=0$), then indeed it stands for the dilaton operator. Thus we can expect the dilaton equation to include a contribution from the central charge. The precise connection between the bosonized ghost and the extra coordinate in the loop variable formalism needs to be clarified in order to find the central charge.

\section{Connection with the Old Covariant Formalism}

In this section we review some results that have been worked out in \cite{BSOC} on the relation between the loop variable method and the old covariant formalism that was alluded to in the last section \footnote{In \cite{BSOC} at level 3 the Q-rules were not used. Here we have used the Q-rules right from the beginning. Thus the discussion on Level 3 is modified a little.} This can be viewed as the first steps towards a proof that in the critical dimension, the two approaches describe the same physical theory.  The discussion is confined to the open string and that too for the first
two massive levels. Hopefully it can be generalized. 
\subsection{Old Covariant Formalism}

In the OC formalism  the physical state constraints 
are given by the action of $L_{+n}, n\geq 0$ and gauge transformations by 
the action of $L_{-n}, n > 0$. In \cite{BSVG} a closed form expression is given
for the following:
\be       \label{OC} 
\eln \gvk |0 \rangle 
\ee
where $\tY_n = {\p _z^n X (z) \over (n-1)!}~, \tilde Y_0 = \tilde Y =X$. We will need it mainly
to linear order in $\la _n$ which can be obtained from:
\be   \label{Vir}
e^{-{1\over 2}\cal Y^T \la  \cal Y }\gvk
\ee
where ${\cal Y}^T = (..,\tY _3 , \tY _2 , \tY _1, -i k_0 , -i k_1 , -2i k _2 , 
-3i k_3,....)$ and $\la $ is a matrix whose elements are given by:
\[ (\la )_{m,n} = \la _{m+n} \] (\ref{Vir}) will be used below.

\subsubsection{Level 2}

{\bf Vertex operators}

The level two vertex operators are obtained from
\be \label{2}
e^{ik_0 .\tY + ik_1 . \tY _1 + ik_2 .\tY _2} |0\rangle =
 e^{ik_0.X}(... -{1\over 2} \kim \kin \p X^\mu \p X^\nu +
 i \ktm \pp X^\mu  +...) |0\rangle
\ee

{\bf Action of $L_{\pm n}$}

       Using (\ref{Vir}) we get
\[
exp ~~ [\la _0 ( {k_0^2\over 2} + ik_1 \tY_1 + 2ik_2 \tY _2) + 
\la _{-1}(k_1.k_0
+2ik_2.\tY_1)
\]
\[ + \la _{-2}(2k_2.k_0 + {1\over 2} k_1.k_1) + \la _1 ( ik_1.\tY_2
+ik_0.\tY_1 )
\]
\be   \label{4}
 + \la _2 (-{1\over 2} \tY_1 .\tY_1 + ik_0 \tY_2 )]
e^{ik_0 .\tY + ik_1 . \tY _1 + ik_2 .\tY _2}|0\rangle
\ee

Since the vertex operators of interest have two $k_1$'s or one $k_2$, the action of $L_{+n}$
on them will give a term with two  $k_1$'s or one $k_2$. $L_{+2}$ will give a term of level zero
and multiplied by $\la _2$ and 
and $L_{+1}$ will give a term of level one and multiplied by $\la _1$.
To get gauge transformations $L_{-1} , L_{-2}$ 
we need to extract the level two terms that have $\la _{-1}$ and $\la _{-2}$ respectively.
We can write down terms quite easily:

\begin{enumerate}
\item ${\bf \la _0 L_0 :}$ 
\be
\la _0 ~[ {k_0^2 \over 2}~
 (-{1\over 2} \kim \kin \p X^\mu \p X^\nu + i \ktm \pp X^\mu )
 -\kim \kin \p X ^\mu \p X^\nu + 2i \ktm  \pp X ^\mu ]~e^{ik_0.X}|0\rangle
\ee

\item ${\bf \la _{-1}L_1 :}$
\be
\la _{-1} [ k_1.k_0 ~ i\kim \p X ^\mu + 2i \ktm \p X ^\mu ]~e^{ik_0.X}|0\rangle
\ee
 
\item ${\bf \la _{-2} L_2 :}$

\be  \label{7}
\la _{-2} [ 2k_2.k_0 + {1\over 2} k_1.k_1 ]~e^{ik_0.X}|0\rangle
\ee

\item ${\bf \la _{1} L_{-1} :}$

\be
\la _1 [ i\kim \pp X ^\mu - \kim \kon \p X^\mu \p X^\nu ]~e^{ik_0.X}|0\rangle
\ee
(It is easy to see that the above is just 
$\la _1 L_{-1}i\kim \p X^\mu ~e^{ik_0.X}|0\rangle $)

 \item ${\bf \la _{2} L_{-2} :}$
\be  
\la _2 [ -{1\over 2} \p X . \p X + i \kom \pp X ^\mu ] ~e^{ik_0.X}|0\rangle
\ee
(This is just $ \la _2 L_{-2} e^{ik_0.X}|0\rangle$)
\end{enumerate}

The $L_0 =1$ equation gives the mass shell condition and the requirements
$L_1, L_2 V|0\rangle =0$   give additional physical state constraints. 

Since $L_n |0\rangle=0, n\geq -1$, the constraints given above
are equivalent to $[L_n, V]=0, n\geq -1$. For the gauge transformations 
$L_{-2} |0\rangle \neq 0$. So $L_{-n}V|0\rangle =   [L_{-n},V]|0\rangle + V l_{-n}|0\rangle$ and differs from the commutator.
In LV formalism one does not include the second term viz.  action on the vacuum. This has
to be accounted for by field redefinitions.

{\bf Liouville Mode}

One can obtain the physical state constraints, which are the action of $L_{+n}$, also by looking at the
Liouville mode dependence. The Liouville mode,, $\rho $ , is related to 
$\la _n$ at linear order by
\be
{d\la \over dt} = \rho
\ee
where $\la (t) = \sum _n \la _n t^{1-n}$ and $\rho (t) = \rho (0)+ 
t \p \rho (0) + {t^2\over 2}\pp \rho (0)+...$. Thus we get
\be  \label{11}
 \la _0 = \rho ,~~ \la _{-1} = {1\over 2} \p \rho ,~~~\la _{-2}={1\over 3!}
\pp \rho ,~~ \la _{-3}= {1\over 4!}\ppp \rho
\ee

This way of looking at the constraints is useful for purposes of
comparison with the LV formalism.

Thus
\[
e^{ik_0 .\tY + ik_1 . \tY _1 + ik_2 .\tY _2}=
:e^{ik_0 .\tY + ik_1 . \tY _1 + ik_2 .\tY _2}: 
\]
\[
e^{{1\over 2}[k_0^2 \langle X X\rangle + 2 k_1.k_0 \langle X \p X \rangle
+ k_1.k_1 \langle \p X \p X \rangle + 2 k_2.k_0 \langle X\pp X \rangle] }
\]
\be   \label{12}
= :e^{ik_0 .\tY + ik_1 . \tY _1 + ik_2 .\tY _2}: 
e^{{1\over 2}[k_0^2 \rho + 2 k_1.k_0 {1\over 2}\p \rho 
+ k_1.k_1 {1\over 6}\pp \rho + 2 k_2.k_0  {1\over 3} \pp \rho] }
\ee
The Liouville mode dependence is obtained using (\ref{11}), (\ref{4}).
This implies $\langle X X\rangle = \rho ,~~ 
\langle X \p X \rangle = {1\over 2}\rho 
,~~ \langle X \pp X \rangle = {1\over 3}\pp \rho ,~~ \langle \p X \p X \rangle =
 {1\over 6}\pp \rho $. These can be derived by other methods also
 \cite{dhoker}.

 In addition to the anomalous dependences, the Liouville mode
also enters at the classical level. This can be obtained by writing covariant derivatives. The vertex operators on the boundary
involve covariant derivatives $\nabla _x $ where $x$ is the coordinate
 along the boundary of the world sheet.  
 The vertex operators on the boundary should be
  : $\int dx   V$ where $V$ is a  one dimensional vector vertex operator
or $\int dx \sqrt g S$ where $S$ is one dimensional scalar.
Note that $g_{xx} = g$ (in one dimension) and $g^{xx} = {1\over g}$
The simplest vertex operator is thus $\nabla _x X = \p _x X$
(since $X$ is a scalar). Further $\nabla ^x X = g^{xx} \nabla _x X$
and using $\nabla _x T^x = {1\over \sqrt g} \p _x ( \sqrt g T^x)$ we get
$\nabla _x \nabla ^x X = {1\over \sqrt g}\p _x \sqrt g g^{xx} \p _x X$
$= {1\over \sqrt g}\p _x {1\over \sqrt g} \p _x X$
is a scalar. Thus $\sqrt g \nabla _x \nabla ^x X = \p _x
 {1\over \sqrt g}  \p _x X$ is the vertex operator with two derivatives.
One can similarly show that 
$\p _x {1\over \sqrt g} \p _x {1\over \sqrt g}\p _x X$
 is the vertex operator with three derivatives. This pattern continues.

The metric on the boundary induced by the metric on the bulk is:
\be
g_{xx} = 2 {\p z \over \p x}{\p \bar z \over \p x}g_{z\bar z}= g_{z\bar z}
\ee

Thus in  conformal coordinates since $g_{z\bar z}= e^{-2\rho}$, we have
${1\over \sqrt g} = e^\rho$. Thus
\be
\int dx~\p X , ~~\int  dx~e^{\rho}(\pp X + \p \rho \p X ), ~~
\int  dx~e^{2\rho}(\ppp X + 3 \pp X \p \rho
+ \p X \pp \rho ),...
\ee
Or if we remove $\int dx\sqrt g$  we get
\be
e^{\rho}\p X , ~~e^{2\rho}(\pp X + \p \rho \p X ), ~~
e^{3\rho}(\ppp X + 3 \pp X \p \rho+ \p X \pp \rho ),...
\ee 
for the vertex operators. The power of $e^\rho$ now counts the dimension
of the unintegrated vertex operators.
 Inserting this into
(\ref{12}) we get:
\be
= :e^{ik_0 . X + ik_1 .e^\rho \p X + ik_2 .e^{2\rho} (\pp X + \p \rho \p X )}: 
e^{{1\over 2}[k_0^2 \rho + 2 k_1.k_0 {1\over 2}\p \rho 
+ k_1.k_1 {1\over 6}\pp \rho + 2 k_2.k_0  {1\over 3} \pp \rho] }
\ee

This expression gives the complete $\rho$ dependence to linear order. 
The coefficient of $\la _{-1} = {1\over 2} \p \rho $
is $2i\ktm \p X ^\mu + k_1.k_0 i \kim \p X ^\mu$
and that of $\la _{-2} = {1\over 6} \pp \rho$ is $({1\over 2} k_1.k_1
+ 2 k_2.k_0)$ as required.  

{\bf Space-time Fields}

We can define fields as usual \cite{BSLV,BSRev} by replacing $\kim \kin$ by $\Phi _{\mu \nu}$
and $\ktm$ by $A_\mu$. Thus the level two boundary action is 
\[
\int dx~ [-\hf \Phi_{\mu\nu} \p X^\mu \p X^\nu + i A_\mu \pp X^\mu]
\]

The gauge parameters are obtained by replacing
$\la _1 \kim$ by $\epsilon ^\mu$ and $\la _2$ by $\epsilon _2$. Then we have
the following:

{\bf Constraints:}
The mass shell constraint fixes $p^2 + 2=0$. In addition we have,

\begin{enumerate}

\item
\be   \label{17}
p_\nu \Phi ^{\nu \mu} + 2 A^\mu =0
\ee

\item
\be          \label{18}
\Phi ^\nu _\nu + 4 p_\nu A^\nu =0
\ee

\end{enumerate}

{\bf Gauge transformations:}

\[
\delta \Phi ^{\mu \nu} = \eta ^{\mu \nu} \epsilon _2 + 
p^{(\mu}\epsilon ^{\nu )}
\]
\be 
\delta A^\mu = p^\mu \epsilon _2 + \epsilon ^\mu
\ee

On mass shell  ($p^2+2=0$), if we set $\eps _2=0$ and $p^\mu \eps _\mu =0$ then the constraints are invariant under the gauge transformations. 
This symmetry (which corresponds to $L_{-1}$) allows us to gauge away all transverse $A_\mu$ that satisfy $p.A=0$. The constraint then says that the longitudinal part is equal to the trace of $\Phi$.  Furthermore there is an additional symmetry:
 the constraints are  invariant under the gauge transformations
with $\epsilon ^\mu = {3\over 2}p^\mu \eps _2$, along with 
the mass shell condition $p^2 +2=0$ and the critical dimension $D=26$. 
Both these  symmetries transformations correspond to adding
 zero norm states: states that are physical as well as pure gauge. The second one  allows us to remove the trace of $\Phi$ or equivalently the longitudinal component of $A_\mu$ as well. This second symmetry thus reduces the number of physical degrees
by one again. The net effect is to remove two polarizations, which is why the light cone gauge description with D-2 polarizations is possible.
It is easy to see that this second gauge transformation corresponds
to the state $(L_{-2} + {3\over 2} L_{-1}^2)e^{ik_0.X}|0\rangle $.
Finally, when all the dust settles, we are left with one symmetric, traceless, transverse 2-tensor, which are described by the light cone oscillators as described earlier in Section 4.

\subsubsection{Level 3}

{\bf Vertex Operators}

The vertex operators in the OC formalism at this level can be written down
as follows:
\[
e^{ik_0 .\tY + ik_1 . \tY _1 + ik_2 .\tY _2 + ik_3 \tY _3} |0\rangle =
\]
\be
( ...+ i{k_3^\mu\over 2!}\ppp X^\mu - \ktm \kin \pp X^\mu \p X^\nu -
i {\kim \kin k_1^\rho \over 3!}
 \p X^\mu \p X^\nu \p X^\rho +...)e^{ik_0.X}|0\rangle
\ee 

{\bf  Action of $L_{\pm n}$} 

Using the same equation (\ref{Vir}) one gets:
\[
exp ~~[\la _3 ( ik_0.\tY _3 - \tY_1 . \tY_2) + \la _2 ( ik_1 \tY_3 +
 ik_0 \tY_2 - {\tY_1.\tY_1\over 2} ) +
 \la _1 (i2k_2\tY_3 + ik_1\tY_2 +ik_0\tY_1)
\]
\[
+ \la _0( i3k_3\tY_3 + 2ik_2\tY_2 + ik_1\tY_1 + {k_0^2\over 2})
+ \la _{-1} (i3k_3\tY_2 + i2k_2\tY_1 + k_1.k_0)
\]
\be   \label{vir1}
+\la _{-2}(i3k_3\tY_1 + 2k_2.k_0 +{k_1.k_1\over 2})
+\la _{-3}( 3k_3.k_0 + 2k_2.k_1)]
e^{ik_0 .\tY + ik_1 . \tY _1 + ik_2 .\tY _2 + ik_3 \tY _3} |0\rangle 
\ee

One can extract as before the action of $L_{+n}$ on the vertex operators of dimension three
by extracting terms of dimension $3-n$ involving $k_3, k_2k_1$ and $k_1k_1k_1$ in \eqref{vir1}. Similarly 
gauge transformations corresponding to $L_{-n}$ are obtained by extracting the level three terms that have $\la_n$ in \eqref{vir1}.

\begin{enumerate}

\item ${\bf \la _0 L_0:}$

\be
\la _0 (3+ {k_0^2\over 2})[ik_{3\mu} \tY_3^\mu  - \ktm \kin \tY_2^\mu \tY_1^\nu
- i \kim \kin \kir \tY_1^\mu\tY_1^\nu \tY_1^\rho ]
\ee

\item ${\bf \la _{-1} L_1:}$

\be
\la _{-1}[(i3k_{3\mu} + i\ktm k_1.k_0 ) \tY_2^\mu -
 (2\ktm \kin + {\kim \kin \over 2}k_1.k_0)\tY _1^\mu \tY_1^\nu ]
\ee

\item ${\bf \la _{-2} L_2:}$

\be
\la _{-2} [i3k_{3\mu} \tY_1^\mu + 2k_2.k_0 i\kim \tY_1^\mu +
 {k_1.k_1\over 2}i\kim \tY_1^\mu]
\ee

\item ${\bf \la _{-3} L_3:}$

\be
\la _{-3} [3\ko.k_3 +2 \kt.\ki]
\ee

\item ${\bf \la _1 L_{-1}:}$

\be
\la _1 [ i2\ktm \tY_3^\mu - \kim \kin \tY_2^\mu \tY_1^\nu -
 \kom \ktn \tY_1^\mu \tY_2^\nu  +{i\over 2} \kom \kin \kir \tY_1^\mu \tY_1^\nu \tY_1^\rho]
\ee

\item ${\bf \la _2 L_{-2}:}$

\be
\la _2 [ i\kim \tY_3^\mu - \kom \kin \tY_2^\mu \tY_1^\nu -
i\kim \tY_1^\mu  {\tY_1.\tY_1 \over 2}]
\ee

\item ${\bf \la _3 L_{-3}:}$

\be
\la _3 [i\kom \tY_3^\mu - \tY_1 .\tY_2]
\ee

\end{enumerate}
{\bf Liouville Mode}

Exactly as in the level two case one can get the Liouville mode dependences
- both the classical and anomalous terms. 

\[
e^{ik_0 .\tY + ik_1 . \tY _1 + ik_2 .\tY _2+ ik_3 . \tY _3}=
:e^{ik_0 .\tY + ik_1 . \tY _1 + ik_2 .\tY _2 + i k_3. \tY_3 }: 
\]
\[
e^{{1\over 2}[k_0^2 \langle X X\rangle + 2 k_1.k_0 \langle X \p X \rangle
+ k_1.k_1 \langle \p X \p X \rangle + 2 k_2.k_0 \langle X\pp X \rangle
+ 2k_3.k_0 \langle {\ppp X\over 2!}  X\rangle +
 2k_2.k_1 \langle \pp X \p X \rangle ] }
\]
This is the anomalous dependence. Using covariant derivatives gives
the classical part also:
\[
= :e^{ik_0 . X + ik_1 .e^\rho \p X + ik_2 .e^{2\rho} (\pp X + \p \rho \p X )
+ i\hf k_3. e^{3\rho}(\ppp X + 3 \pp X \p \rho+ \p X \pp \rho ) }: 
\]
\be   \label{12}
e^{{1\over 2}[k_0^2 \rho + 2 k_1.k_0 {1\over 2}\p \rho 
+ k_1.k_1 {1\over 6}\pp \rho + 2 k_2.k_0  {1\over 3} \pp \rho
+ 2k_3.k_0 {\ppp \rho \over 8} + 2k_2.k_1 {\ppp \rho \over 12}] }
\ee

We have used $\langle \ppp X X\rangle = {\ppp \rho \over 4}$ and
$\langle \pp X \p X \rangle = {\ppp \rho \over 12}$. Using \eqref{11} giving the relation  
between $\la _n$ and $\p ^n \rho$ one can check that this is the same as
the results given above. This form is useful for comparison with the loop variable formalism
where analogous terms are present.

{\bf Space-time Fields}

We introduce space-time fields as before by replacing $\kim \kin \kir$
by $\Phi ^{\mu \nu \rho}$, $ \ktm \kin$ by $B^{\mu \nu} + C^{\mu \nu}$
where $B$ is symmetric and $C$ is antisymmetric, and $k_3^\mu$ by $A^\mu$.
Thus the boundary action is 
\[
\int dx~[-{i\over 3!} \Phi_{\mu\nu\rho} \tY_1^\mu \tY_1^\nu \tY_1^\rho - (B_{\mu \nu} + C_{\mu \nu}) \tY_2^\mu  \tY_1^\nu + i A_3  \tY_3^\mu]e^{i\ko.\tY}
\]

For the gauge parameters we let $\la _3$ be $\epsilon _3$, $\la _2 \kim$
be $\eps _{12}^\mu$, $\la _1 \ktm$ be $\eps _{21}^\mu$ and $\la _1 \kim \kin$
be $\eps _{111}^{\mu \nu}$. We then have:

{\bf Constraints:}

The mass shell constraint $L_0=1$ becomes $p^2+4=0$. In addition,

\begin{enumerate}

\item
{\bf $L_1$}
\be
 p^\nu (B_{\mu \nu} + C_{\mu \nu}) + {3}A_\mu =0
\ee
\item
{\bf $L_1$}
\be
{p^\rho \Phi _{\rho\mu \nu}\over 4} + B_{\mu \nu} =0
\ee

\item
{\bf $L_3$}
\be
{B^\nu _{~\nu} \over 12} + {p^\nu A_\nu \over 8}=0
\ee

\item
{\bf $L_2$}
\be
{\Phi _{~\nu \mu}^\nu \over 12} + {A_\mu \over 2} + {p^\nu (B_{\nu \mu}+C_{\nu\mu})\over 3}
=0 
\ee

\end{enumerate}

{\bf Gauge Transformations:}

\[
\delta \Phi ^{\mu \nu \rho} = \eps _{12}^{(\mu }\eta ^{\nu \rho )} + 
p^{(\mu}\eps _{111}^{\nu \rho )}
\]
\[\delta (B^{\mu \nu}+C^{\mu \nu}) = p^\mu \eps _{12}^\nu
+ p^\nu \eps_{21}^\mu + \eps _{111}^{\mu \nu} + \eps _3 \eta ^{\mu \nu}
\]
\be
\delta A^\mu =p^\mu \eps _3  + \eps _{12}^\mu + 2\eps _{21}^\mu
\ee
(Symmetrization indicated in the first line involves
adding  two other orderings giving three permutations for each term.) 

 {\bf Invariance of Constraints under Gauge Transformations:}

One can show that the constraints are invariant under the gauge transformation under some conditions.
Invariance of constraint 1 imposes the following condition on the gauge parameters using:
\br  \label{I}
p.\eps_{12}+ 4 \eps _3 &=&0\nonumber \\
p^\nu \eps_{111\mu\nu} + 3\eps _{12\mu}+ (p^2+6)\eps_{21\mu}&=&0
\er
Invariance of constraint 2 requires:
\br	\label{II}
p^2+4 &=&0\nonumber \\
 p^\nu \eps_{111\mu\nu}+3 \eps_{12\mu} + 2\eps_{21\mu} &=&0\nonumber\\
p.\eps_{12}+ 4 \eps _3 &=&0
\er

Clearly for $p^2=-4$ the two sets \eqref{I} and \eqref{II} are equivalent.

Invariance of constraint 3 requires (using the above conditions) the following additional condition between gauge parameters.
\be
{2D+ 3p^2-20 \over 24} \eps_3 + {8(p.\eps_{21})\over 24} + {\eps^\nu_{~\nu}\over 24}=0
\ee
Finally using the above conditions and setting $D=26$ and $p^2+4=0$ one finds that the fourth constraint is satisfied.

Using $\eps_{111\mu\nu}$ one can gauge away $B_{\mu\nu}$, and using one of the vector parameters $\eps_{12\mu}$ one can gauge away
$A_\mu$. (The second vector gauge parameter is then fixed by the conditions \eqref{I}.) Then constraint 1 says that $C_{\mu\nu}$ is transverse and constraint 2 says that $\Phi_{\mu\nu\rho}$ is transverse. Using these we also see from constraint 4 that $\Phi$ is traceless.
Thus we have a traceless and transverse symmetric two tensor and an antisymmetric two tensor which is the correct count for level 3 open strings. 

\subsection{Loop Variable Formalism}

We now consider the system of constraints and gauge transformations in the loop variable approach. The generalized Liouville mode ($\SI$)
dependence is given in \eqref{LV} in Section 3.2. 

We consider the level two and three operators in turn.
\subsubsection{Level 2}

{\bf Vertex Operators}

\[
e^{{(k_0^2+q_0^2)\over 2}\SI}[i\ktm \ytm + i q_2 \theta _2 -
 {\kim \kin \over 2} \yim \yin -
{q_1 q_1 \over 2} \theta _1 \theta _1 -  \kim q _1 \yim \theta _1  
\]
\[
+i(\kim \yim + q_1 \theta _1 )(k_1.k_0 + q_1 q_0){1\over 2} {\p \SI \over \p x_1}
+(k_2.k_0 + q_2q_0){1\over 2}{\p \SI \over \p x_2} +
\]
\[ -i {\kim \kin \over 2} q_1 \yim \yin \theta _1 - i {q_1q_1\over 2} \kim \theta _1 \theta _1 \yim \]
\be
 {(k_1.k_1 + q_1q_1)\over 2}{1\over 2}
({\pp \SI \over \p x_1^2} - {\p \SI\over  \p x_2})]e^{ik_0.Y}  
\ee

We have  written the dimensionally reduced version.
 Weyl Invariance is independence of $\SI$. The coefficients of
$\SI$ and its derivatives have to be set to zero. There are the constraints.
It will be seen that field redefinitions will make these equivalent to the
constraints of the OC formalism (\ref{17}) and (\ref{18}). This implies that
the classical Liouville mode dependence is included here indirectly
through terms involving $q_n$.

{\bf Space-time Fields, Gauge Transformations and Constraints}

\begin{itemize}
\item{\bf Space-time Fields:}

The fields are obtained by setting $\kim \kin \approx S_{11}^{\mu \nu}$,
 $\ktm \approx S_2^\mu$,
$\kim q_1 q_0 \approx S_{11}^\mu$, $q_1q_1 \approx S_{11}$, 
and $q_2q_0\approx S_2$.
The gauge parameters are $\la _2 \approx  \Lambda _2$, 
$\la _1 \kim \approx \Lambda _{11}^\mu$,
$\la _1 q_1 q_0 \approx \Lambda _{11}$. 

\item{\bf Gauge Transformations:}

\[
\delta S_{11}^{\mu \nu}  = p^{(\mu}\Lambda _{11}^{\nu )},~~ \delta S_2^\mu =
 \Lambda _{11}^\mu + p^\mu \Lambda _2,~~
\delta S_{11}^\mu = \Lambda _{11}^\mu + p^\mu \Lambda _{11},
\]
\be 
\delta S_2 = \Lambda _2 q_0^2 + \Lambda_{11} ,~~ \delta S_{11} =2\Lambda _{11} 
\ee

Now one can make the following identifications:
 $q_1 q_1 \approx q_2 q_0$, $\kim q_1 \approx \ktm q_0$,
 and $\la _1 q_1 \approx \la _2 q_0$.
This gives: $S_{11}^\mu = 2 S_2^\mu$, $S_{11} \approx S_2$ and 
$\Lambda _{11} \approx 2\Lambda _2$ and the gauge transformations are consistent 
with these identifications.

\item{\bf Constraints:}

 The coefficient of $\SI$ gives the usual mass shell condition $p^2+q_0^2 =0$.
Note that the $(mass)^2$ equals  the dimension of the operator, but the $\SI$
dependence representing this (and also all other $\SI$ dependences) 
comes from an anomaly rather than from the 
classical dependence as in the OC formalism.
\begin{enumerate}
\item Coefficient of ${\pp \SI \over \p x_1 ^2}$
\be  \label{c1}
k_1 . k_1 + q_1 q_1 = 0 ~~\Rightarrow   S_{11 \mu}^\mu + S_2 =0
\ee

\item  Coefficient of ${\p \SI \over \p x_2}$
\be    \label{ c_{2}}
k_2.k_0 + q_2 q_0 =0 ~~ \Rightarrow p_\mu S_2^\mu + S_2 =0
\ee

\item  Coefficient of ${\p \SI \over \p x_1 } \yim$

\be  \label{c3}
(k_1.k_0 + q_1 q_0)\kim =0 ~~\Rightarrow p_\nu S_{11}^{\mu \nu} + 2 S_2^\mu =0
\ee

\end{enumerate}

The constraint proportional to $ \theta _1$ is seen to be
 a linear combination of the above.
 The equations of motion
are obtained by setting the variational derivative of 
$\SI$ equal to zero, and are gauge invariant.

\end{itemize}
\subsubsection{Level 3}

{\bf Vertex Operator}
The complete Level 3 gauge covariantized vertex operator is:

\[
e^{{(k_0^2+q_0^2)\over 2}\SI}[ik_3^\mu Y_3^\mu + i q_3 \theta _3 -
 \ktm \kin  \ytm \yin -
\]
\[
q_1 q_2 \theta _1 \theta _2 -  \kim q_2  \yim \theta _2 -
 \ktm q_1 \ytm \theta _1 -i {\kim \kin k_1^{\rho}\over 3!} 
\yim \yin Y _1^\rho -i{(q_1)^3\over 3!}(\theta _1)^3
\]
\[
+i(\ktm \ytm + q_2 \theta _2 )(k_1.k_0 + q_1 q_0){1\over 2} {\p \SI \over \p x_1}
\]
\[
+i(\kim \yim + q_1 \theta _1 )[(k_2.k_0 + q_2 q_0) {1\over 2} {\p \SI \over \p x_2}
 + {1\over 2}(k_1.k_1 + q_1 q_1){1\over 2}({\pp \SI \over \p x_1 \p x_1} -
 {\p \SI\over  \p x_2})]
\]
\be 	\label{VO3}
+(k_3.k_0 + q_3q_0){1\over 2}{\p \SI \over \p x_3} + (k_2.k_1 + q_2q_1)
{1\over 2}({\pp \SI \over \p x_1 \p x_2} - {\p \SI\over  \p x_3})]e^{ik_0.Y}  
\ee

{\bf Space-time Fields, Gauge Transformations and Constraints}
\begin{itemize}
\item{\bf Space-time Fields}

This was worked out in Section 4 \eqref{spin3truncated} and we reproduce it here for convenience:
\br   \label{spin3truncated}
 \lan \kim \kin \kir \ran &=& S_{111\mu\nu\rho}\nonumber \\
 \lan k_{2[\mu}k_{1\nu]} \ran &= &A_{21[\mu\nu]} \nonumber \\
 \lan k_{2(\mu} k_{1\nu)} = \kim \kin q_1 \ran &=& S_{21(\mu\nu)}\nonumber \\
 \lan k_{3\mu} q_0^2= \kim q_1^2 = \hf(\kim q_2 + \ktm q_1)q_0 \ran &=& S_{3\mu}q_0^2\nonumber \\
 \lan \kim q_2 \ran &=& S_{12\mu}\nonumber \\
 \lan q_3 q_0^2=q_2 q_1 q_0 = q_1^3 \ran &=& S_3 q_0^2
 \er
\item{\bf Gauge Parameters}
 \br
 \lan \li q_1 q_1 =\hf (\lt q_1 +\li q_2)q_0  = \lambda _3 q_0^2\ran &=& \Lambda _3 q_0^2\nonumber \\
 \lan \hf(\li q_2 - \lt q_1) \ran &=& \Lambda _Aq_0\nonumber \\
 \lan \li q_1 \kim = \hf(\lt \kim +\li \ktm ) q_0 \ran &=& \hf(\Lambda _{12\mu} + \Lambda _{21\mu})q_0 = q_0\Lambda _{S\mu}\nonumber \\
 \lan \hf(\lt \kim -\li \ktm )\ran &=& \hf(\Lambda _{21\mu}-\Lambda_{12\mu})=\Lambda _{A\mu}\nonumber \\
 \lan \li \kim \kin \ran &=& \Lambda _{111\mu\nu}
 \er

Note that there is also a tracelessness condition:
\be
\Lambda_{111~\mu}^{~~~\mu}+\Lambda_3 \qo^2=0
\ee

\item{\bf Gauge Transformations}
\br
 \delta S_3 &=& 3\Lambda _3\qo\nonumber \\
 \delta S_{3\mu} &=& 2 \Lambda _{S\mu} + \kom \Lambda _3\nonumber \\
 \delta S_{12\mu}&=&\qo[2 \Lambda_{S\mu}+\Lambda_{A\mu} +(\Lambda_A+\Lambda_3)\kom]\nonumber \\
 \delta (S_{12\mu}-S_{3\mu} q_0  ) \equiv \delta S_{A\mu}\qo&=& \Lambda _{A\mu}q_0 + k_{0\mu} \Lambda _A \qo\nonumber \\
 \delta S_{\mu\nu} &=& \Lambda _{111\mu\nu} + k_{0(\mu } \Lambda _{S\nu)}\nonumber \\
 \delta A_{\mu\nu}&=& k_{0[\mu}\Lambda _{A\nu]}\nonumber \\
 \delta S_{111\mu\nu\rho} &=& k_{0(\mu }\Lambda_{111\nu\rho)}
 \er

\item{\bf Constraints}

Level 3 terms involving  $\SI$ derivatives in \eqref{VO3} give the constraints. In writing them below, the Q-rules \eqref{Q3} have been used.

\begin{enumerate}
\item
\be    
k_3.k_0 + q_3q_0 =0 ~~\Rightarrow p^\nu S_{3\nu} + S_3 \qo=0 
\ee
\item
\be
k_2.k_1 + q_2 q_1 =0 ~~\Rightarrow S_{21\mu}^\mu + S_{3}\qo=0 
\ee
\item
\be
(k_1.k_1 + q_1 q_1 )\kim =0~~\Rightarrow S_{111\mu \nu}^{\nu } + S_{3\mu} \qo^2=0 
\ee
\item
\[
(k_2.k_0 + q_2 q_0)\kim =0 ~~ \Rightarrow p^\nu (S_{21\nu \mu}+ A_{21\nu\mu}) +
 S_{12\mu}\qo =0 
\]
\be
\Rightarrow p^\nu (S_{21\nu \mu}+ A_{21\nu\mu}) +
 (S_{3\mu}+S_{A\mu})\qo =0 
\ee
\item
\be
(k_1.k_0 + q_1q_0)\kim \kin =0~
~\Rightarrow p^\rho S_{111\mu \nu \rho} + \qo^2 S_{21\mu \nu} =0
\ee
\item
\[ 
(k_1.k_0 +q_1.q_0)\ktm =0~~\Rightarrow p^\nu (S_{21\mu \nu}+ A_{21\mu \nu}) + 2S_{3\mu} \qo^2 - S_{12\mu}\qo=0
\]
\be
\Rightarrow p^\nu (S_{21\nu \mu}- A_{21\nu \mu}) + (S_{3\mu} - S_{A\mu})\qo^2=0
\ee
\item
\[
(k_2.k_0 + q_2 q_0) q_1 q_0 =0 ~~\Rightarrow  p^\nu (2S_{3\nu}\qo - S_{12\nu})+ S_{3}\qo^2=0
\]
\be
\Rightarrow  p^\nu (S_{3\nu} - S_{A\nu})\qo+ S_{3}\qo^2=0
\ee
\end{enumerate} 
Note that the last constraint comes from ${\p \SI \over \p x_2}\theta _1$.  Combining this with the first constraint gives $p^\nu S_{A\nu}=0$.
This is not a new constraint: Combining constraints 4 and 6 we get using the antisymmetry of $A_{21\mu\nu}$:
\[
p^\nu A_{21\nu \mu}+ S_{A\mu}=0 ~\Rightarrow p^\nu p^\mu A_{21\nu \mu}+ p^\mu S_{A\mu}~\Rightarrow p^\mu S_{A\mu}=0
\]

Constraint 6 says that only one of the vectors is independent. Constraint 1 says the scalar is not independent. If we remove one vector and one scalar we have the same number of fields as in the OC formalism.

\item {\bf Invariance of Constraints Under Gauge Transformations:}
It can be checked that the constraints are invariant under the gauge transformations provided some transversality constraints are satisfied by the gauge parameter. However there is no constraint on  $D$. The mass shell constraint becomes $p^2+\qo^2=0$ with arbitrary $\qo$.  One finds the following conditions.
 \br
 p.\Lambda_S + \Lambda_3 \qo^2&=&0\nonumber \\
 \ko ^\nu \Lambda_{111\mu \nu}+ \Lambda _{S\mu}\qo^2&=&0\nonumber\\
 \Lambda _{111~\nu}^{~~~\nu}+ \Lambda_3\qo^2 &=&0\nonumber \\
 p^\nu \Lambda_{A\nu}+ \Lambda_A \qo^2&=&0
 \er

The third condition on $\Lambda _{111\mu\nu}$ is nothing but the D+1 dimensional tracelessness constraint on the gauge parameter which is always there - even for gauge invariance of the equations. The other three can be understood as generalized (i.e in D+1 dimensions)  transversality. Thus the conditions below are equivalent to the ones above if one uses the Q-rules. 
 \[
p^\nu \la _1 \kim \kin =0~~;~~~p^\mu \la _1 \ktm =0~~~; ~~p^\mu \la _2 \kim =0~;~~~\la _1 \ki \circ \ki=0~;~\mu:0~-~D
\]
Here as before $\circ$ denotes a D+1 dimensional dot product.

One can gauge away $S_{21\mu\nu}$ and this automatically ensures the transversality of $S_{111\mu\nu\rho}$ by the constraint 5. Gauging away $S_3\mu$ also gets rid of the trace of $S_{111\mu\nu\rho}$ by constraint 3. It also gets rid of $S_3$ by constraint 1. Thus we have a transverse traceless  3 tensor as required. The antisymmetric 2 tensor is transverse once $S_{A\mu}$ is gauge away. This completes the counting of degrees of freedom.
\end{itemize}

\subsection{Mapping from OC Formalism to LV Formalism}

\subsubsection{Level 2}

{\bf Mapping of fields}

The mapping is given by:

\[
\Phi ^{\mu \nu } = S_{11}^{\mu \nu} + (\eta ^{\mu \nu} + 
{5\over 2}p^\mu p^\nu)S_2
\]
\be   \label{53}
A^\mu = S_2 ^\mu + 2 p^\mu S_2
\ee

{\bf Mapping Gauge Transformations}

If one makes a LV gauge transformation with parameters $\Lambda _{11}^\mu$
 and $\Lambda _2$ one obtains:

\[
\delta \Phi ^{\mu \nu} = p^{(\mu}\Lambda _{11}^{\nu )} + (\eta ^{\mu \nu} +
 {5\over 2} p^\mu p^\nu )4\Lambda _2
\]
\be         \label{55}
\delta A^\mu = \Lambda _{11}^\mu + 9 ~p^\mu \Lambda _2
\ee

The relative values of the different terms in (\ref{53}),
and therefore in the gauge
transformation (\ref{55}), are fixed by requiring 
that the gauge transformation be 
generated by some combination of  $L_{-n}$'s. 
One can check that (\ref{55}) corresponds to a gauge transformation by
 $4\Lambda _2(L_{-2} + {5\over 4} L_{-1}^2) + \Lambda _{11}^\mu L_{-1}$. 

{\bf Mapping Constraints}

  Now consider the constraint
(\ref{18}). We see that 
\be   \label{45}
p_\nu \Phi ^{\nu \mu} + 2A^\mu = p_\nu S_{11}^{\mu \nu} + 2 S_2^\mu +
 (5-{5\over 2}q_0^2)p^\mu S_2
\ee
Only for $q_0^2=2$ does it become the LV constraint
 $p_\nu S_{11}^{\mu \nu} + 2 S_2^\mu$.
Furthermore

\be   \label{46} 
4p.A + \Phi ^\mu _\mu = 4p.S_2 + S_{11\mu}^\mu  + [(D-{5\over 2}q_0^2) -8q_0^2]S_2
\ee
This should equal
\be      \label{47}
4(p.S_2 + S_2) + S_{11\mu}^\mu + S_2 
\ee
This fixes $D=26$ (using $q_0^2=2$). Thus we see that the while the LV
 equations are gauge invariant
in any dimension, when we require equivalence with OC formalism the critical 
dimension is picked out. 
{\bf Equivalence of OC and LV Formalisms}

Now we can see that the LV formalism is equivalent to the OC formalism: 
Start with a vertex operator in the OC formalism with fields that obey
 (\ref{45}, \ref{46}). 
This implies
that the corresponding LV vertex operator obeys the same constraint 
(\ref{45}). Similarly
(\ref{46}) implies (\ref{47}). This is the sum of two constraints 
of the LV formalism. 
Since the LV formalism is gauge invariant one can choose a gauge
(using invariance under $\Lambda _2$ transformations), where
$S_2$ to equal $-p.S_2$. This implies (by the constraint) that $S_{11\mu}^\mu
+ S_2 =0$. 
 Thus if the fields obey the physical state constraints of the OC formalism,
 then using the gauge invariance, we see that the LV constraints are also
satisfied. In the reverse direction it is easier because we just have to take
a linear combination of two LV constraints (\ref{c1}-\ref{c3})
  to get an OC constraint.  

We can go further in analyzing the constraints. After obtaining $p.S_2+ S_2=0$
using a $\Lambda _2$ transformations, there is a further invariance involving
both $\Lambda _{11}^\mu $ and $\Lambda _2$ with 
$p.\Lambda _{11} + q_0^2\Lambda _2 =0$. This transformation preserves all the
 constraints. (We also have to use the mass shell condition $p^2+q_0^2=0$.) 
Using this invariance we can set $S_2=0$ while preserving 
$p.S_2 +S_2=0=S_{11\mu}^\mu + S_2$. This then implies that 
$p.S_2=S_{11\mu}^\mu =0$. A very similar analysis done on the OC side
using the constraint $4p.A + \Phi ^\mu _\mu =0$ and the gauge transformations
with $\eps ^\mu + {3\over 2 }p^\mu \eps$ that preserves the constraint
(provided $D=26,~~q_0^2=2$):
we can use it to set $p.A$ to zero and so $\Phi ^\mu _\mu=0$.  
Thus on both sides we have a transverse vector and a traceless tensor
obeying $p_\nu \Phi ^{\nu \mu} + 2 A^\mu =0$.

 There are also
terms involving $\theta _n$ on the LV side. But if we focus on the equations
of motion involving only vertex operators with $\yn$, then the two systems
are identical. Since the physical states of the string are conjugate
to these vertex operators (that have only $\yn$), this is all we need to 
describe the physics of string theory. 	

Finally we can use transverse gauge transformations involving $\eps ^\mu$
(i.e with $p.\eps =0$) to gauge away the transverse vector $A^\mu$ 
(and the same thing can be done on the LV side to gauge away $S_2^\mu$). This
leaves a tensor $\Phi ^{\mu \nu}$ which is transverse -
 $p_\nu \Phi ^{\nu \mu} =0$ - and traceless. This is the right number of
 degrees of freedom for the first massive state of the bosonic open string. 

This concludes the demonstration of the equivalence of the OC formalism and LV
formalism for Level 2 at the free level. 

\subsubsection{Level 3}

We have seen in earlier subsections that in both formalisms the physical degrees of freedom are the same as that of the open string at level 3. It is thus clear that there exists a map from one set of fields to the other. We have also seen that the OC formalism requires D=26 and $p^2=-4$, whereas the LV formalism is valid in any dimension and any $\qo^2$. Thus one expects that the map from one to the other, with both constraints and gauge transformations being mapped respectively, will work only in $D=26$ and for $\qo^2=4$. We saw this explicitly in the last section for level 2. Level 3 is more complicated. We work out some details - just enough to see the critical dimension emerging.

We start by defining a fairly general map between the fields of the OC and LV formalisms:
\br
\Phi_{\mu\nu\rho} &=& f_1 S_{111\mu\nu\rho} + f_2 S_{3(\mu}\eta_{\nu\rho)} +f_3 S_{A(\mu}\eta_{\nu\rho)}+f_4 p_{(\mu}S_{\nu\rho)}\nonumber \\
B_{\mu\nu}& =& b_1 S_{\mu\nu}+b_2 p_{(\mu}S_{3\nu)} +b_3 p_{(\mu}S_{A\nu)} +b_4 S_3 \eta_{\mu\nu}\nonumber \\
C_{\mu\nu}  &=&c_1 A_{\mu\nu}+ c_2 p_{[\mu}S_{3\nu]} + c_3 p_{[\mu}S_{A\nu]} \nonumber \\
A_\mu&=& a_1 S_{3\mu} + a_2 S_{A\mu} + a_3 p_\mu S_3
\er
This is not the most general map - derivatives of the scalar field can be added in some places. We now write down the constraints in the OC formalism and require that they be satisfied when the LV constraints are used. This gives us linear equations in the coefficients. These equations also involve $D$ and $\qo$. This is not enough to fix $D$ or $\qo$. We then require that the gauge transformations can be consistently mapped from one to the other (i.e. that a consistent map between the gauge parameters should exist). This gives several more equations. We use the tensor and vector gauge parameters in the analysis. The scalar parameters are not included in this analysis. The resultant equations and solution are given in Appendix D \eqref{append}. We see that $D$ and $\qo$ get fixed exactly as in the level 2 case.
We have not attempted to fix all the coefficients - it is quite tedious and not very illuminating for our purposes.

Thus in this section we have shown that at Level 2 and 3, the LV formalism and OC formalism describe the same degrees of freedom once we analyze the constraints and gauge transformations. For level 2 we also gave an explicit map of the fields. The map works only in the critical dimension and with the correct string spectrum. We expect this to continue to be true at higher levels also and also for closed strings.

Once the theory is gauge fixed and constraints imposed, the vertex operators are identical in the two theories. Thus the S-matrix is also the same. The theories describe the same physics. Thus we conclude that the gauge invariant ERG equations obtained here must describe
string theory in the critical dimension. In other dimensions the theory described by the LV formalism is gauge invariant and hence plausibly consistent classically, but may or may not be equivalent to a non critical string theory. Quantum mechanical consistency is an open question.

Note that the loop variable formalism gives results right away in a convenient form - as a massless theory dimensionally reduced. Neither the OC formalism nor the BRST formalism does this. The Q-rules have to be imposed to get the spectrum in the LV formalism to match that of strings. These rules are also consistent with dimensional reduction.  The higher dimensional origin of this theory is intriguing. It is tempting to speculate that this is related to M-theory.

\section{Conclusions and Open Questions} 
     The loop variable approach is an attempt to construct a formalism where all the gauge symmetries of string theory are present.The gauge symmetry associated with the massless graviton is general coordinate invariance. Making this manifest as done in standard General Relativity makes the formalism manifestly background independent. As a by product it gives us gauge invariant and generally covariant (interacting) equations of motion for massive fields in an arbitrary gravitational background. 
     
     The main tool used is the ``loop variable" - which has all the vertex operators of the string collected into one non local loop variable. The equations are the Exact Renormalization Group equations for the word sheet theory, which impose conformal invariance.  An infinite number of  ``proper time"  variables are present in this formalism which allows us to work in the space of all gauge transformations. This makes the equations gauge invariant. The main guiding principles are world sheet conformal invariance and space time gauge invariance. World sheet reparametrization invariance is not imposed. 
          
     The extra coordinate that has all the auxiliary fields necessary for a gauge invariant description is the moral equivalent of the ghost coordinate in the BRST formalism although the precise connection is not clear. The origin of the critical dimension and mass spectrum are not clear. These are free parameters and can be chosen to match string theory. It is possible that quantum consistency will force this on us. But at the classical level there do not seem to be any such constraints. 
     
     The Q-rules need to be extended to higher levels. The present method is very tedious and leads to a highly over determined set
of linear equations that miraculously seem to have a solution. This suggests that some underlying pattern lies unnoticed. Furthermore the fact
that the rules can be made consistent with dimensional reduction is also not {\it a priori} clear and again points to some underlying structure.

     Finally and perhaps most importantly the gauge transformations have a simple form of local (along the string) scale transformations of the generalized loop variable momenta. This was in fact the original motivation for this formalism. The possible geometrical space time  interpretation in terms of a {\em space time} renormalization group  (with a finite cutoff) as the symmetry group of string theory is described in \cite{BSLV}. The speculation described there is that string theory should be thought of as a method of regularizing a field theory by a Lorentz invariant cutoff and therefore should be equivalent to a dynamical space time lattice.

The gauge transformations for the open string have an Abelian form which means the interactions do not modify it. The interactions
are in terms of gauge invariant field strengths - generalizations of Maxwell field strengths. This is in contrast to BRST string field theory
\cite{Wi2}. For closed strings this continues to be the case for the massive fields. But for the massless field this field strength is not gauge invariant unless the gauge transformation is modified to a non Abelian form. This then becomes general coordinate transformations. 

One can speculate on the possible space time interpretation of the other gauge symmetries of the closed string.  For instance in Sec 7.2.1 and  \cite{BSERGclosed1} it was noted that the gauge parameters for the graviton and antisymmetric tensor seem to have the natural interpretation of the real and imaginary parts of a complex parameter. This leads naturally to a speculation that space time coordinates could be complex. Such a speculation has been made earlier also \cite{WTop,AW}

The massive modes have higher gauge symmetries. These are realized in a ``broken" form with Stuckelberg fields. However we should remember that a massive spin 2 field with a Stuckelberg realization is what is naturally described by the formalism. To make the graviton massless we had to extend the gauge transformations to a ``non Abelian" form. It is possible that extending the higher gauge symmetries to a non Abelian form will allow the higher spin modes to become massless as well. This would then describe a different theory - or a different phase of the same theory - perhaps a more symmetric form.

We hope to return to these questions soon.

\begin{appendices}
\section{Riemann Normal Coordinates and Tensors}
\label{appena}
\setcounter{equation}{0}
We review some facts about Riemann Normal Coordinates \cite{AGFM,Pet,Eis}.

\subsection{Defining the Coordinate System}

\begin{enumerate}

\item   In a manifold (with metric), first pick a coordinate system. Call it $x^i$ and let the metric be $g_{ij}$ (We also occasionally use $a$ or $\mu$ for the indices). Pick an origin O at a point labelled as $x_0$. Draw geodesics through O.
The geodesics are labelled by their unit tangent vectors at O, $\xi^i$ and parametrized by $s$ the proper distance along the geodesic.  The equation for the geodesic is 
\be	\label{GE}
{d^2x^a\over ds^2} + \Gamma ^a_{bc}\dot x^b\dot x^c =0
\ee
One can also write a Taylor expansion:
\be	\label{TE}
x^a(s) = x_0^a + s{dx^a(s)\over ds}|_{s=0} + {s^2\over 2!}{d^2x^a(s)\over ds^2}|_{s=0}+....
\ee
Clearly ${dx^a(s)\over ds}|_{s=0}=\xi^a$. Similarly
\[
{d^2x^a(s)\over ds^2}|_{s=0}=\Gamma ^a_{bc}(x_0)\dot x^b(0)\dot x^c(0)=\Gamma ^a_{bc}(x_0)\xi^b\xi^c
\]

By differentiating \eqref{GE} we get
\br
{d^3x^a(s)\over ds^3} &=& -{d\over ds}[\Gamma ^a_{bc}\dot x^b\dot x^c] = -(\p_d \Gamma ^a_{bc})\dot x^b\dot x^c\dot x^d - \Gamma ^a_{bc}
{d\over ds}[\dot x^b\dot x^c]\nonumber \\
&=&-\underbrace{[\p_d \Gamma ^a_{bc}-\Gamma ^a_{bi}\Gamma ^i_{dc}-\Gamma ^a_{ic}\Gamma ^i_{db}]}_{\Gamma ^a_{bcd}(s)}\dot x^b\dot x^c\dot x^d\nonumber \\
&=&-{1\over 3!}\Gamma ^a_{(bcd)}(s)\dot x^b\dot x^c\dot x^d\equiv \tilde \Gamma ^a_{bcd}(s)\dot x^b\dot x^c\dot x^d
\er
The geodesic equation \eqref{GE} has been used and we have symmetrized on the indices in the last step.

A similar calculation for ${d^4x^a(s)\over ds^4}$ gives
\br
{d^4x^a(s)\over ds^4} &=& -{d\over ds}[\tilde \Gamma ^a_{bcd}\dot x^b\dot x^c\dot x^d] = -(\p_e \tilde \Gamma ^a_{bcd})\dot x^b\dot x^c\dot x^d\dot x^e - \tilde \Gamma ^a_{bcd}
{d\over ds}[\dot x^b\dot x^c\dot x^d]\nonumber \\
&=&-\underbrace{[\p_e \tilde\Gamma ^a_{bcd}-\tilde \Gamma ^a_{bid}\Gamma ^i_{ec}-\tilde \Gamma ^a_{icd}\Gamma ^i_{eb}-\tilde\Gamma ^a_{bci}\Gamma ^i_{ed}]}_{\Gamma ^a_{bcde}(s)}\dot x^b\dot x^c\dot x^d\dot x^e\nonumber \\
&=&-{1\over 4!}\Gamma ^a_{(bcde)}(s)\dot x^b\dot x^c\dot x^d\dot x^e\equiv \tilde \Gamma ^a_{bcde}(s)\dot x^b\dot x^c\dot x^d\dot x^e
\er
Plug these into \eqref{TE} to get
\be   \label{TE1}
x^a(s) = x_0^a + s\xi ^a + {s^2\over 2!}\Gamma^a_{bc}(0)\xi^b\xi^c+  {s^3\over 3!}\tilde \Gamma^a_{bcd}(0)\xi^b\xi^c\xi^d+ {s^4\over 4!}\tilde \Gamma^a_{bcde}(0)\xi^b\xi^c\xi^d\xi^e+...
\ee

It can also be shown that
\be	\label{Gamma}
 \underbrace{\Gamma^a_{bcd}}_{\ydiagram{1} \otimes \ydiagram{2}}=   \underbrace{\tilde\Gamma^a_{bcd}}_{ \ydiagram{3}} 
+{1\over 3}\underbrace{(R^a_{~bdc}+R^a_{~dbc})}_{ \ydiagram{2,1}}
 \ee
 
 Since the Riemann tensor has some antisymmetric indices, it drops out of the Taylor expansion \eqref{TE1}.

\item Consider an arbitrary point,P,  labelled $x$ in the original coordinate system. We give it a new label: Consider the geodesic starting from the origin,O,  and going through P. Let the unit tangent vector be $\xi^i$. 
Let $y^i = s\xi^i$. Choose $y^i$ to label this point. The $y$ defines a coordinate system called Riemann Normal Coordinates (RNC). The relation between $x$ and $y$ is obtained by rewriting \eqref{TE1} in terms of $y$.
\be
x^a(y) = x_0^a + y^a + {1\over 2!}\Gamma^a_{bc}(0)y^b y^c+  {1\over 3!}\tilde \Gamma^a_{bcd}(0)y^by^cy^d+ {1\over 4!}\tilde \Gamma^a_{bcde}(0)y^by^cy^d y^e+...
\ee

\end{enumerate}

\subsection{Properties of $\Gamma$}

Let us put bars to denote quantities in the RNC. (For eg. $\bar \Gamma ^a_{bc}$). Then if we write \eqref{TE1} for geodesics
in the RNC $y$, then only the first term survives : $y^a = s \xi^a$. Thus all the $\tilde {\bar \Gamma}^a_{bc..} (0)\xi^b\xi^c...=0$ at the origin of the coordinate system. But at the origin $\xi^a$ can point in any direction. Therefore it must be true that {\em at the origin}:
\be   \label{const}
\bar \Gamma ^a_{bc}(0)=\tilde{\bar \Gamma}^a_{bcd}(0)= \tilde{\bar \Gamma}^a_{bcde}(0)=...=0
\ee

Furthermore, all along the geodesic, in the RNC, ${d^ny^a\over ds^n}=0$ - because the geodesics through the origin are straight lines. Thus from the geodesic equation \eqref{GE} and its derivatives we can conclude that {\em all along the geodesic} specified by $\xi^a$:
\be
\tilde{\bar \Gamma}^a_{bcde...}(y) \xi^b \xi^c\xi^d \xi^e...=0
\ee

That \eqref{const} is a complete specification of the freedom of coordinate transformation is obvious from counting parameters in the coordinate transformation $x'(x)$. If we write this as a Taylor series
\[
x'^a =  {\p x'^a\over \p x^b}|_{x=0} x^b + {1\over 2!}{\pp x'^a\over \p x^b\p x^c}|_{x=0}x^bx^c+{1\over 3!}{\ppp x'^a\over \p x^b\p x^c\p x^d}|_{x=0}x^bx^cx^d+...+
\]
At each order the coefficients of the Taylor expansion have a tensor structure with one upper index $a$, followed by completely symmetrized lower indices $b,c,d...$. This is exactly the index structure of the $\tilde{\Gamma}^a_{bcd...}$.  Thus their numbers are exactly the same in any dimension.

\subsection{Expansion of Tensors}

\subsubsection{Scalar}

Consider first the Taylor expansion of a scalar field about a point O labelled by $x_0$. Let $\Delta x^\mu = x^\mu -x_0^\mu$. Then
\be   \label{ScalarTaylor}
\phi (x) = \phi(x_0) + \Delta x^\mu \p_\mu \phi (x_0) + {1\over 2!} \Delta x^\mu \Delta x^\nu \p_\mu \p _\nu \phi (x_0)+...
\ee

Let us rewrite ordinary derivatives in terms of covariant derivatives:
\br
\p_\mu \phi &=& \nabla _\mu \phi \nonumber \\
\p_\mu \p_\nu \phi &=& \n _\mu (\n _\nu \phi) + \Gamma ^\rho_{\mu\nu} \n _\rho \phi
\er
Similarly after some algebra
\be   \label{3d}
\p_\rho \p_\mu \p_\nu \phi = \n _\rho\n_\mu \n_\nu \phi + \Gamma^\la_{(\rho\mu}\n_{|\la|}\n_{\nu)} \phi +[\tilde{\Gamma}^\la_{\mu\nu\rho}+\Gamma^\la_{(\mu |\sigma|}\Gamma^\s_{\rho\nu)}]\n_\la \phi
\ee
In a general coordinate system $\Delta x^\mu$ is not a tensor, hence this is not a covariant expansion in terms of tensors - as the explicit
presence of $\Gamma$'s shows.

The solution is well known: work with $y^\mu = s \xi ^\mu$. In this case the relation between $x$ and $y$ is as given above:
\be  \label{RNC}
x^a(y) = x_0^a + y^a + {1\over 2!}\Gamma^a_{bc}(0)y^b y^c+  {1\over 3!}\tilde \Gamma^a_{bcd}(0)y^by^cy^d+ {1\over 4!}\tilde \Gamma^a_{bcde}(0)y^by^cy^d y^e+...
\ee
We use bars over quantities to indicate that they are written in a RNC. Scalars do not transform so we do not need a bar.
In the RNC, all the $\bar \Gamma$'s in \eqref{3d} vanish at the origin so one obtains a covariant expansion, because the $y^\mu$ are geometric objects -  vectors - at the origin:

\[
\phi(y)= \phi (0) + y^\mu \bar \n_\mu \phi(0) +{1\over 2!} y^\mu y^\nu \bar \n_\mu \bar \n_\nu \phi(0) +{1\over 3!}y^\mu y^\nu y^\rho \bar \n_\mu \bar \n_\nu \bar \n_\rho \phi(0)...
\]
\be   \label{TaylorRNC}
= \phi (0) + y^\mu \p_\mu \phi(0) +{1\over 2!} y^\mu y^\nu \p_\mu \p_\nu \phi(0) +{1\over 3!}y^\mu y^\nu y^\rho \p_\mu \p_\nu \p_\rho \phi(0)...
\ee
i.e. in the RNC, for a scalar, the ordinary Taylor expansion is an expansion in covariant derivatives. The LHS is a scalar at $y$. Each term on the RHS is a scalar {\em at the origin} $y=0$. Nevertheless, since both sides are scalars, one can transform the first equation, which is manifestly covariant,  to any other coordinate system, $x$, with the understanding that $y^\mu$ transforms as a vector into:
\[
\delta x^\nu = {\p x^\nu \over \p y^\mu}|_{y=0}y^\mu
\]
This $\delta x^\nu \neq \Delta x^\mu (\equiv x^\mu-x_0^\mu)$.

\subsubsection{Vector}

One can perform the same set of steps for a vector:

\[
S_\mu(x_0+\Delta x)=S_\mu(x_0) + \Delta x^\rho \p_\rho S_\mu(x_0) + \hf \Delta x^\rho \Delta x^\s \p_\rho \p_\s S_\mu(x_0) +...
\]
\[
= S_\mu(x_0) + \Delta x^\rho [ \n_\rho S_\mu(x_0) + \Gamma ^\nu_{\rho\mu}S_\nu(x_0)]+\hf \Delta x^\rho \Delta x^\s [\n_\rho \n_\s S_\mu(x_0)+
\Gamma^\la _{\s \rho}\n_\la S_\mu(x_0) + 2 \Gamma^\la_{\rho\mu}\n _\s S_\la(x_0)]
\]
\be	\label{vect}
+\hf \Delta x^\rho \Delta x^\s [\tilde{\Gamma}^\nu_{\s \mu \rho}(x_0)+{1\over 3}R^\nu_{~\s \rho \mu}(x_0)+ \Gamma^\nu_{(\s|\la|}\Gamma^\la_{\mu\rho)}(x_0)]S_\nu(x_0)+...
\ee

The equation is manifestly not covariant because of the presence of the explicit $\Gamma$'s. Once again we can rewrite this equation in the RNC, $y$. All the $\bar \Gamma$'s are zero (at the origin, $y=0$) and we get
\be
\bar S_\mu(y)=\bar S_\mu(0) + y^\rho \bar \n_\rho \bar S_\mu(0) + \hf y^\rho y^\s [\bar \n_\rho \bar \n_\s \bar S_\mu(0) +{1\over 3}\bar R^\nu_{~\s \rho \mu}(0)\bar S_\nu(0)]+...
\ee
The main difference with the scalar is the appearance of the Riemann tensor and its derivatives. 

Note that the LHS is a vector at the general point $y$ whereas the RHS is a sum of vectors at the {\em origin} $y=0$. Thus this equation is valid only in the RNC. If we want transform to a different coordinate system the LHS and RHS transform differently, because they are vectors at {\em different} points.

We can write for the LHS
\[
\bar S_\nu (y) =  {\p x^\mu \over \p y_\nu}|_{y} S_\mu(x)
\]
and for the RHS we can write similarly
\[
\bar S_\nu (0) =  {\p x^\mu \over \p y_\nu}|_{0}  S_\mu(x_0)
\]
Plugging in these transformation matrices, one can relate $S_\mu(x)$ to $S_\mu(x_0)$ and its derivatives.

\subsubsection{Tensor}

\[
\bar W_{\al _1 ....\al _p}(y) 
= \bar W_{\al _1 ....\al _p}(0) ~+~
\bar W_{\al _1 ....\al _p , \mu}(0)y^\mu ~+~
\]
\[
{1\over 2!}\{\bar W_{\al _1 ....\al _p ,\mu \nu}(0) 
~-~{1\over 3}
\sum _{k=1}^p \bar R^\beta _{~\mu \al _k \nu}(0) 
\bar W_{\al _1 ..\al _{k-1}\beta \al _{k+1}..\al _p}(0)\}
 y^\mu y^\nu
~+~
\]
\[   
{1\over 3!}\{\bar W_{\al _1 ....\al _p ,\mu \nu \rho}(0) - 
\sum _{k=1}^p \bar R^\beta _{~\mu \al _k \nu}(0) 
\bar W_{\al _1 ..\al _{k-1}\beta \al _{k+1}..\al _p, \rho}(0)
\]
\be \label{Taylor}
-
{1\over 2}\sum _{k=1}^p \bar R^\beta _{~\mu \al _k \nu ,\rho }(0) 
\bar W_{\al _1 ..\al _{k-1}\beta \al _{k+1}..\al _p}(0)\}y^\mu y^\nu y^\rho +...
\ee
The commas denote covariant derivatives. The bars remind us that we are in an RNC.

\subsection{Expansion of Vertex Operators}

We have seen that it is easier to work with scalars. In the world sheet action the terms are all scalars, because the tensor indices
are contracted with those of the vertex operators. For eg $S_\mu(X(z,)) \p_z X^\mu(z)$. We use $X(z), Y(z)$ to denote the world sheet fields instead of $x$ and $\bar Y$ instead of RNC $y$.
 So the combined object is a scalar (in space time).
So we study the expansion of vertex operators. This is required when one performs an OPE of vertex operators in the interaction term of the ERG. It is important to do this covariantly.

The value of $X$ is parametrized by $z$ (and $\zb$ for closed strings).   So a Taylor expansion in $z$ bcomes a Taylor expansion in $X$. Since $z$ is a space time scalar we are guaranteed that the expansion coefficients will be scalars.In loop variables, we have $Y(z,\xn)$ and the Taylor expansion is in $\xn$. We denote these variables generically by $z^\al$. Thus let us choose the origin, $x_0$, of the RNC system such that $X(0)=x_0$.
Then
\be   \label{delx}
X^i(z)-X^i(0)=\Delta X^i = z^\al \p_\al X^i(0) + {z^\al z^\beta\over 2!} \p_\al \p_\beta X^i(0)  +{z^\al z^\beta z^\gamma\over 3!} \p_\al \p_\beta \p_\gamma X^i(0)  +....
\ee
Now $ X^i_\al\equiv \p_\al X^i $ is a vector: 
\[
{\p X'^i(z)\over \p z^\al}= {\p X'^i(z)\over \p X^j(z)}{\p X^j(z)\over \p z^\al}
\]
However ${\pp X^i\over \p z^\al \p z^\beta}$ is not a tensor. Define a covariant derivative:
\be   \label{CD}
D_\beta X_\al^i(z) \equiv \p_\beta X_\al^i(z) + \Gamma^i_{ab}(X(z))X_\beta^a(z) X_\al ^b(z)
\ee
(We will for convenience use $a,b, c..$ for the dummy indices and $i,j,k..$ for the uncontracted indices.)
Then
\be
\p_\beta X^i_\alpha(z) = D_\beta X^i_\alpha(z) - \Gamma ^i_{ba}(X(z))X^b_\beta (0)X^a_\alpha(0)
\ee
 Clearly in RNC, at the origin, $\bar \Gamma ^i_{ba}(X(0))=0$,
\be
\p_\beta \bar X^a_\al(0) = \bar D_\beta \bar X^a_\al(0)
\ee
\br
\p_\gamma \p_\beta X_\alpha ^i(z) &=&D_\gamma D_\beta X_\alpha ^i- \Gamma^i_{ca}X^c_\gamma D_\beta X^a_\alpha - \p_\gamma [\Gamma ^i_{ba} X^b_\beta X^a_\alpha] \nonumber \\
&=&D_\gamma D_\beta X_\alpha ^i- \Gamma^i_{ca}X^c_\gamma D_\beta X^a_\alpha -(\p_\gamma \Gamma ^i_{ba}) X^b_\beta X^a_\alpha
-\Gamma ^i_{ba} \p_\gamma[X^b_\beta X^a_\alpha]\nonumber \\
&=&D_\gamma D_\beta X_\alpha ^i- \Gamma^i_{ca}X^c_\gamma D_\beta X^a_\alpha -(\p_\gamma \Gamma ^i_{ba}) X^b_\beta X^a_\alpha
-\Gamma ^i_{ba}D_\gamma[X^b_\beta X^a_\alpha]-\Gamma ^i_{ba}[-\Gamma^b_{cd}X_\gamma^c X_\beta ^d X_\alpha^a - \Gamma ^a_{cd}
X^c_\gamma X^b_\beta X^d_\alpha ]\nonumber\\
&=&D_\gamma D_\beta X_\alpha ^i -\Gamma^i_{ca}X^c_\gamma D_\beta X^a_\alpha-\Gamma ^i_{ba}D_\gamma[X^b_\beta X^a_\alpha]-(\p_c \Gamma ^i_{ba})X^c_\gamma X^b_\beta X^a_\alpha-\Gamma ^i_{ba}[-\Gamma^b_{cd}X_\gamma^c X_\beta ^d X_\alpha^a - \Gamma ^a_{cd}
X^c_\gamma X^b_\beta X^d_\alpha ]\nonumber\\
&=&D_\gamma D_\beta X_\alpha ^i(z) -\Gamma^i_{ca}(X(z))X^c_\gamma D_\beta X^a_\alpha(z)-\Gamma ^i_{ba}(X(z))D_\gamma[X^b_\beta X^a_\alpha(z)]
-\Gamma^i_{bac}(X(z))X^c_\gamma X^b_\beta X^a_\alpha (z)\nonumber
\er 
We can now symmetrize the RHS in $\alpha, \beta , \gamma$ because the LHS is symmetric, and write
\be
\p_\gamma \p_\beta X_\alpha ^i={1\over 3!}\{D_{(\gamma} D_\beta X_{\alpha )} ^i -\Gamma^i_{ca}X^c_{(\gamma} D_\beta X^a_{\alpha )}-\Gamma ^i_{ba}D_{(\gamma}[X^b_\beta X^a_{\alpha )}]
-\tilde \Gamma^i_{bac}X^c_{( \gamma} X^b_\beta X^a_{\alpha )}\}
\ee 
For the last term we have used the fact that $X^c_{( \gamma} X^b_\beta X^a_{\alpha )}$ is also symmetric now in $a,b,c$.  It is clear from the above pattern that symmetrized vertex operators will involve the $\tilde \Gamma^a_{bcde...}$ as defined in the last subsection.

If we now specialize to RNC, where $\bar {\tilde{ \Gamma}}^a_{bcde...}=0$, we find that, again at the origin,
\be
\p_\gamma \p_\beta \bar X_\alpha ^i(0)={1\over 3!}\bar D_{(\gamma} \bar D_\beta \bar X_{\alpha )}^i(0)
\ee

Let us do one more:
\br
\p_\delta \p_\gamma \p_\beta X^i_\al &=& {1\over 4!}\{ D_{(\delta} D_\gamma D_\beta X_{\al)}^i - \Gamma^i_{da}X^d_{(\delta}D_\gamma D_\beta X_{\al )}^a \}\nonumber \\
& &-{1\over 8}\{ [\tilde{\Gamma}^i_{cad}-{1\over 6}(R^i_{~dac}+R^i_{~cad})]X_{(\delta}^dX_\gamma^c D_\beta X_{\al)}^a + \Gamma_{ca}^iD_{(\delta}X_\gamma^cD_\beta X_{\al)}^a\}\nonumber\\
& &-{1\over 4!}\{\tilde{\Gamma}^i_{abcd}X^d_{(\delta}X^c_\gamma X^b_\beta X^a_{\al)} + \tilde{\Gamma}^i_{abc} D_{(\gamma}(X^c_{\gamma}X^b_\beta X^a_{\al)})\}
\er

Note that we have used \eqref{Gamma} to write $\Gamma^i_{cad}$ in terms of $\tilde{\Gamma}^i_{cad}$ and the Riemann tensor. Once again specializing to RNC, at the origin, we get
\[
\p_\delta \p_\gamma \p_\beta \bar X^i_\al(0) = {1\over 4!}\bar D_{(\delta} \bar D_\gamma \bar D_\beta \bar X_{\al)}^i(0) +{1\over 48}(\bar R^i_{~dac}(0)+\bar R^i_{~cad}(0))]X_{(\delta}^d \bar X_\gamma^c \bar D_\beta \bar X_{\al)}^a(0)
\]
Thus the Taylor expansion in an RNC is
\[
\bar X^i(z)= z^\al \bar X^i_\al(0)+  {z^\al z^\beta\over 2!}D_\beta \bar X^a_\al(0) +{z^\al z^\beta z^\gamma\over 3!}{1\over 3!}D_{(\gamma} D_\beta \bar X_{\alpha )}^i(0)
 \]
\be	\label{VOP}
+{z^\al z^\beta z^\gamma z^\delta \over 4!}\{{1\over 4!}D_{(\delta} D_\gamma D_\beta \bar X_{\al)}^i(0) +{1\over 48}(\bar R^i_{~dac}(0)+\bar R^i_{~cad}(0))]\bar X_{(\delta}^d\bar X_\gamma^c D_\beta \bar X_{\al)}^a(0)\}
\ee

\subsection{Expansion of Scalar Combination of Tensor and Vertex Operator}

\subsubsection{Scalar}

The simplest case is again a scalar $\phi(X(z))$ which we can expand in powers of $z^\al$. We start with
\be   
\phi (x) = \phi(x_0) + \Delta x^\mu \p_\mu \phi (x_0) + {1\over 2!} \Delta x^\mu \Delta x^\nu \p_\mu \p _\nu \phi (x_0)+...
\ee

and substitute for the ordinary derivatives expressions such as \eqref{3d} :
\[   
\phi (X(z)) = \phi(X(0)) + \Delta X^\mu \n_\mu \phi (X(0)) + {1\over 2!} \Delta X^\mu \Delta X^\nu (\n_\mu \n _\nu \phi (X(0))+ \Gamma ^\rho_{\mu\nu}(X(0)) \n _\rho \phi(X(0))
\]
\be
+ {1\over 3!} \Delta X^\mu \Delta X^\nu \Delta X^\rho\{\n _\rho\n_\mu \n_\nu \phi (X(0)) + \Gamma^\la_{(\rho\mu}(X(0))\n_{|\la|}\n_{\nu)} \phi(X(0)) +[\tilde{\Gamma}^\la_{\mu\nu\rho}+\Gamma^\la_{(\mu |\sigma|}\Gamma^\s_{\rho\nu)}(X(0))]\n_\la \phi(X(0))\}+...
\ee

This equation does not look covariant because $\Delta X^\mu$ is not a covariant object. For $\Delta X^\mu$ we substitute \eqref{delx} which is a sum of non covariant terms multiplied by powers of $z^\al$. But the final expression has a scalar on the LHS and the RHS is an expansion in $z^\al$ which is a space time scalar. Accordingly each term in the sum must be a scalar. We in fact do find that all the explicit non covariant $\Gamma$'s cancel amongst themselves and the result is manifestly covariant. We give the first few terms:
\br   
\phi(X(z))&=&\phi(X(0))+z^\al X_\al^i \n_i\phi (X(0)) + {z^\al z^\beta\over 2!}(D_\beta X_\al^i\n_i\phi(X(0))+ X_\al^i X_\beta ^j \n_i \n_j \phi(X(0)))\nonumber \\
& &+ z^\al z^\beta z^\gamma\{ [{1\over 3!}{1\over 6}D_{(\al}D_\beta X_{\gamma)}^i]\n_i\phi(X(0))+ {1\over 4}X_\al ^{(i}D_\beta X_\gamma^{j)} \n_i\n_j \phi(X(0)) \nonumber\\
& &+ {1\over 3!} X_\al^i X_\beta ^j X_\gamma ^k \n_i\n_j\n_k\phi(X(0)) \}\nonumber
\er
\be   \label{8}
+....~~~~~~~~~~~~~~~~~~~~~~~~~~~~~~~~~~~~~~~~~~~~~~~~~~~~~~~~~~~~~~~~~~~~~~~~~~~~~
\ee
This calculation has been done in a general coordinate system. Nevertheless, it could have been done in an RNC, where neither the expansion for $\phi$, nor the expansion for $\Delta X$ has any $\Gamma$'s - both expansions look manifestly covariant. If we combine these expansions, we must necessarily obtain the same equation. Thus
\be   \label{phi}
\phi(\bar Y(z))= \phi (0) + \bar Y^i(z)\bar  \n_i \phi(0) +{1\over 2!} \bar Y^i(z) \bar Y^j(z)\bar  \n_i \bar \n_j \phi(0) +{1\over 3!}\bar Y^i(z) \bar Y^j(z) \bar Y^k(z) \bar \n_i \bar \n_j \bar \n_k \phi(0)...
\ee
and
\[
\bar Y^i(z)= z^\al \bar Y^i_\al(0)+  {z^\al z^\beta\over 2!}\bar D_\beta \bar Y^a_\al(0) +{z^\al z^\beta z^\gamma\over 3!}{1\over 3!}\bar D_{(\gamma} \bar D_\beta \bar Y_{\alpha )}^i(0)
 \]
\be   \label{Y}
+{z^\al z^\beta z^\gamma z^\delta \over 4!}\{{1\over 4!}\bar D_{(\delta} \bar D_\gamma \bar D_\beta Y_{\al)}^i(0) +{1\over 48}(\bar R^i_{~dac}(0)+\bar R^i_{~cad}(0))]\bar Y_{(\delta}^d\bar Y_\gamma^c \bar D_\beta \bar Y_{\al)}^a(0)\}
\ee
As explained above, while these equations look covariant, they are in fact valid only in RNC. Nevertheless
it is easy to check that when \eqref{phi} and \eqref{Y} are combined we do get the same expansion \eqref{8}  but with bars - i.e. in the RNC. But now this expression is a sum of {\em scalars} on the RHS and a {\em scalar} on the LHS (albeit at different points) so we can remove the bars and use this equation in any coordinate system.

\subsubsection{Vector}

Let us now consider the vector: $S_i(X(z))\p_\al X^i(z)=S_i(X(z))X_\al^i(z)$. We use \eqref{vect} for the expansion of the vector in powers of $\Delta X^i$ and \eqref{delx} for $\Delta X^i$. Both have non covariant terms but the non covariant terms cancel amongst themselves and the result is a manifestly covariant expansion in powers of $z^\al$.
\br
S_i (X(z))X_\al^i(z)&=& S_i (X(0))X_\al^i(0) + z^\beta [ \hf S_i(X(0))D_{(\beta}X_{\al)}^i(0) + \n_jS_i(X(0))X_\al^iX_\beta ^j(0)]\nonumber \\
& & {z^\beta z^\al\over 2!}[\hf \n_j S_i (X(0)) D_{(\beta}X_{\al)}^i + ( {\n_{(i}\n_{j)}S_i(X(0))\over 2} + {1\over 3} R^l_{~jki}(X(0))S_l(X(0)))X_\beta ^k X_\gamma ^j X_\al^i(0) + \nonumber \\
& &+ {D_{(\al}D_\beta X_{\gamma)}^i(0)S_i(X(0))\over 6} + (\n_jS_i(X(0)))D_{(\beta} X_{\al)})^i X_\gamma^j(0)]\nonumber
\er
\be		\label{Vec}
+...~~~~~~~~~~~~~~~~~~~~~~~~~~~~~~~~~~~~~~~~~~~~~~~~~~~~~~~~~~~~~~~~~~~~~~~~~~~~~~~~~
\ee

As with the scalar, if one works in RNC, one gets the manifestly covariant expression directly, without having to worry about cancellations
amongst the non covariant terms.

As mentioned above, these expansions are required when one performs OPE in the interaction term of the ERG.

\subsection{Covariantizing $\bar Y^\mu(z)$}

$\bar Y^\mu(z)$ is the RNC space time coordinate field that occurs in the world sheet theory. As explained above at some length
it is a geometric object defined at the origin of the RNC and is tangent to the geodesic that starts from the origin, O,  of the RNC and goes through a given point, P,  with RNC coordinate $\bar Y^\mu$. It is thus a coordinate as well as a vector at the origin O.  One may well ask if there is a geometric object at P, that coincides with this in the RNC. Note that $\bar Y^\mu$ is a tangent to the geodesic at P as well,
because in the RNC the geodesics are straight lines. Thus one can also think of $\bar Y^\mu$ as a geometric object at P - viz the tangent vector to the geodesic at P, and transform it as a vector, into any other coordinate system. This is useful because one needs a covariant definition of the Green function, that in the RNC is
$\lan \bar Y^\mu(z) \bar Y^\nu(w)\ran$. 

 We elaborate on this idea: As before let $X$ be a general coordinate system. At a point O (with coordinates $X_0$) we set the origin of an RNC system $\bar Y^\mu$. The point O has coordinate $\bar Y^\mu=0$. For a general point P with coordinate $X$, we consider a geodesic
 that starts from O and goes through P. Let the tangent vector to this geodesic at O be $\vec \xi _P$ and the proper distance along this geodesic to P be $t_P$. Then $\bar Y^\mu=t_P\xi_P ^\mu$. $\vec \xi_P$ is a geometric object - a vector  at O, {\em not} at P. So $\bar Y^\mu$ transforms as a vector at O. One would like an object that is a vector at P. So let us define the tangent vector field, $\xi^\mu(P)$ (or $\xi ^\mu (X_P)$) of unit norm vectors tangent to the geodesics through O at the (general) point P.  They obey
 \[
 \xi ^\nu \nabla _\nu \xi ^\mu = \xi ^\nu {\p \xi ^\mu \over \p X^\nu} + \Gamma ^\nu_{\mu \rho} \xi ^\mu \xi^\rho=0
 \]
 In the RNC this equation becomes
 \[
\bar \xi ^\nu \nabla _\nu \bar \xi ^\mu = \bar \xi ^\nu {\p \bar \xi ^\mu \over \p X^\nu} + \bar \Gamma ^\nu_{\mu \rho} \bar \xi ^\mu \bar \xi^\rho=0
\] 
But we know that in the RNC all along the geodesic,
\[
\bar \Gamma ^\nu_{\mu \rho} \bar \xi ^\mu \bar \xi^\rho=0
\]

This follows from the fact that $\bar Y^\mu$ satisfies the geodesic equation 
\[
{d^2\bar Y^\mu\over dt^2}+ \bar \Gamma ^\nu_{\mu \rho} {d \bar Y ^\mu\over dt} {d\bar Y^\rho\over dt}=0
\] 
and since ${d^2\bar Y^\mu\over dt^2}=0$ we get $ \bar \Gamma ^\nu_{\mu \rho} {d \bar Y ^\mu\over dt} {d\bar Y^\rho\over dt}=
 \bar \Gamma ^\nu_{\mu \rho} \bar \xi^\mu \bar \xi^\rho=0$.
  
Thus we get   
\[
\bar \xi ^\nu {\p \bar \xi ^\mu \over \p X^\nu}=0
\]
which means $\bar \xi^\mu$ is constant along a geodesic. A solution to this is thus the constant (along a geodesic) vector field $\bar \xi^\mu (\bar Y_P) = \bar \xi_P$. This is just the obvious fact that in the RNC geodesics are straight lines through the origin, so the tangent vector field is a constant (along a geodesic) vector field. We thus see that in the RNC $\bar Y^\mu_P = t_P\bar \xi^\mu (\bar Y_P) $ is not only a coordinate, it is also a vector field, i.e. the two objects coincide.  This will not be the case in a general coordinate system.

Thus when we change coordinates to $Y$, the vector field $\bar Y^\mu(\bar Y_P) = t_P \bar \xi^\mu(\bar Y_P)$ transforms like a vector field {\em at P} to a new vector field, 
\be 	\label{Vecfld}
y^\mu(P) \equiv t_P \xi ^\mu (Y_P) = t_P{\p Y^\mu \over \p \bar Y^\nu}|_P \bar \xi^\nu (P)
\ee
 whereas the coordinate $\bar Y^\mu _P$ becomes $Y^\mu _P$.
 
Thus when we see an expression involving $\bar Y^\mu = t_P \bar \xi^\mu$ in the RNC, there are two distinct geometric objects that it can correspond to in a general coordinate system: $y^\mu(O)\equiv
t_P\xi ^\mu(O)$ or $y^\mu(P)\equiv t_p \xi^\mu (P)$. (This is in addition to the original interpretation as a {\em coordinate}, in which case it just becomes $X^\mu$ in a new coordinate system.)

\section{Free Equation}
\label{appenb}
\setcounter{equation}{0}
\subsection{}

The details of the calculation of the Level 2 (graviton) and Level 4 free equation are given here.

We have to evaluate the second derivative, which is given by the action of a functional derivative on (\ref{FnlDII}):
\[
=\int dz'\int dz''~\dot G(z',z'')~
\]
\[\eta^{\mu\nu}\int du~[{\p\over \p Y^\nu(u)}\delta(u-z') +[{\p\over \p x_1}\delta(u-z')]{\p \over \p Y^\nu_{1;0}(u)}+
\]\[
[{\p\over \p \bar x_1}\delta(u-z')]{\p \over \p Y^\nu_{0;\bar1}(u)}+[{\p\over \p x_2}\delta(u-z')]{\p \over \p Y^\nu_{2;0}(u)}+\]
\[[{\p\over \p \bar x_2}\delta(u-z')]{\p \over \p Y^\nu_{0;\bar 2}(u)}+
 [{\pp\over \p x_1\p \bar x_1}\delta(u-z')]{\p \over \p Y^\nu_{1;\bar 1}(u)}+...]\]
 \[
 \Big\{ \underbrace{{\p {\cal L} [Y(u),Y_{n,\mb}(u)]\over \p Y^\mu(u)} \delta (u-z'')}_I-\underbrace{\p_{x_1}{\p {\cal L} [Y(u),Y_{n;\mb}(u)]\over \p Y_{1;0}^\mu(u)} \delta (u-z'')}_{II}
\]
\be - \underbrace{\p_{\bar x _1} {\p {\cal L}[Y(u),Y_{n;\mb}(u)] \over \p Y_{0;\bar 1} ^\mu(u) }
\delta (u-z'')}_{III} +  \underbrace{\p _{ x_{1}}\p_{\bar x_{1}}{\p {\cal L}[Y(u),Y_{n,\mb}(u)] \over \p Y_{1,\bar 1}^\mu (u) }
 \delta (u-z'')}_{IV} \Big\}
\ee

Let us evaluate the action of the derivatives on each of the four terms labeled I,II,III and IV. The result has to be symmetric in $z'\leftrightarrow z''$ and also for every term, there is also a corresponding complex conjugate term. This fact reduces the number of independent terms to be evaluated.  (Our notation is: $\xn$ refers to $u$, $\xn '$ refers to $z'$ and $\xn ''$ refers to $z''$. Thus for instance, ${\p \delta(u-z')\over \p \xn}=-{\p \delta(u-z')\over \p \xn '}$)
\begin{enumerate}
\item 
\[
\int du~\eta^{\mu \nu}{\p\over \p Y^\nu(u)}{\p {\cal L} [Y(u),Y_{n,\mb}(u)]\over \p Y^\mu(u)} \delta(u-z')\delta (u-z'')
\]
\be=
-\ko^2 {\cal L} (z')\delta(z'-z'')
\ee
\item
\[
\int du~\eta^{\mu \nu}\Big([{\p\over \p x_1}\delta(u-z')]{\p \over \p Y^\nu_{1;0}(u)}{\p {\cal L} [Y(u),Y_{n,\mb}(u)]\over \p Y^\mu(u)} \delta (u-z'')\Big) +\Big( z'\leftrightarrow z''\Big) 
\]
\[=~\eta^{\mu \nu}\Big(-{\p\over \p x'_1}[\delta(z''-z')i\ko.iK_{1;0}{\cal L}[z'']] \Big) +\Big( z'\leftrightarrow z''\Big)
\]
\[=\eta^{\mu \nu}\Big(-[{\p\over \p x'_1}+{\p\over \p x''_1}][\delta(z''-z')i\ko.iK_{1;0}{\cal L}[z'']] \Big)
\]

We restore the integrals over $z',z''$, and use $G(z',z'')= \lan Y(z')Y(z'')\ran$ and integrate by parts on $x',x''$ to get
\[{d\over d ln~ a}\int dz'dz''~({\p\over \p x'_1}+{\p\over \p x''_1})\lan Y(z')Y(z'')\ran [\delta(z''-z')i\ko.iK_{1;0}{\cal L}[z'']]=
\]
\[=
{d\over d ln~ a}\int dz'~[{\p\over \p x'_1}\lan Y(z')Y(z')\ran] i\ko.iK_{1;0}{\cal L}[z']]\]
\[
=-\int dz'~\dot G(z',z') i\ko.iK_{1;0}{\p\over \p x'_1}[{\cal L}[z']]
\]
We have integrated by parts again in the last step.

Finally we can add the complex conjugate to obtain:
\be
=\int dz'~\dot G(z',z')\Big( \ko.K_{1;0}{\p\over \p x'_1}[{\cal L}[z']]+\ko.K_{0;\bar 1}{\p\over \p \bar x'_1}[{\cal L}[z']]\Big)
\ee
\item
\[
\eta ^{\mu \nu}\int dz'\int dz''~\dot G(z',z'')\int du~[{\p\over \p x_1}\delta(u-z')][{\p\over \p x_1}\delta(u-z'')]{\pp{\cal L}[u]\over \p Y^\mu_{1;0}(u)\p Y^\nu_{1;0}(u)} 
\]
\[=
{d\over d ln~ a}\eta ^{\mu \nu}\int dz'\int dz'' \lan Y_{1;0}(z')Y_{1;0}(z'')\ran \delta(z'-z''){\pp{\cal L}[z']\over \p Y^\mu_{1;0}(z')\p Y^\nu_{1;0}(z')} 
\]
\[=
{d\over d ln~ a}\eta ^{\mu \nu}\int dz'[\hf ({\pp \over \p x_1^{'2}}-{\p \over \p x'_{2}})\lan Y(z')Y(z')\ran]{\pp{\cal L}[z']\over \p Y^\mu_{1;0}(z')\p Y^\nu_{1;0}(z')} 
\]
\[=
\eta ^{\mu \nu}\int dz'\dot G(z',z')\hf ({\pp \over \p x_1^{'2}}+{\p \over \p x'_{2}}){\pp{\cal L}[z']\over \p Y^\mu_{1;0}(z')\p Y^\nu_{1;0}(z')} 
\]
\be
=-\int dz'\dot G(z',z') K_{1;0}.K_{1;0}\hf ({\pp \over \p x_1^{'2}}+{\p \over \p x'_{2}}){\cal L}[z']
\ee
\item The complex conjugate is:
\be
-\int dz'\dot G(z',z') K_{0;\bar 1}.K_{0;\bar1}\hf ({\pp \over \p \bar x_1^{'2}}+{\p \over \p \bar x'_{2}}){\cal L}[z']
\ee
\item
\[
\int dz' dz'' ~\dot G(z',z'')\Big( \eta^{\mu \nu}\int du~[{\p\over \p x_2}\delta(u-z')]\delta(u-z''){\pp{\cal L}[u]\over \p Y^\mu_{2;0}(u)\p Y^\nu(u)} ~+~z'\leftrightarrow z'' \Big)
\]
\[
= \int dz' dz'' ~\dot G(z',z'')\Big( \eta^{\mu \nu}~[-{\p\over \p x'_2}\delta(z''-z')]{\pp{\cal L}[z'']\over \p Y^\mu_{2;0}(z'')\p Y^\nu(z'')} ~+~z'\leftrightarrow z'' \Big)
\]
\[
= \int dz' dz'' ~{d \over d ln~a} [{\p\over \p x'_2}+ {\p\over \p x''_2}]G(z',z'')\Big( \eta^{\mu \nu}~[\delta(z''-z')]{\pp{\cal L}[z'']\over \p Y^\mu_{2;0}(z'')\p Y^\nu(z'')} \Big)
\]
\be
= \int dz'  ~\dot G(z',z')\Big(K_{2;0}.\ko~({\p\over \p x_2}{\cal L}[z'']) \Big)
\ee
\item
Complex conjugate gives:
\be
= \int dz'  ~\dot G(z',z')\Big(K_{0;\bar 2}.\ko~({\p\over \p \bar x_2}{\cal L}[z'']) \Big)
\ee
\item
\[
\int dz' dz'' ~\dot G(z',z'')\Big( \eta^{\mu \nu}\int du~[{\pp\over \p x_1 \p \bar x_1}\delta(u-z')]\delta(u-z''){\pp{\cal L}[u]\over \p Y^\mu_{1;\bar 1}(u)\p Y^\nu(u)} ~+~z'\leftrightarrow z'' \Big)
\]
\[ 
={d\over d~ln~a}\int dz' dz'' ~ \lan Y (z')_{1;\bar 1} Y(z'')+Y (z') Y_{1;\bar 1}(z'')\ran \Big( \eta^{\mu \nu} \delta(z'-z''){\pp{\cal L}[z']\over \p Y^\mu_{1;\bar 1}(z')\p Y^\nu(z')} \Big)
\]
\be  \label{8}
={d\over d~ln~a}\int dz' dz'' ~ \lan Y (z')_{1;\bar 1} Y(z'')+Y (z') Y_{1;\bar 1}(z'')\ran \Big(( iK_{1;\bar 1}.i\ko ) \delta(z'-z''){\cal L}[z'] \Big)
\ee
\item
\[
\int dz' dz'' ~\dot G(z',z'')\Big( \eta^{\mu \nu}\int du~[{\p\over \p \bar x_1 }\delta(u-z')][{\p \over \p x_1}\delta(u-z'')]{\pp{\cal L}[u]\over \p Y^\mu_{1;0}(u)\p Y^\nu_{0;\bar 1}(u)} ~+~z'\leftrightarrow z'' \Big)
\]
\be \label{9}
=
{d\over d~ln~a}\int dz' dz'' ~ \lan Y_{0:\bar 1}(z')Y_{1;0}(z'')+Y_{0:\bar 1}(z'')Y_{1;0}(z')\ran\Big( \delta(z'-z'')(iK_{1;0}.iK_{0;\bar 1}){\cal L}[z'] \Big)
\ee
\end{enumerate}
(\ref{8}) and (\ref{9}) can be added to give 
\[
=
{d\over d~ln~a}\int dz' ~ [{\pp\over \p x'_1\p \bar x'_1}\lan Y(z')Y(z')\ran]\Big( (iK_{1;0}.iK_{0;\bar 1}){\cal L}[z'] \Big)
\]
\be =
\int dz' ~ \dot G(z',z')\Big( (iK_{1;0}.iK_{0;\bar 1})[{\pp\over \p x'_1\p \bar x'_1}{\cal L}[z']] \Big)
\ee
{\em provided} the following constraint is imposed :
\be  \label{Kconstraint}
K_{1;0}.K_{0;\bar 1}~ {\cal L}= K_{1;\bar 1}.\ko ~ {\cal L}
\ee
The constraint is gauge covariant since both sides have identical gauge transformation properties.
Since $K_{1;\bar 1}$ is an auxiliary field (i.e. not physical) we are free to impose this constraint. In fact since $K_{1;\bar 1}.\ko$
contains $q_{1;\bar 1} q_0$  (for  $q_0\neq 0$) , this can be treated as an algebraic constraint on $q_{1;\bar 1}$.

The massless case (Graviton) is discussed in Section 3 and Section 4.

Similar constraints on $K_{n;\mb}$ occur at every level. We will refer to them as K-constraints. They are described in the next Appendix \eqref{appenc}.

\subsection{Level (1,1)}

The terms calculated above are sufficient to extract the coefficient of the graviton multiplet vertex operators at level ($1;\bar 1$), $Y_{1;0}^\mu Y_{0;\bar 1}^\nu$ and the next massive level, $Y_{1;0}^\mu Y_{1;0}^\nu Y_{0;\bar 1}^\rho Y_{0;\bar 1}^\sigma \e$, a vertex operator in closed string theory at level $(2;\bar 2)$. We revert to the notation
$k_1=K_{1;0}, k_{\bar 1}=K_{0;\bar 1},...$ below.

We get for level $(1,\bar 1)$:
\be   \label{grav}
[-\ko^2 \kim k_{\bar 1\nu} + \ko .\ki \kom k_{\bar 1\nu} + \ko. k_{\bar 1}\kon \kim - \ki.k_{\bar 1}\kom \kon]Y_{1;0}^\mu Y_{0;\bar 1}^\nu=0
\ee
\be   \label{K11b}
[-\ko^2 K_{1;\bar 1\mu} + \ko.\ki k_{\bar 1\mu} +\ko.\kib \kim -\ki.\kib \kom]Y_{1;\bar 1}^\mu=0
\ee

If we use the constraint \eqref{Kconstraint}, then the two equations above are not independent:  $\kom$ dotted into \eqref{K11b} is equal to the trace of \eqref{grav}.

Furthermore we can dimensionally reduce  \eqref{K11b} to get
an equation that looks exactly like \eqref{K11b} but with the dot product going over $D$ dimensions. The remaining term in the dot product
cancels if we use the definition 
 \be	\label{defn}
K_{1;\bar 1\mu }= {q_1\over \qo}k_{\bar 1\mu} + {q_{\bar 1}\over \qo} \kim - {\qi q_{\bar 1}\over \qo^2} \kom
\ee

Note that according to this definition, when $\mu=D$,  $Q_{1;\bar 1}$ is fixed to:
\[
\qo Q_{1;\bar 1}= \qi \qib
\]
Thus $\qo Q_{1;\bar 1}$ also stands for the dilaton. For the other values of the indices the situation is more complicated.
In the closed string there is the extra condition that the number of $\qi$'s must equal the number of $\qib$'s in any term - this is the analog of the condition that there be no $\qi$ in the open string expressions. This means
that the first two terms in the definition of $K_{1;\bar 1\mu }$ are not allowed - either they have to be set to zero, or set equal to some other operator in a way that is consistent with gauge transformation. $\qi \qib$ is the dilaton. Thus we can think of \eqref{K11b} as a way
of defining $K_{1;\bar 1\mu }$ in terms of the other physical fields.

Now another subtlety here is that $\qo=0$ at this (massless) level. Thus 
it is not clear what to make of \eqref{defn}. Since this variable does not occur in the graviton equation, after using the K-constraint, we will not worry about this problem. At higher level this variable occurs in conjunction with other $\kn,q_n$ so we can apply the Q-rules to get rid of them. Also $\qo\neq 0$ at higher levels.  This problem occurs only for level 1.

\subsection{Level (2,2)}

At level $(2,\bar 2)$ for $Y_{1;0}^\mu Y_{1;0}^\nu Y_{0;\bar 1}^\rho Y_{0;\bar 1}^\sigma \e$ we get:

\[
-{1\over 4} \ko^2 (k_1 . Y_1)^2(k_{\bar 1}.Y_{\bar 1})^2 + \hf \ko.\ki (\ko . Y_1)(\ki .Y_1)(k_{\bar 1}.Y_{\bar 1})^2+\hf \ko.k_{\bar 1} (\ko . Y_{\bar 1})(k_{\bar 1} .Y_{\bar1})(k_{ 1}.Y_{1})^2
\]
\be   \label{EOMlevel4}
- {\ki .\ki\over 4} (\ko . Y_1)^2(k_{\bar 1}.Y_{\bar 1})^2 - {k_{\bar 1} .k_{\bar 1}\over 4} (\ko . Y_{\bar 1})^2(k_{ 1}.Y_{\ 1})^2 - \ki .k_{\bar 1} (\ko . Y_1)(\ko . Y_{\bar 1})(\ki .Y_1)(k_{\bar 1} .Y_{\bar1})
\ee

This can easily be seen to be gauge invariant under $\kim \rightarrow \kim + \li \kom$ and $k_{\bar 1\mu} \rightarrow k_{\bar 1\mu} +  \la _{\bar 1} \kom$, after using the tracelessness condition on the gauge parameter, $\li \ki .k_{\bar 1}=0=\li k_{\bar1}.k_{\bar 1}$ and the same for its complex conjugate.

\section{Q-rules for Level 5}
\label{appenq}
\setcounter{equation}{0}

We use the notation $QQ[...]$ for the Q-rules. 

{\bf Four Index}
\br
 {QQ}\left[k_{\mu } k_{1\nu } k_{1\rho } k_{1\sigma } q_1\right]
&=&\frac{1}{4} \left(k_{2 \mu } k_{1\nu } k_{1\rho } k_{1\sigma }+k_{1\mu } k_{2 \nu } k_{1\rho } k_{1\sigma }+k_{1\mu } k_{1\nu } k_{2 \rho } k_{1\sigma }+k_{1\mu } k_{1\nu } k_{1\rho } k_{2 \sigma }\right) q_0\nonumber\\
 {QQ}\left[k_{1\mu } k_{1\nu } k_{1\rho } q_1 \lambda _1\right]&=&\frac{1}{4} q_0 \left(\left(k_{2 \mu } k_{1\nu } k_{1\rho }+k_{1\mu } k_{2 \nu } k_{1\rho }+k_{1\mu } k_{1\nu } k_{2 \rho }\right) \lambda _1+k_{1\mu } k_{1\nu } k_{1\rho } \lambda _2\right)
\er

{\bf Three Index}

\br
 {QQ}\left[k_{2 \mu } k_{1\nu } k_{1\rho } q_1\right]&=&{b_3} k_{3 \mu } k_{1\nu } k_{1\rho } q_0+{a_3} k_{1\mu } k_{2 \nu } k_{2 \rho } q_0+\frac{1}{2} {a_{3p}} k_{2 \mu } \left(k_{2 \nu } k_{1\rho }+k_{1\nu } k_{2 \rho }\right) q_0\nonumber \\ 
&+&\frac{1}{2} {b_{3p}} k_{1\mu } \left(k_{3 \nu } k_{1\rho }+k_{1\nu } k_{3 \rho }\right) q_0+{c_3} k_{1\mu } k_{1\nu } k_{1\rho } q_2
\nonumber\\
 {QQ}\left[k_{1\nu } k_{1\rho } q_1 \lambda _2\right]&=&\frac{1}{2} {b_{3p}} k_{3 \nu } k_{1\rho } q_0 \lambda _1+{a_3} k_{2 \nu } k_{2 \rho } q_0 \lambda _1+\frac{1}{2} {b_{3p}} k_{1\nu } k_{3 \rho } q_0 \lambda _1+{c_3} k_{1\nu } k_{1\rho } q_2 \lambda _1\nonumber \\
&+&\frac{1}{2} {a_{3p}} k_{2 \nu } k_{1\rho } q_0 \lambda _2+\frac{1}{2} {a_{3p}} k_{1\nu } k_{2 \rho } q_0 \lambda _2+{b_3} k_{1\nu } k_{1\rho } q_0 \lambda _3\nonumber\\
 {QQ}\left[k_{2 \mu } k_{1\rho } q_1 \lambda _1\right]&=&{b_3} k_{3 \mu } k_{1\rho } q_0 \lambda _1+\frac{1}{2} {a_{3p}} k_{2 \mu } k_{2 \rho } q_0 \lambda _1+\frac{1}{2} {b_{3p}} k_{1\mu } k_{3 \rho } q_0 \lambda _1+{c_3} k_{1\mu } k_{1\rho } q_2 \lambda _1+\frac{1}{2} {a_{3p}} k_{2 \mu } k_{1\rho } q_0 \lambda _2\nonumber \\
&+&{a_3} k_{1\mu } k_{2 \rho } q_0 \lambda _2+\frac{1}{2} {b_{3p}} k_{1\mu } k_{1\rho } q_0 \lambda _3
\er

{\bf Three Index 2 q's}

\br
 {QQ}\left[k_{1\mu } k_{1\nu } k_{1\rho } q_1^2\right]&=&{b_{32}} \left(k_{2 \mu } k_{2 \nu } k_{1\rho }+k_{2 \mu } k_{1\nu } k_{2 \rho }+k_{1\mu } k_{2 \nu } k_{2 \rho }\right) q_0^2+{c_{32}} \left(k_{3 \mu } k_{1\nu } k_{1\rho }+k_{1\mu } k_{3 \nu } k_{1\rho }+k_{1\mu } k_{1\nu } k_{3 \rho }\right) q_0^2\nonumber \\ &+&{a_{32}} k_{1\mu } k_{1\nu } k_{1\rho } q_0 q_2
\nonumber\\
 {QQ}\left[k_{1\mu } k_{1\nu } k_{1\rho } q_1 \lambda _1\right]&=&\frac{1}{4} q_0 \left(\left(k_{2 \mu } k_{1\nu } k_{1\rho }+k_{1\mu } k_{2 \nu } k_{1\rho }+k_{1\mu } k_{1\nu } k_{2 \rho }\right) \lambda _1+k_{1\mu } k_{1\nu } k_{1\rho } \lambda _2\right)
\er

{\bf Two index}

\br
 {QQ}\left[k_{2 \mu } k_{2 \nu } q_1\right]&=&\frac{1}{2} {d_2} \left(k_{3 \mu } k_{2 \nu }+k_{2 \mu } k_{3 \nu }\right) q_0\nonumber \\&+&\frac{1}{2}  { c_{2}} \left(k_{4 \mu } k_{1\nu }+k_{1\mu } k_{4 \nu }\right) q_0+\frac{1}{2}  { b_{2}} \left(k_{2 \mu } k_{1\nu }+k_{1\mu } k_{2 \nu }\right) q_2+ { a_{2}} k_{1\mu } k_{1\nu } q_3\nonumber\\
 {QQ}\left[k_{3 \mu } k_{1\nu } q_1\right]&=& { D_{2a}} \left(k_{3 \mu } k_{2 \nu }-k_{2 \mu } k_{3 \nu }\right) q_0+ { D_{2s}} \left(k_{3 \mu } k_{2 \nu }+k_{2 \mu } k_{3 \nu }\right) q_0+ { C_{2a}} \left(k_{4 \mu } k_{1\nu }-k_{1\mu } k_{4 \nu }\right) q_0\nonumber\\&+& { C_{2s}} \left(k_{4 \mu } k_{1\nu }+k_{1\mu } k_{4 \nu }\right) q_0+ { B_{2a}} \left(k_{2 \mu } k_{1\nu }-k_{1\mu } k_{2 \nu }\right) q_2+ { B_{2s}} \left(k_{2 \mu } k_{1\nu }+k_{1\mu } k_{2 \nu }\right) q_2+ { A_{2p}} k_{1\mu } k_{1\nu } q_3\nonumber\\
 {QQ}\left[k_{2 \nu } q_1 \lambda _2\right]&=&\frac{1}{2}  { c_{2}} k_{4 \nu } q_0 \lambda _1+\frac{1}{2}  { b_{2}} k_{2 \nu } q_2 \lambda _1+ { a_{2}} k_{1\nu } q_3 \lambda _1+\frac{1}{2}  { d_{2}} k_{3 \nu } q_0 \lambda _2+\frac{1}{2}  { b_{2}} k_{1\nu } q_2 \lambda _2+\frac{1}{2}  { d_{2}} k_{2 \nu } q_0 \lambda _3+\frac{1}{2}  { c_{2}} k_{1\nu } q_0 \lambda _4\nonumber\\
 {QQ}\left[k_{1\nu } q_1 \lambda _3\right]&=&- { C_{2a}} k_{4 \nu } q_0 \lambda _1+ { C_{2s}} k_{4 \nu } q_0 \lambda _1- { B_{2a}} k_{2 \nu } q_2 \lambda _1+ { B_{2s}} k_{2 \nu } q_2 \lambda _1+ { A_{2p}} k_{1\nu } q_3 \lambda _1- { D_{2a}} k_{3 \nu } q_0 \lambda _2\nonumber\\&+& { D_{2s}} k_{3 \nu } q_0 \lambda _2+ { B_{2a}} k_{1\nu } q_2 \lambda _2+ { B_{2s}} k_{1\nu } q_2 \lambda _2+ { D_{2a}} k_{2 \nu } q_0 \lambda _3+ { D_{2s}} k_{2 \nu } q_0 \lambda _3+ { C_{2a}} k_{1\nu } q_0 \lambda _4+ { C_{2s}} k_{1\nu } q_0 \lambda _4\nonumber\\
 {QQ}\left[k_{3 \mu } q_1 \lambda _1\right]&=& { C_{2a}} k_{4 \mu } q_0 \lambda _1+ { C_{2s}} k_{4 \mu } q_0 \lambda _1+ { B_{2a}} k_{2 \mu } q_2 \lambda _1+ { B_{2s}} k_{2 \mu } q_2 \lambda _1+ { A_{2p}} k_{1\mu } q_3 \lambda _1+ { D_{2a}} k_{3 \mu } q_0 \lambda _2\nonumber\\&+& { D_{2s}} k_{3 \mu } q_0 \lambda _2- { B_{2a}} k_{1\mu } q_2 \lambda _2+ { B_{2s}} k_{1\mu } q_2 \lambda _2- { D_{2a}} k_{2 \mu } q_0 \lambda _3+ { D_{2s}} k_{2 \mu } q_0 \lambda _3- { C_{2a}} k_{1\mu } q_0 \lambda _4+ { C_{2s}} k_{1\mu } q_0 \lambda _4 \nonumber\\
 & &
\er

{\bf Two index with $q^2$}
\br
 {QQ}\left[q_1 q_2 k_{1,\mu } k_{1,\nu }\right]&=& { d_{22}} \left(k_{3 \mu } k_{2 \nu }+k_{2 \mu } k_{3 \nu }\right) q_0^2+ { a_{22}} q_0 q_3 k_{1,\mu } k_{1,\nu }+ { b_{22}} q_0 q_2 \left(k_{2 \nu } k_{1,\mu }+k_{2 \mu } k_{1,\nu }\right)+ { c_{22}} q_0^2 \left(k_{4 \nu } k_{1,\mu }+k_{4 \mu } k_{1,\nu }\right)\nonumber\\
 \er
 \br
 QQ[k_{2 \mu } q_1^2 k_{1,\nu }]&=&q_0 [  d_{23a} (k_{3 \mu } k_{2 \nu }-k_{2 \mu } k_{3 \nu }) q_0+  d_{23s} (k_{3 \mu } k_{2 \nu }+k_{2 \mu } k_{3 \nu } q_0+  a_{23} q_3 k_{1,\mu } k_{1,\nu }\nonumber\\
 &+&  b_{23a} q_2 (-k_{2 \nu } k_{1,\mu }+k_{2 \mu } k_{1,\nu })+  b_{23s}q_2 (k_{2 \nu } k_{1,\mu }+k_{2 \mu } k_{1,\nu })+  c_{23a} q_0 (-k_{4 \nu } k_{1,\mu }+k_{4 \mu } k_{1,\nu }) \nonumber\\&+&  c_{23s} q_0 (k_{4 \nu } k_{1,\mu }+k_{4 \mu } k_{1,\nu }) ]
   \nonumber\\
 \er
 \br
 {QQ}\left[q_1^2 \lambda _1 k_{1,\mu } k_{1,\nu }\right]&=& { b_{32}} k_{2 \mu } k_{2 \nu } q_0^2 \lambda _1+ { c_{32}} k_{3 \nu } q_0^2 \lambda _1 k_{1,\mu }+ { b_{32}} k_{2 \nu } q_0^2 \lambda _2 k_{1,\mu }+ { c_{32}} k_{3 \mu } q_0^2 \lambda _1 k_{1,\nu }+ { b_{32}} k_{2 \mu } q_0^2 \lambda _2 k_{1,\nu }\nonumber\\&+& { a_{32}} q_0 q_2 \lambda _1 k_{1,\mu } k_{1,\nu }+ { c_{32}} q_0^2 \lambda _3 k_{1,\mu } k_{1,\nu }\nonumber\\
 {QQ}\left[q_1^2 \lambda _2 k_{1,\nu }\right]&=&q_0 (- { c_{23a}} k_{4 \nu } q_0 \lambda _1+ { c_{23s}} k_{4 \nu } q_0 \lambda _1- { b_{23a}} k_{2 \nu } q_2 \lambda _1+ { b_{23s}} k_{2 \nu } q_2 \lambda _1- { d_{23a}} k_{3 \nu } q_0 \lambda _2+ { d_{23s}} k_{3 \nu } q_0 \lambda _2\nonumber \\&+& { d_{23a}} k_{2 \nu } q_0 \lambda _3+ { d_{23s}} k_{2 \nu } q_0 \lambda _3+ { a_{23}} q_3 \lambda _1 k_{1,\nu }+ { b_{23a}} q_2 \lambda _2 k_{1,\nu }+ { b_{23s}} q_2 \lambda _2 k_{1,\nu }+ { c_{23a}} q_0 \lambda _4 k_{1,\nu }\nonumber\\&+& { c_{23s}} q_0 \lambda _4 k_{1,\nu })\nonumber\\
 {QQ}[k_{2 \mu } q_1^2 \lambda _1]&=&q_0 ( { c_{23a}} k_{4 \mu } q_0 \lambda _1+ { c_{23s}} k_{4 \mu } q_0 \lambda _1+ { b_{23a}} k_{2 \mu } q_2 \lambda _1+ { b_{23s}} k_{2 \mu } q_2 \lambda _1+ { d_{23a}} k_{3 \mu } q_0 \lambda _2+ { d_{23s}} k_{3 \mu } q_0 \lambda _2 \nonumber \\&-& { d_{23a}} k_{2 \mu } q_0 \lambda _3+ { d_{23s}} k_{2 \mu } q_0 \lambda _3+ { a_{23}} q_3 \lambda _1 k_{1,\mu }- { b_{23a}} q_2 \lambda _2 k_{1,\mu }+ { b_{23s}} q_2 \lambda _2 k_{1,\mu }- { c_{23a}} q_0 \lambda _4 k_{1,\mu }\nonumber\\&+& { c_{23s}} q_0 \lambda _4 k_{1,\mu })\nonumber\\ & &
\er

{\bf Two index $q^3$}
\br
 {QQ}\left[q_1^3 k_{1,\mu } k_{1,\nu }\right]&=& { d_{24}} \left(k_{3 \mu } k_{2 \nu }+k_{2 \mu } k_{3 \nu }\right) q_0^3+ { a_{24}} q_0^2 q_3 k_{1,\mu } k_{1,\nu }+ { b_{24}} q_0^2 q_2 \left(k_{2 \nu } k_{1,\mu }+k_{2 \mu } k_{1,\nu }\right)+ { c_{24}} q_0^3 \left(k_{4 \nu } k_{1,\mu }+k_{4 \mu } k_{1,\nu }\right)\nonumber\\
 {QQ}\left[q_1^3 \lambda _1 k_{1,\nu }\right]&=& { c_{24}} k_{4 \nu } q_0^3 \lambda _1+ { b_{24}} k_{2 \nu } q_0^2 q_2 \lambda _1+ { d_{24}} k_{3 \nu } q_0^3 \lambda _2+ { d_{24}} k_{2 \nu } q_0^3 \lambda _3+ { a_{24}} q_0^2 q_3 \lambda _1 k_{1,\nu }+ { b_{24}} q_0^2 q_2 \lambda _2 k_{1,\nu }\nonumber\\&+& { c_{24}} q_0^3 \lambda _4 k_{1,\nu }\nonumber \\ & &
\er
{\bf One index}

\br
 {QQ}\left[k_{4 \mu } q_0 q_1\right]&=& { a_{1}} k_{5,m} q_0^2+ {b1} k_{3 \mu } q_0 q_2+ {c1} k_{2 \mu } q_0 q_3+ {d1} q_{4} q_0 k_{1,\mu }+ {e1} q_2^2 k_{1,\mu }\nonumber\\
 {QQ}\left[k_{3 \mu } q_1^2\right]&=& { a_{11}1} k_{5,m} q_0^2+ { b_{11}} k_{3 \mu } q_0 q_2+ { c_{11}} k_{2 \mu } q_0 q_3+ { d_{11}} q_{4} q_0 k_{1,\mu }+ { e_{11}} q_2^2 k_{1,\mu }\nonumber\\
 {QQ}\left[q_1 q_3 k_{1,\mu }\right]&=& { a_{13}} k_{5,m} q_0^2+ { b_{13}} k_{3 \mu } q_0 q_2+ { c_{13}} k_{2 \mu } q_0 q_3+ { d_{13}} q_{4} q_0 k_{1,\mu }+ { e_{13}} q_2^2 k_{1,\mu }\nonumber\\
 {QQ}\left[k_{2 \mu } q_1^3\right]&=& { a_{14}} k_{5 \mu } q_0^3+ { b_{14}} k_{3 \mu } q_0^2 q_2+ { c_{14}} k_{2 \mu } q_0^2 q_3+ { e_{14}} q_0 q_2^2 k_{1,\mu }+ { d_{14}} q_0^2 q_4 k_{1,\mu }\nonumber\\
 {QQ}\left[q_1^4 k_{1,\mu }\right]&=& { a_{16}} k_{5 \mu } q_0^4+ { b_{16}} k_{3 \mu } q_0^3 q_2+ { c_{16}} k_{2 \mu } q_0^3 q_3+ { e_{16}} q_0^2 q_2^2 k_{1,\mu }+ { d_{16}} q_0^3 q_4 k_{1,\mu }\nonumber\\
 {QQ}\left[q_1^4 \lambda _1\right]&=& { e_{16}} q_0^2 q_2^2 \lambda _1+ { d_{16}} q_0^3 q_4 \lambda _1+ { c_{16}} q_0^3 q_3 \lambda _2+ { b_{16}} q_0^3 q_2 \lambda _3+ { a_{16}} q_0^4 \lambda _5\nonumber\\
 {QQ}\left[q_1 q_2 \lambda _2\right]&=& { e_{12}} q_2^2 \lambda _1+ { d_{12}} q_0 q_4 \lambda _1+ { c_{12}} q_0 q_3 \lambda _2+ { b_{12}} q_0 q_2 \lambda _3+ {a12} q_0^2 \lambda _5\nonumber\\
 {QQ}\left[q_0 q_1 \lambda _4\right]&=& {e1} q_2^2 \lambda _1+ {d1} q_0 q_4 \lambda _1+ {c1} q_0 q_3 \lambda _2+ {b1} q_0 q_2 \lambda _3+ {a1} q_0^2 \lambda _5\nonumber\\
 {QQ}\left[q_1^2 \lambda _3\right]&=& { e_{11}} q_2^2 \lambda _1+ { d_{11}} q_0 q_4 \lambda _1+ { c_{11}} q_0 q_3 \lambda _2+ { b_{11}} q_0 q_2 \lambda _3+ { a_{11}} q_0^2 \lambda _5\nonumber\\
 {QQ}\left[q_1 q_3 \lambda _1\right]&=& { e_{13}} q_2^2 \lambda _1+ { d_{13}} q_0 q_4 \lambda _1+ { c_{13}} q_0 q_3 \lambda _2+ { b_{13}} q_0 q_2 \lambda _3+ { a_{13}} q_0^2 \lambda _5\nonumber\\
 {QQ}\left[q_1^3 \lambda _2\right]&=& { e_{14}} q_0 q_2^2 \lambda _1+ { d_{14}} q_0^2 q_4 \lambda _1+ { c_{14}} q_0^2 q_3 \lambda _2+ { b_{14}} q_0^2 q_2 \lambda _3+ { a_{14}} q_0^3 \lambda _5\nonumber\\
 {QQ}\left[q_1^2 q_2 \lambda _1\right]&=& { e_{15}} q_0 q_2^2 \lambda _1+ { d_{15}} q_0^2 q_4 \lambda _1+ { c_{15}} q_0^2 q_3 \lambda _2+ { b_{15}} q_0^2 q_2 \lambda _3+ { a_{15}} q_0^3 \lambda _5
\er

{\bf No index}

\br
 {QQ}\left[q_1 q_4\right]&=& { B_1} q_2 q_3+ { A_1} q_0 q_5\nonumber\\
 {QQ}\left[q_1 q_2^2\right]&=& { B_2} q_0 q_2 q_3+ {A2} q_0^2 q_5\nonumber\\
 {QQ}\left[q_1^2 q_3\right]&=& { B_3} q_0 q_2 q_3+ {A3} q_0^2 q_5\nonumber\\
 {QQ}\left[q_1^3 q_2\right]&=&
 { B_4} q_0^2 q_2 q_3+ {A4} q_0^3 q_5\nonumber\\
 {QQ}\left[q_1^5\right]&=& { B_5} q_0^3 q_2 q_3+ { A_5} q_0^4 q_5
\er

{\bf Final solution in terms of $ b_{2}, C_{2a},c_{23a},b_{22}$}

\[
\left\{ { a_{3p}}=-\frac{2}{5} (-2+ { b_{3}}), { b_{3p}}=2 (-1+ { b_{3}}), { c_{3}}=-\frac{3}{5} (-3+4  { b_{3}}), { a_{3}}=\frac{2- { b_{3}}}{5}\right\}
\]
\[
\left\{ { c_{2}}=\frac{1}{4} (-2-5  { b_{2}}), { d_{2}}=\frac{6- { b_{2}}}{4}, { a_{2}}=\frac{ { b_{2}}}{2}, { b_{3}}=\frac{-6+7  { b_{2}}}{-8+6  { b_{2}}}\right\}
\]
\[
\left\{ { B_{2a}}=\frac{1}{2}- { C_{2a}}, { D_{2a}}=\frac{1}{2}, { C_{2s}}=\frac{1}{8} (2-5  { b_{2}}), { D_{2s}}=\frac{2- { b_{2}}}{8}, { A_{2p}}=\frac{ { b_{2}}}{2}, { B_{2s}}=\frac{ { b_{2}}}{2}\right\}
\]
\[
\left\{ { a_{32}}=\frac{2}{5}-\frac{12  { c_{32}}}{5}, { b_{32}}=\frac{1}{5}-\frac{ { c_{32}}}{5}\right\}
\]
\[
\left\{ { c_{22}}=\frac{1}{8}-\frac{5  { b_{22}}}{4}, { d_{22}}=\frac{1}{8}-\frac{ { b_{22}}}{4}, { c_{32}}=-\frac{3- { b_{2}}}{2 (-4+3  { b_{2}})}+\frac{5  { b_{22}}}{-4+3  { b_{2}}}, { a_{22} a_{22}}=\frac{1}{2}+ { b_{22}}\right\}
\]
\[
\left\{ { b_{23a}}=\frac{1}{2}- { c_{23a}}, { d_{23a}}=\frac{1}{2}, { d_{23s}}=\frac{1}{8} (3- { b_{2}}-2  { b_{22}})\right\}\]
\[\left\{ { c_{23s}}=\frac{1}{8} (3-5  { b_{2}}-10  { b_{22}}), { a_{23}}=\frac{1}{2} (-1+ { b_{2}}+2  { b_{22}}), { b_{23s}}=\frac{1}{2} ( { b_{2}}+2  { b_{22}})\right\}
\]
\[
\left\{ { c_{24}}=\frac{1}{8} (7-30  { b_{22}}), { d_{24}}=\frac{3}{8}-\frac{3  { b_{22}}}{4}, { a_{24}}=-\frac{1}{2}+3  { b_{22}}, { b_{24}}=-\frac{1}{2}+3  { b_{22}}\right\}
\]
\[
\left\{ { a_{1}}=\frac{10+8  { b_{22}} (3-2  { C_{2a}})+4  { C_{2a}}-3  { b_{2}} (5+2  { C_{2a}})}{7+ { b_{22}} (6-4  { C_{2a}})+6  { C_{2a}}+ { b_{2}} (3+6  { C_{2a}})}, {b1}=\frac{(1+2  { b_{22}}) (-3+2  { C_{2a}})+ { b_{2}} (3+6  { C_{2a}})}{7+ { b_{22}} (6-4  { C_{2a}})+6  { C_{2a}}+ { b_{2}} (3+6  { C_{2a}})}\right\}\]
\[\left\{ {c1}=\frac{1+9  { b_{2}}-6  { b_{22}}+2  { C_{2a}}+6  { b_{2}}  { C_{2a}}+4  { b_{22}}  { C_{2a}}}{7+3  { b_{2}}+6  { b_{22}}+6  { C_{2a}}+6  { b_{2}}  { C_{2a}}-4  { b_{22}}  { C_{2a}}}, {d1}=\frac{-7+2  { C_{2a}}+ { b_{22}} (-6+4  { C_{2a}})+ { b_{2}} (-3+6  { C_{2a}})}{7+ { b_{22}} (6-4  { C_{2a}})+6  { C_{2a}}+ { b_{2}} (3+6  { C_{2a}})}\right\}\]
\[ {e1}=-\frac{(2+3  { b_{2}}) (-3+2  { C_{2a}})}{7+ { b_{22}} (6-4  { C_{2a}})+6  { C_{2a}}+ { b_{2}} (3+6  { C_{2a}})}
\]
\[
\left\{ { a_{11}}=\frac{4 (2+4  { C_{2a}}-3  { b_{2}} (1+2  { C_{2a}})+ { b_{22}} (-6+4  { C_{2a}}))}{7+ { b_{22}} (6-4  { C_{2a}})+6  { C_{2a}}+ { b_{2}} (3+6  { C_{2a}})}, { b_{11}}=\frac{2 (3+3  { b_{2}}+6  { b_{22}}-2  { C_{2a}}+6  { b_{2}}  { C_{2a}}-4  { b_{22}}  { C_{2a}})}{7+ { b_{22}} (6-4  { C_{2a}})+6  { C_{2a}}+ { b_{2}} (3+6  { C_{2a}})}\right\}\]
\[\left\{ { c_{11}}=\frac{2 (-1+ { b_{22}} (6-4  { C_{2a}})-2  { C_{2a}}+ { b_{2}} (3+6  { C_{2a}}))}{7+ { b_{22}} (6-4  { C_{2a}})+6  { C_{2a}}+ { b_{2}} (3+6  { C_{2a}})}, { d_{11}}=-\frac{16  { C_{2a}}}{7+ { b_{22}} (6-4  { C_{2a}})+6  { C_{2a}}+ { b_{2}} (3+6  { C_{2a}})}\right\}\]
\[ { e_{11}}=\frac{-5+ { b_{22}} (6-4  { C_{2a}})+14  { C_{2a}}+ { b_{2}} (3+6  { C_{2a}})}{7+ { b_{22}} (6-4  { C_{2a}})+6  { C_{2a}}+ { b_{2}} (3+6  { C_{2a}})}
\]
\[
 { a_{12}}={(2+8  { c_{23a}}+8  { b_{22}} (-3+6  { c_{23a}}-8  { C_{2a}})-3  { b_{2}} (1+4  { c_{23a}}-2  { C_{2a}})-4  { C_{2a}})\over (7+ { b_{22}} (6-4  { C_{2a}})+6  { C_{2a}}+ { b_{2}} (3+6  { C_{2a}}))}\]
\[ { b_{12}}={(-(1+2  { b_{22}}) (-5+8  { c_{23a}}-10  { C_{2a}})+ { b_{2}} (3+6  { C_{2a}}))\over (7+ { b_{22}} (6-4  { C_{2a}})+6  { C_{2a}}+ { b_{2}} (3+6  { C_{2a}}))}\]
\[ { c_{12}}={(3-2  { c_{23a}}+ { b_{2}} (3+6  { c_{23a}})+4  { C_{2a}}+ { b_{22}} (14-20  { c_{23a}}+24  { C_{2a}}))\over(7+ { b_{22}} (6-4  { C_{2a}})+6  { C_{2a}}+ { b_{2}} (3+6  { C_{2a}}))}\]
\[ { d_{12}}={(-2 (7+3  { b_{2}}+6  { b_{22}})  { c_{23a}}+2 (5+3  { b_{2}}+10  { b_{22}})  { C_{2a}})\over (7+ { b_{22}} (6-4  { C_{2a}})+6  { C_{2a}}+ { b_{2}} (3+6  { C_{2a}}))}\]
\[ { e_{12}}={(-3+6  { b_{22}}+16  { c_{23a}}+12  { b_{2}}  { c_{23a}}-2 (7+6  { b_{2}}+2  { b_{22}})  { C_{2a}})\over (7+ { b_{22}} (6-4  { C_{2a}})+6  { C_{2a}}+ { b_{2}} (3+6  { C_{2a}}))}
\]
\[
 { a_{13} }={(6+3  { b_{2}} (-3+4  { c_{23a}}-2  { C_{2a}})+4  { C_{2a}}-8 ( { c_{23a}}+6  { b_{22}}  { c_{23a}}+2  { b_{22}}  { C_{2a}}))\over(7+ { b_{22}} (6-4  { C_{2a}})+6  { C_{2a}}+ { b_{2}} (3+6  { C_{2a}}))}\]
\[ { b_{13}}=\frac{-6+8  { c_{23a}}-4  { C_{2a}}+4  { b_{22}} (-1+4  { c_{23a}}+2  { C_{2a}})}{7+ { b_{22}} (6-4  { C_{2a}})+6  { C_{2a}}+ { b_{2}} (3+6  { C_{2a}})}\]
\[ { c_{13}}={(2 (1+ { c_{23a}}+ { C_{2a}}+3  { b_{2}} (1- { c_{23a}}+ { C_{2a}})+2  { b_{22}} (1+5  { c_{23a}}+ { C_{2a}})))\over (7+ { b_{22}} (6-4  { C_{2a}})+6  { C_{2a}}+ { b_{2}} (3+6  { C_{2a}}))}\]
\[
 { d_{13}}=\frac{2 ((7+3  { b_{2}}+6  { b_{22}})  { c_{23a}}+(1+3  { b_{2}}+2  { b_{22}})  { C_{2a}})}{7+ { b_{22}} (6-4  { C_{2a}})+6  { C_{2a}}+ { b_{2}} (3+6  { C_{2a}})}\]
\[
 { e_{13}}={(5+6  { b_{2}}+6  { b_{22}}-16  { c_{23a}}-12  { b_{2}}  { c_{23a}}+2  { C_{2a}}-4  { b_{22}}  { C_{2a}})\over (7+3  { b_{2}}+6  { b_{22}}+6  { C_{2a}}+6  { b_{2}}  { C_{2a}}-4  { b_{22}}  { C_{2a}})}
\]
\[
 { a_{14}}={(13+24  { c_{23a}}+2  { b_{22}} (-33+72  { c_{23a}}-74  { C_{2a}})-6  { b_{2}} (1+6  { c_{23a}}-2  { C_{2a}})+2  { C_{2a}})\over (7+ { b_{22}} (6-4  { C_{2a}})+6  { C_{2a}}+ { b_{2}} (3+6  { C_{2a}}))}\]
\[
 { b_{14}}={(2 ( { b_{2}} (3+6  { C_{2a}})+4 (1-3  { c_{23a}}+2  { C_{2a}}+ { b_{22}} (3-6  { c_{23a}}+6  { C_{2a}}))))\over (7+ { b_{22}} (6-4  { C_{2a}})+6  { C_{2a}}+ { b_{2}} (3+6  { C_{2a}}))}\]
\[
 { c_{14}}={(-5-6  { c_{23a}}+3  { b_{2}} (1+6  { c_{23a}}-2  { C_{2a}})-2  { C_{2a}}+30  { b_{22}} (1-2  { c_{23a}}+2  { C_{2a}}))\over (7+ { b_{22}} (6-4  { C_{2a}})+6  { C_{2a}}+ { b_{2}} (3+6  { C_{2a}}))}\]
\[
 { d_{14}}={(-6 (7+3  { b_{2}}+6  { b_{22}})  { c_{23a}}+4 (4+3  { b_{2}}+12  { b_{22}})  { C_{2a}})\over (7+ { b_{22}} (6-4  { C_{2a}})+6  { C_{2a}}+ { b_{2}} (3+6  { C_{2a}}))}\]
\[
 { e_{14}}={(-9+18  { b_{22}}+12 (4+3  { b_{2}})  { c_{23a}}-2 (13+12  { b_{2}}+6  { b_{22}})  { C_{2a}})\over (7+ { b_{22}} (6-4  { C_{2a}})+6  { C_{2a}}+ { b_{2}} (3+6  { C_{2a}}))}
\]
\[
 { a_{15}}={(11-8  { c_{23a}}+3  { b_{2}} (-1+4  { c_{23a}}-2  { C_{2a}})+14  { C_{2a}}-6  { b_{22}} (7+8  { c_{23a}}+6  { C_{2a}}))\over (7+ { b_{22}} (6-4  { C_{2a}})+6  { C_{2a}}+ { b_{2}} (3+6  { C_{2a}}))}\]
\[
 { b_{15}}=\frac{4 (-1+2  { c_{23a}}-2  { C_{2a}}+ { b_{22}} (2+4  { c_{23a}}+4  { C_{2a}}))}{7+ { b_{22}} (6-4  { C_{2a}})+6  { C_{2a}}+ { b_{2}} (3+6  { C_{2a}})}\]
\[
 { c_{15}}={(-1+2  { c_{23a}}-2  { C_{2a}}+ { b_{2}} (3-6  { c_{23a}}+6  { C_{2a}})+2  { b_{22}} (11+10  { c_{23a}}+6  { C_{2a}}))\over(7+ { b_{22}} (6-4  { C_{2a}})+6  { C_{2a}}+ { b_{2}} (3+6  { C_{2a}}))}\]
\[
 { d_{15}}=\frac{2 ((7+3  { b_{2}}+6  { b_{22}})  { c_{23a}}-4  { C_{2a}}+8  { b_{22}}  { C_{2a}})}{7+ { b_{22}} (6-4  { C_{2a}})+6  { C_{2a}}+ { b_{2}} (3+6  { C_{2a}})}\]
\[
 { e_{15}}={(1-16  { c_{23a}}+6  { b_{22}} (3-2  { C_{2a}})+10  { C_{2a}}+ { b_{2}} (3-12  { c_{23a}}+6  { C_{2a}}))\over(7+ { b_{22}} (6-4  { C_{2a}})+6  { C_{2a}}+ { b_{2}} (3+6  { C_{2a}}))}
\]
\[
 { a_{16}}={-(25+3  { b_{2}}-102  { b_{22}}+26  { C_{2a}}+6  { b_{2}}  { C_{2a}}-124  { b_{22}}  { C_{2a}})\over(-7-3  { b_{2}}-6  { b_{22}}-6  { C_{2a}}-6  { b_{2}}  { C_{2a}}+4  { b_{22}}  { C_{2a}})}\]
\[\left\{ { b_{16}}=\frac{4 (-1+6  { b_{22}}-2  { C_{2a}}+12  { b_{22}}  { C_{2a}})}{7+3  { b_{2}}+6  { b_{22}}+6  { C_{2a}}+6  { b_{2}}  { C_{2a}}-4  { b_{22}}  { C_{2a}}}, { c_{16}}=-\frac{8 (-1+6  { b_{22}}- { C_{2a}}+6  { b_{22}}  { C_{2a}})}{-7-3  { b_{2}}-6  { b_{22}}-6  { C_{2a}}-6  { b_{2}}  { C_{2a}}+4  { b_{22}}  { C_{2a}}}\right\} 
\]
\[\left\{ { d_{16} }=\frac{8 (- { C_{2a}}+6  { b_{22}}  { C_{2a}})}{7+3  { b_{2}}+6  { b_{22}}+6  { C_{2a}}+6  { b_{2}}  { C_{2a}}-4  { b_{22}}  { C_{2a}}}, { e_{16}}=-\frac{2 (3-18  { b_{22}}-2  { C_{2a}}+12  { b_{22}}  { C_{2a}})}{7+3  { b_{2}}+6  { b_{22}}+6  { C_{2a}}+6  { b_{2}}  { C_{2a}}-4  { b_{22}}  { C_{2a}}}\right\}
\]
\[
\left\{ { A_1}=-\frac{-2+3  { b_{2}}}{1+6  { b_{22}}}, { B_1}=-\frac{1-3  { b_{2}}-6  { b_{22}}}{1+6  { b_{22}}}\right\}
\]
\[
\left\{ { A_2}=-\frac{2 (-1+6  { b_{22}})}{1+6  { b_{22}}}, { B_2}=-\frac{1-18  { b_{22}}}{1+6  { b_{22}}}\right\}
\]
\[
\left\{ { A_3}=-\frac{-4+3  { b_{2}}+12  { b_{22}}}{1+6  { b_{22}}}, { B_3}=\frac{3 (-1+ { b_{2}}+6  { b_{22}})}{1+6  { b_{22}}}\right\};
\]
\[
\left\{ { A_4}=-\frac{-7+30  { b_{22}}}{1+6  { b_{22}}}, { B_4}=\frac{6 (-1+6  { b_{22}})}{1+6  { b_{22}}}\right\};
\]
\[
\left\{ { A_5}=-\frac{-13-3  { b_{2}}+54  { b_{22}}}{1+6  { b_{22}}}, { B_5}=-\frac{3 (4+ { b_{2}}-20  { b_{22}})}{1+6  { b_{22}}}\right\};
\]

{\bf Dimensional Reduction compatibility fixes the parameters to be:}

\[\{ { b_{2}}=-18/7, { C_{2a}}=16/7, { c_{23}a}=16/7\}; { b_{22}}=(2+ { b_{2}})/8
\]

This fixes all the constants. Once again the set of equations is a highly overdetermined set and it is somewhat surprising that a solution exists at all. It would be interesting to find the underlying logic.

\section{K-constraints}
\label{appenc}
\setcounter{equation}{0}

We derive the K-constraints that occur in the free equations.

For the free part of the equation we do not need the individual $K_{[n]_i;[\mb]_j\mu}$. We can write ${\cal L}$ in terms of $Y_{n;\mb}^\mu$. Thus the coefficient of $Y_{n;\mb}^\mu$ is $\sum_{i,j}K_{[n]_i;[\mb]_j\mu} = \tilde  K_{n;\mb\mu}$ as defined in (\ref{intrel}).

The general case involves combining the following two terms:
\[
\int dz'\int dz''\dot G(z',z'') {\p \over \p \xn}{\p\over \p Y_m(u)}\delta(u-z') {\p \over \p x_\mb}{\p\over \p Y_\mb(u)}\delta(u-z''){\cal L}~+z'\leftrightarrow z''
\]
\be  \label{A12}
=
\int dz'{d\over d\tau}( [\lan Y_n (z')Y_\mb (z')\ran + \lan Y_\mb(z') Y_n(z')\ran](ik_n.ik_\mb) {\cal L}
\ee
and
\[
\int dz' \dot G(z',z'') {\pp\over \p \xn \p x_\mb} {\p\over \p Y_{n;\mb}}\delta(u-z'){\p\over \p Y}\delta(u-z''){\cal L} ~+z'\leftrightarrow z''
\]
\be	\label{A13}
=
\int dz' {d\over d\tau} [\lan Y_{n;\mb}(z') Y(z')\ran + \lan Y(z') Y_{n;\mb}(z')\ran](i\tilde K_{n;\mb}.i\ko) {\cal L} 
\ee

Now {\em if} 
\be	\label{Knm}
(ik_n.ik_\mb) {\cal L}=(i\tilde K_{n;\mb}.i\ko) {\cal L}
\ee
{\em then} we can combine the two terms, (\ref{A12}) and (\ref{A13}),   and write 
\[
\int dz' [{d\over d\tau} {\pp\over \p \xn \p x_{\mb}} \lan Y(z')Y(z')\ran] (-k_n.k_\mb){\cal L}
\]
\be
=\int dz' ~\dot G(z',z') (-k_n.k_\mb){\pp\over \p \xn \p x_{\mb}}{\cal L}
\ee

Since the $\tilde K_{n;\mb\mu}$ are made of the usual loop variables and no new degrees are involved, the K-constraints (\ref{Knm}) would seem
to reduce the number of independent degrees of freedom. However we also have the option of adding  {\em one new} loop variable $k_{n;\mb\mu}$ , (with $\mu$ chosen to be $D$, so we can call it $q_{n;\mb}$ )to $\tilde K_{n;\mb\mu}$ so that the constraint plays the role of determining this variable. $q_{n;\mb}$ should be defined to have the same gauge transformation as $\tilde K_{n;\mb\mu}$, viz: 

\[
q_{n;\mb}\rightarrow q_{n;\mb}+ \la _p q_{n-p;\mb} + \bar \la _p q_{n;\mb-p}
\]
Then the constraint does not affect the degrees of freedom count. 

We have 
\[
\tilde K_{n;\mb\mu} = \bar q_n k_{\mu\mb}+ \bar q_\mb k_{n\mu} - \bar q_n \bar q_\mb \kom
\]
\[
\tilde Q_{n;\mb}= {q_n\over q_0}q_\mb + {q_\mb\over q_0} q_n - {q_n\over q_0}{q_\mb\over q_0}q_0 +q_{n;\mb}={q_n q_\mb\over q_0}+q_{n;\mb}
\]

The constraint (\ref{Knm}) becomes
\[
q_{n;\mb}= k_n.k_{\mb} - \bar q_n k_\mb.\ko + \bar q_\mb k_n.\ko - \bar q_n \bar q_\mb \ko^2
\]
thus fixing $q_{n;\mb}$ in terms of the others. This idea was made use of in the discussion on massless spin 2 level. 

\section{LV to OC map}
\label{append}
\setcounter{equation}{0}

We give below the relations between the coefficients defining the map from the LV fields to the OC fields that were introduced  in Section 8 and reproduced below for convenience.
\br
\Phi_{\mu\nu\rho} &=& f_1 S_{111\mu\nu\rho} + f_2 S_{3(\mu}\eta_{\nu\rho)} +f_3 S_{A(\mu}\eta_{\nu\rho)}+f_4 p_{(\mu}S_{\nu\rho)}\nonumber \\
B_{\mu\nu}& =& b_1 S_{\mu\nu}+b_2 p_{(\mu}S_{3\nu)} +b_3 p_{(\mu}S_{A\nu)} +b_4 S_3 \eta_{\mu\nu}\nonumber \\
C_{\mu\nu}  &=&c_1 A_{\mu\nu}+ c_2 p_{[\mu}S_{3\nu]} + c_3 p_{[\mu}S_{A\nu]} \nonumber \\
A_\mu&=& a_1 S_{3\mu} + a_2 S_{A\mu} + a_3 p_\mu S_3
\er

They are obtained by requiring that the constraints and gauge transformations map into each other. For the gauge transformation only the tensor and vector parameters were included in the analysis. These are enough to fix $D$ and $\qo$. 

The relation between the gauge parameters in the two formalisms turn out to be:
\br
(b_1+2b_2)\Lambda_{S\mu} + b_3 \Lambda_{A\mu} &=& \eps _{S\mu} \nonumber \\
(c_1+c_3) \Lambda_{A\mu}+ 2 c_2 \Lambda_{S\mu} &=& \eps_{A\mu}\nonumber \\
b_1 \Lambda_{111\mu\nu} &=& \eps_{111\mu\nu}
\er
We have not worked out the dependence on the scalar parameter in the above. The various equations obtained are:

\br
{-q_0^2 f_1 + f_2 (D + 2) - q_0^2 f_4)\over12} + {(-b_1 q_0^2 - (b_2 + c_2) q_0^2)\over 3} + 
  {a_1\over 2} &=& 0\nonumber \\
  {f_3 (D + 2)\over 12} + {(-c_1 q_0^2 - (b_3 + c_3) q_0^2)\over 3} + {a_2\over 2} &=& 0\nonumber \\
{-f_4 q_0\over 12} + {(-(b_2 - c_2) q_0 + b_4)\over 3 }+ {a_3\over 2} &=& 0 \nonumber \\ 
b_1 q_0^2 - 3 a_1 + (b_2 - c_2) q_0^2 &=& 0 \nonumber \\
3 a_2 - (b_3 - c_3) q_0^2 + c_1 q_0^2 &=& 0\nonumber \\
3 a_3 + b_4 - (b_2 + c_2) q_0 &=& 0\nonumber \\
{f_1 q_0^2\over 4} - b_1 + {f_4 q_0^2 \over 4} &=& 0\nonumber \\
 b_3 +{ f_3\over 4} &=& 0\nonumber \\
 b_4 - {f_2 q_0 \over 4} &=& 0\nonumber \\
b_2 + {f_2\over 4} -{ q_0^2 f_4 \over 4} &=& 0\nonumber \\
 -3 a_3 q_0^2 + 2 D b_4 - 3 a_1 q_0 - 2 b_1 q_0 - 4 b_2 q_0 &=& 0\nonumber \\
3 (b_1 + 2 b_2) - 2 c_2 - 2 a_1 &=& 0\nonumber \\
3 b_3 - (c_1 + c_3) - a_2 &=& 0\nonumber \\
c_1 + c_3 - b_3 - f_3 &=& 0\nonumber \\
2 c_2 + b_1 + 2 b_2 - 2 f_2 &=& 0\nonumber \\
f_1 + f_4 - b_1 &=& 0
  \er
  
  The solution is 
  \[D=26, \qo =2
  \]
  \[
  a_2 = 0,~~ a1 = -((44 b_2)/3),~~ b_4 = -2 b_2,~~ f_3 = 0,~~ a_3 = (
 16 b_2)/9,~~  f2 = -4 b_2,~~ 
 \]
 \[f_1 = -((34 b_2)/
  3),~~ c_1 = -c_3,~~ b_3 = 0,~~ b_1 = -((34 b_2)/3),~~ 
  \]
  \be c_2 = (
 2 b_2)/3,~~ f_4 = 0,~~ 
 \ee
 
 We see that $D$ and $\qo$ are fixed by this analysis. To fix the remaining coefficients one needs to include the scalar gauge transformation parameters in the analysis. 
\end{appendices}

\end{document}